\documentclass[useamsfonts]{pasj01}
\definecolor{royalblue}{rgb}{0.25, 0.41, 0.88}
\definecolor{brickred}{rgb}{0.8, 0.25, 0.33}
\usepackage{graphicx}
\usepackage{latexsym}
\usepackage{url}
\usepackage{savesym}
\savesymbol{leftroot}
\savesymbol{uproot}
\savesymbol{iint}
\savesymbol{iiint}
\savesymbol{iiiint}
\savesymbol{idotsint}
\usepackage{amsmath,physics}

\makeatletter
\DeclareFontFamily{U}{mathb}{}
\DeclareFontShape{U}{mathb}{m}{n}{
  <-5.5> mathb5
  <5.5-6.5> mathb6
  <6.5-7.5> mathb7
  <7.5-8.5> mathb8
  <8.5-9.5> mathb9
  <9.5-11.5> mathb10
  <11.5-> mathbb12
}{}
\DeclareRobustCommand{\sqcdot}{%
  \mathbin{\text{\usefont{U}{mathb}{m}{n}\symbol{"0D}}}%
}
\makeatother

\renewcommand\labelenumi{(\roman{enumi})}
\renewcommand\theenumi\labelenumi
\newcommand{\hscPipe}{\texttt{hscPipe}}

\newcommand{\xlrv}[1]{#1}

\newcommand{\hmrv}[1]{#1}

\newcommand{\reGauss}{\texttt{reGauss}}
\newcommand{\galsim}{\texttt{GalSim}}

\newcommand{\res}{\mathcal{R}}
\newcommand{\dNNz}{\texttt{dNNz}}
\newcommand{\DEmP}{\texttt{DEmP}}
\newcommand{\mizuki}{\texttt{mizuki}}

\makeatletter
\let\old@ssect\@ssect 
\DeclareFontFamily{U}{mathb}{}
\DeclareFontShape{U}{mathb}{m}{n}{
  <-5.5> mathb5
  <5.5-6.5> mathb6
  <6.5-7.5> mathb7
  <7.5-8.5> mathb8
  <8.5-9.5> mathb9
  <9.5-11.5> mathb10
  <11.5-> mathbb12
}{}
\makeatother

\usepackage{natbib}
\usepackage[pdfpagelabels=false]{hyperref}
\hypersetup{colorlinks=True,linkcolor={royalblue},citecolor={royalblue},urlcolor={brickred},breaklinks=true}

\makeatletter
\def\@ssect#1#2#3#4#5#6{%
  \NR@gettitle{#6}
  \old@ssect{#1}{#2}{#3}{#4}{#5}{#6}
}
\makeatother

\begin{document}


\title{The three-year shear catalog of the Subaru Hyper Suprime-Cam
SSP Survey}
\author{Xiangchong Li\altaffilmark{1,2,11}}
\author{Hironao Miyatake\altaffilmark{3,4,5,1,6}}
\author{Wentao Luo\altaffilmark{7,1}}
\author{Surhud More\altaffilmark{8,1}}
\author{Masamune Oguri\altaffilmark{9,2,1}}
\author{Takashi Hamana\altaffilmark{10}}
\author{Rachel Mandelbaum\altaffilmark{11}}
\author{Masato Shirasaki\altaffilmark{10,12}}
\author{Masahiro Takada\altaffilmark{1}}
\author{Robert Armstrong\altaffilmark{13}}
\author{Arun Kannawadi\altaffilmark{14}}
\author{Satoshi Takita\altaffilmark{15,10}}
\author{Satoshi Miyazaki\altaffilmark{10,16}}
\author{Atsushi J. Nishizawa\altaffilmark{4}}
\author{Andrés A. Plazas Malagón\altaffilmark{14}}
\author{Michael A. Strauss\altaffilmark{14}}
\author{Masayuki Tanaka\altaffilmark{10,16}}
\author{Naoki Yoshida\altaffilmark{1,2,9}}

\altaffiltext{1}{Kavli Institute for the Physics and Mathematics of the
Universe (Kavli IPMU, WPI), UTIAS, The
University of Tokyo, Chiba 277-8583, Japan}
\altaffiltext{2}{Department of Physics, The University of Tokyo, Tokyo
113-0033, Japan}
\altaffiltext{3}{Kobayashi-Maskawa Institute for the Origin of Particles and
the Universe (KMI), Nagoya University, Nagoya, 464-8602, Japan}
\altaffiltext{4}{Institute for Advanced Research, Nagoya University, Nagoya,
464-8601, Japan}
\altaffiltext{5}{Division of Physics and Astrophysical Science, Graduate School
of Science, Nagoya University, Nagoya 464-8602, Japan}
\altaffiltext{6}{Jet Propulsion Laboratory, California Institute of Technology,
Pasadena, CA 91109, USA}
\altaffiltext{7}{CAS Key Laboratory for Research in Galaxies and Cosmology,
University of Science and Technology of China, Hefei, Anhui 230026, China}
\altaffiltext{8}{The Inter-University Center for Astronomy and Astro-physics,
Post bag 4, Ganeshkhind, Pune, 411007, India}
\altaffiltext{9}{Research Center for the Early Universe, University of Tokyo,
Tokyo 113-0033, Japan}
\altaffiltext{10}{National Astronomical Observatory of Japan, Mitaka, Tokyo
181-8588, Japan}
\altaffiltext{11}{McWilliams Center for Cosmology, Department of Physics,
Carnegie Mellon University, Pittsburgh, PA 15213, USA}
\altaffiltext{12}{The Institute of Statistical Mathematics, Tachikawa, Tokyo
190-8562, Japan}
\altaffiltext{13}{Lawrence Livermore National Laboratory, Livermore, CA 94550,
USA}
\altaffiltext{14}{Department of Astrophysical Sciences, Princeton University, 4
Ivy Lane, Princeton, NJ 08544, USA}
\altaffiltext{15}{Institute of Astronomy, University of Tokyo, 2-21-1 Osawa,
Mitaka, Tokyo 181-0015, Japan}
\altaffiltext{16}{Department of Astronomy, School of Science, Graduate
University for Advanced Studies (SOKENDAI), 2-21-1, Osawa, Mitaka, Tokyo
181-8588, Japan}

\email{xiangchl@andrew.cmu.edu}
\KeyWords{Cosmology, Weak Gravitational Lensing, Catalog}

\maketitle

\begin{abstract}
We present the galaxy shear catalog that will be used for the three-year
cosmological weak gravitational lensing analyses using data from the Wide layer
of the Hyper Suprime-Cam (HSC) Subaru Strategic Program (SSP) Survey. The
galaxy shapes are measured from the $i$-band imaging data acquired from 2014 to
2019 and calibrated with image simulations that resemble the observing
conditions of the survey based on training galaxy images from the Hubble Space
Telescope in the COSMOS region. The catalog covers an area of 433.48 deg$^2$ of
the northern sky, split into six fields. The mean $i$-band seeing is 0.59
arcsec. With conservative galaxy selection criteria (e.g., $i$-band magnitude
brighter than 24.5), the observed raw galaxy number density is 22.9
arcmin$^{-2}$, and the effective galaxy number density is 19.9 arcmin$^{-2}$.
The calibration removes the galaxy property-dependent shear estimation bias to
a level: $|\delta m|<9\times 10^{-3}$. The bias residual $\delta m$ shows no
dependence on redshift in the range $0<z\leq 3$. We define the requirements for
cosmological weak lensing science for this shear catalog, and quantify
potential systematics in the catalog using a series of internal null tests for
systematics related to point-spread function modelling and shear estimation. A
variety of the null tests are statistically consistent with zero or within
requirements, but (i) there is evidence for PSF model shape residual
correlations; and (ii) star-galaxy shape correlations reveal additive
systematics. Both effects become significant on $>1$ degree scales and will
require mitigation during the inference of cosmological parameters using cosmic
shear measurements.
\end{abstract}

\section{Introduction}
\label{sec:Intro}

In the current standard structure formation paradigm (the $\Lambda$CDM model),
dark matter and dark energy constitute a large fraction (about 95\%) of the
total energy density of the Universe
\citep{2020A&A...641A...6P,2012ApJ...746...85S,SDSS-gglensDR7-Mandelbaum2013}.
Unveiling the nature of these two mysterious components, dark matter and dark
energy, is one of the most tantalizing problems in cosmology and physics, and
is one of the major goals for ongoing and upcoming wide-area galaxy surveys
(see \citealt{2013PhR...530...87W} for a review). Among different cosmological
probes, weak gravitational lensing provides us with a unique means of measuring
matter distribution (including dark matter) in the universe
\citep[e.g.][]{2018PASJ...70S..27M}, via the deflection of light due to the
gravitational potential field in cosmic structures along the line-of-sight,
which both magnifies and distorts galaxy shapes -- the so-called cosmological
weak lensing or cosmic shear (see \citealt{revRachel17} for a review). Since the
initial detections of cosmic shear
\citep{2000MNRAS.318..625B,2000A&A...358...30V,2001ApJ...552L..85R}, weak
lensing now has become one of the indispensable methods for precision
cosmology.

The standard method to measure cosmic shear is based on the auto-correlation of
galaxy shape distortions. When combined with photometric redshift information
of individual galaxies via their multi-color photometry, known as ``cosmic
shear tomography'', the cosmic shear correlation functions are very powerful at
measuring scale-dependent amplitudes and time evolution of matter clustering in
large-scale structure. These measurements are in turn used to place powerful
constraints on the present-day amplitudes of matter fluctuations, the matter
density (mostly dark matter), and the nature of dark energy \citep[see
e.g.,][]{2017MNRAS.465.1454H,2018PhRvD..98d3528T,cosmicShear_HSC1_Chiaki2019,
HSC1-cs-real,KiDS1000-CS,DESY3-cosmicshearS,DESY3-cosmicShearA}. The
galaxy-shear cross-correlation function, or galaxy-galaxy weak lensing, can be
combined with galaxy clustering to observationally disentangle galaxy bias
uncertainty and thus obtain useful constraints on the cosmological parameters
\citep[see
e.g.,][]{SDSS-gglensDR7-Mandelbaum2013,
2015ApJ...806....2M,2018PhRvD..98d3526A,2021A&A...646A.140H,HSC1-ggLensClustII}.
Furthermore, when combined with the redshift-space distortion effect due to
peculiar velocities of lens galaxies, properties of gravity (i.e.\ gravity
theory) on cosmological scales can be tested
\citep[e.g.][]{2016MNRAS.456.2806B,2017MNRAS.465.4853A}.

The current generation wide-area multi-color surveys that have weak lensing
among their primary science cases are: the Kilo-Degree
Survey\footnote{\url{http://kids.strw.leidenuniv.nl}} \citep[KiDS;][]{KIDS13},
the Dark Energy Survey\footnote{\url{https://www.darkenergysurvey.org}}
\citep[DES;][]{DES16}, and the survey that is the subject of this paper: the
Hyper Suprime-Cam survey\footnote{\url{https://hsc.mtk.nao.ac.jp/ssp/}}
\citep[HSC;][]{2018PASJ...70S...1M,HSC-SSP2018}. The unique aspect of the HSC
survey is its combination of depth and high-resolution imaging that gives it a
longer redshift baseline than the others. Hence the weak lensing information
obtained from the HSC survey is complementary to those of the KiDS and DES
surveys that probe weak lensing effects at lower redshifts, but over a wider
area than the current HSC survey does. In addition, the excellent image quality
in HSC should enable us to pin down sources of systematic uncertainties in weak
lensing shear. In the coming decade, three ultimate imaging surveys will become
available and promise to place further stringent constraints on cosmological
parameters including the nature of dark energy.  Those are the Euclid satellite
mission\footnote{\url{https://sci.esa.int/web/euclid}} \citep{Euclid2011}, Vera
C. Rubin Observatory's Legacy Survey of Space and Time
\footnote{\url{https://www.lsst.org}} \citep[LSST;][]{LSSTOverviwe2019}, and
the Nancy Grace Roman Space
Telescope\footnote{\url{https://roman.gsfc.nasa.gov}} \citep{WFIRST15}. Since
the HSC data is the deepest among the ongoing surveys, the HSC survey can be
considered as a precursor survey for LSST since they are both ground-based data
and share similarities in the depth and image quality. Hence it is important
and timely to assess and figure out whether the quality and issues of the HSC
data can meet requirements to use the weak lensing measurements for cosmology,
compared to the statistical errors of the current HSC data.

However, weak lensing shear is a tiny effect typically causing one percent
ellipticities in the observed galaxy images, which are smaller than the
root-mean-square (RMS) of intrinsic galaxy shapes. Thus the shear is only
measurable in a statistical sense. Hence an accurate weak lensing measurement
requires exquisite characterization of individual galaxy images as well as
control and calibrations of all observational effects such as atmospheric
effects (point-spread function and background noise) and  the detector noise.
It is important to ensure that residual systematic errors are well below the
statistical error floor so that any physical constraints obtained from the weak
lensing measurements are not biased.
Observationally there are several sources of systematic effects inherent in
characterizing galaxy shapes, even in a statistical sense:
(i) ``noise bias'' due to the non-linear impact of noise on shear estimation
\citep{noiseBiasRefregier2012,Z11};
(ii) ``model bias'' due to imperfect
assumptions about galaxy morphology
\citep[e.g.,][]{modelBias-Bernstein10};
(iii) ``weight bias'' caused by shear-dependent weighting
\citep[e.g.,][]{KIDS-shapeCalib-Conti2017};
(iv) ``selection bias'' originating from an improper treatment of selection
effects around cuts \citep[e.g.,][]{SDSS-shape-Mandelbaum2005};
(v) systematics related to blending of galaxy light profiles
\citep[e.g.,][]{FPFS-Li2018,metaDet-Sheldon2020};
(vi) mis-estimation of the point-spread function \citep[PSF; e.g.,][]{LuPSF17};
and (vii) other systematics from detector non-idealities -- e.g., ``tree rings'',
``edge distortions'' \citep{Plazas2014}, and brighter-fatter effects
\citep{Antilogus2014} -- and from the atmosphere -- e.g., differential
chromatic refraction \citep[DCR;][]{Plazas2012}.
There are other astrophysical uncertainties such as photometric redshift
errors, intrinsic alignments of galaxy shapes and the impact of baryonic
effects \citep{revRachel17}. In this paper we focus on the observational
effects in galaxy shape characterizations for weak lensing measurements.

Because of the systematics mentioned above, it is necessary to validate the
shear catalog generation pipeline using image simulations. To develop
simulations representative of the real data, the issue that arises here is how
to maximally represent the real observational conditions and the galaxy
properties in the HSC data. Much effort has been made to produce
simulations that faithfully represent the image characteristics that affect
shear estimation
\citep{HSC1-GREAT3Sim,KIDS-ImgSim-Kannawadi2019,DESY3-BlendshearCalib-MacCrann2021}.
Shear estimators must be calibrated if the biases discovered with image
simulations exceed the systematic error requirements of the weak lensing
survey. In addition, internal ``null tests'' related to galaxy and star shapes
within the shear catalog are important to uncover the signatures of the
aforementioned systematics
\citep[e.g.,][]{HSC1-shape,KiDS1000-catalog,DESY3-catalog}.

In this paper, we describe the process to generate the three-year shear catalog
for weak lensing statistics from the HSC-SSP S19A internal data release
(released in September 2019). First, we measure galaxy shapes using the
re-Gaussianization method \citep[\reGauss{};][]{Regaussianization}, and
calibrate the shear estimation bias using HSC-like galaxy image simulations
following the formalism of \citet{HSC1-GREAT3Sim}. We then calculate the
requirements for cosmological analysis based on the survey parameters. We
subsequently proceed with data quality control with ``null tests'' on the
catalog following \citet{HSC1-shape}, which include tests related to PSF
modelling, cross-correlations of galaxy shapes with random positions, star
positions and star shapes, and tests related to weak lensing mass maps.

The structure of the paper is outlined as follows. In
Section~\ref{sec:dataPipe}, we present the S19A internal HSC data release, and
outline the updates in the pipeline used to process the S19A data. In
Section~\ref{sec:catalog}, we calibrate \reGauss{} galaxy shapes with realistic
image simulations and characterize the three-year HSC shear catalog. In
Section~\ref{sec:requirement}, we define the requirements for the shape catalog
on the PSF modelling and shear inference to ensure that the three-year weak
lensing science is minimally affected by the systematics we listed above. In
Section~\ref{sec:psftest}, we perform various systematic tests associated to
the PSF modelling to ensure the quality of PSF reconstruction and correction.
Finally, we conduct null tests on the shear catalog in
Section~\ref{sec:nullTest}, and summarize in Section~\ref{sec:summary}.

\section{HSC Data and Pipeline}
\label{sec:dataPipe}
The HSC instrument \citep{HSC-hardware-Furusawa2018,HSC-hardware-Miyazaki2018}
is a wide-field optical imager mounted on the $8.2$-meter Subaru Telescope. The
HSC-SSP \citep[][]{HSC-SSP2018} is a deep multi-band imaging survey with a
target area of $1400$ $\rm{deg}^2$ on the northern sky.  The HSC pipeline
\citep{HSC1-pipeline} is a fork of Rubin's LSST Science Pipelines
\citep{LSSTpipe-Bosch2019}; the fork is being developed to process the data
from the HSC-SSP survey, while an updated version of Rubin's LSST Science
Pipelines will be used for LSST.

The first public data release\footnote{see
https://hsc-release.mtk.nao.ac.jp/doc/ for HSC-SSP data releases.} of HSC data
\citep[PDR1,][]{HSC1-data} was based on the S15B internal data release
(released in January 2016) and included images and catalogs processed with
$\hscPipe~{\rm v}4$ \citep{HSC1-pipeline}. The first-year HSC shear catalog
\citep{HSC1-shape} was based on the S16A internal data release (released in
August 2016) and was also processed with $\hscPipe~{\rm v}4$.

The second public data release (PDR2) of images and catalogs was based on the
S18A internal data release (released in June 2018) processed with
$\hscPipe~{\rm v}6$ \citep{HSC2-data}.  There were major updates on the
pipeline from $\hscPipe~{\rm v}4$ to $\hscPipe~{\rm v}6$ as summarized in
\citet{HSC2-data}.

The shear catalog introduced in this paper is based on the S19A internal data
release (released in September 2019) acquired from March 2014 to April 2019.
The S19A images are processed with $\hscPipe~{\rm v}7$. Here we briefly
summarize the new features of $\hscPipe~{\rm v}7$ updated from $\hscPipe~{\rm
v}4$ that are important for weak lensing measurements. In addition, we
summarize the changes in the observing strategy. As our first-year shear
catalog helped to identify areas where progress was needed in the image
processing pipeline, we expect this paper to provide a snapshot of the current
state of the software pipeline, and to help in identifying further areas for
progress.

\begin{figure}
\begin{center}
\includegraphics[width=0.45\textwidth]{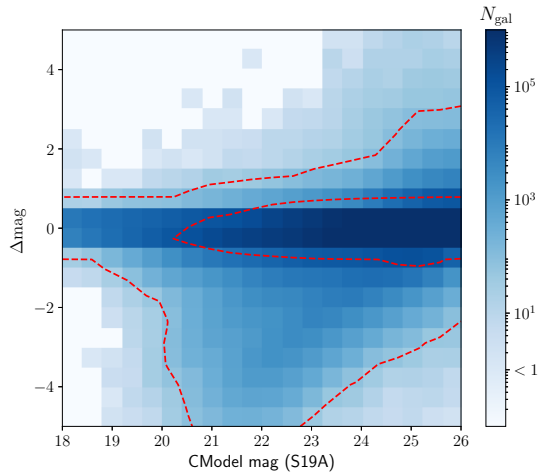}
\end{center}
\caption{
    $2$D histogram of the $i$-band magnitude difference and the S19A CModel
    magnitude.  The magnitude difference ($\Delta$mag) is defined as the S19A
    CModel minus the S16A CModel magnitude. Galaxies are matched between S19A
    and S16A within the first-year HSC weak lensing full depth full color
    region \citep{HSC1-shape} within $0\farcs5$. The contours are for galaxy
    numbers of $10^{2}$ and $10^{5}$, respectively.
    }
    \label{fig:dMagHist}
\end{figure}

\subsection{Improvements in PSF modelling}
\label{subsec:PSF_improvement}

The HSC pipeline uses a repackaged version of PSFEx \citep{PSFEx11} to estimate
point-spread function (PSF) models on single exposures, and the PSF models on
coadds are estimated using the PSF models from each exposure, while accounting
for the warping kernel used for image coaddition \citep{HSC1-pipeline}.

The PSFs on single exposures are modelled by PSFEx using a pixellated basis
function, and in principle the over-sampled PSF model can be shifted by
sub-pixel offsets using sinc interpolation. However, the Lanczos kernels,
employed by the original version of PSFEx in $\hscPipe~{\rm v}4$ to
approximate the sinc kernel caused problems for images with the ``very best
seeing''. As shown in Fig.~9 of \cite{HSC2-data}, the sizes of PSF models are
less than the sizes of observed stars by $0.4$\% for regions with seeing FWHM
of around $0\farcs5$.

For the second data release, as described in Section~4.6 in \citet{HSC2-data},
the pipeline resampled the PSF models by interpreting the PSF models as a
constant over each sub-pixel, rather than a continuous function sampled at the
pixel center. This mitigated the PSF model errors for images with the ``very
best seeing'', reducing the fractional size residual between PSF models and
observed stars from $\sim 0.4$\% to $\sim 0.1$\%. This new interpolation scheme
is subsequently applied in the S19A image processing.

\subsection{Improvements to the warping kernel}

In the coaddition process, each single CCD image is convolved with a warping
kernel to transform discrete (pixellated) images into continuous images. The
warped images are subsequently resampled onto a common coordinate system.

For the data releases before S19A, a third-order Lanczos kernel was used to
warp CCD images before coadding the images. As reported in Section~6.4 of
\citet{HSC2-data}, the sizes of observed PSFs on coadds are 0.4\% larger than
that of reconstructed PSF models. \citet{HSC2-data} showed that the amplitude
of PSF size residuals decreases when the order of the warping kernel is
increased to fifth-order.

A systematic bias on galaxy shape measurements stemming from such a $0.4\%$
fractional size residual in PSF size was not significant when compared to the
first-year weak lensing science requirements \citep{HSC1-shape}. However, for
the three-year weak lensing shear catalog, the survey area has significantly
increased and the science requirements are consequently much tighter (see
Section~\ref{sec:requirement}). Therefore, we switch to using the fifth-order
Lanczos warping kernel. The tests quantifying PSF model fidelity are presented
in Section~\ref{sec:psftest}.


\subsection{Background subtraction}
\label{subsec:bgSub}
\xlrv{
For the HSC first-year data release (DR1), the pipeline performed a local
background subtraction at the single exposure level with a $128\times128$
($\sim 22 \times 22\arcsec$) pixel-mesh on each CCD individually. To estimate
the sky background, the pipeline averaged pixels in each pixel-mesh ignoring
detected pixels. Then the background was modelled with $2$D Chebyshev
polynomials. After coadding single exposures into coadds, the pipeline
performed a background subtraction with a larger ($\rm 4k\, \times 4k$, or
$11\arcmin \times 11\arcmin$) pixel-mesh (see \citealt{HSC1-pipeline} for more
details) after masking out the detections on coadds. This background
subtraction scheme was found to cause over-subtraction around bright objects
since it subtracts flux from the wings of bright extended objects along with
the sky background \citep{HSC1-pipeline}.

In the second-year data release (DR2), the background subtraction scheme was
updated as follows: At the single exposure level, the pipeline performed a
global joint estimation of the background using all the CCDs across the focal
plane to reduce the aforementioned over-subtraction. In addition, the pipeline
estimates and subtracts the ``sky frame'' --- the mean response of the
instrument to the sky for a particular filter. The sky frame is estimated from
a clipped-mean of the pixel-mesh with detected objects masked out from many
observations with large dithers (see \citealt{HSC2-data} for more details). The
pipeline then applied the same background subtraction scheme as before on
coadds. This background subtraction scheme preserves the extended wings of
bright objects; however, it influences the CModel measurement, which measures
the flux by fitting the galaxy's surface brightness profile with an exponential
and a de Vaucouleurs \citep{deVauProfile1948} profile separately. The preserved
wings of neighboring bright objects and background residuals lead to larger
estimates of galaxy CModel radii and increase the CModel flux estimates,
especially for faint sources near bright objects.

With the intent to mitigate the under-subtraction problem and improve the
performance of CModel measurements, a local background subtraction with a
$128\times128$ (local) pixel-mesh is applied on coadds in S19A. In addition, we
use an improved global background subtraction scheme during single exposure
image processing to remove global sky background and ``sky frame'' (see
\citealt{HSC3-data} for more details). This background subtraction scheme
reduces the aforementioned background residuals caused by the background
subtraction scheme in the second data release. However, the CModel magnitude
estimates in S19A are still brighter than in S16A due to the influence of
background residuals in S19A. As illustrated by the $2$D histogram of the
$i$-band CModel magnitude difference between S19A and S16A as a function of the
S19A magnitude in Fig.~\ref{fig:dMagHist}, the histogram is skewed to negative
$\Delta$mag. Fig.~\ref{fig:dMagHist} indicates that objects appear brighter in
S19A. In addition, we find that the galaxies with negative magnitude difference
cluster around bright objects (e.g., bright stars and bright galaxies). The
details are summarized in the HSC third data release paper \citep{HSC3-data}.
}

\begin{figure*}[ht!]
\begin{center}
\includegraphics[width=0.85\textwidth]{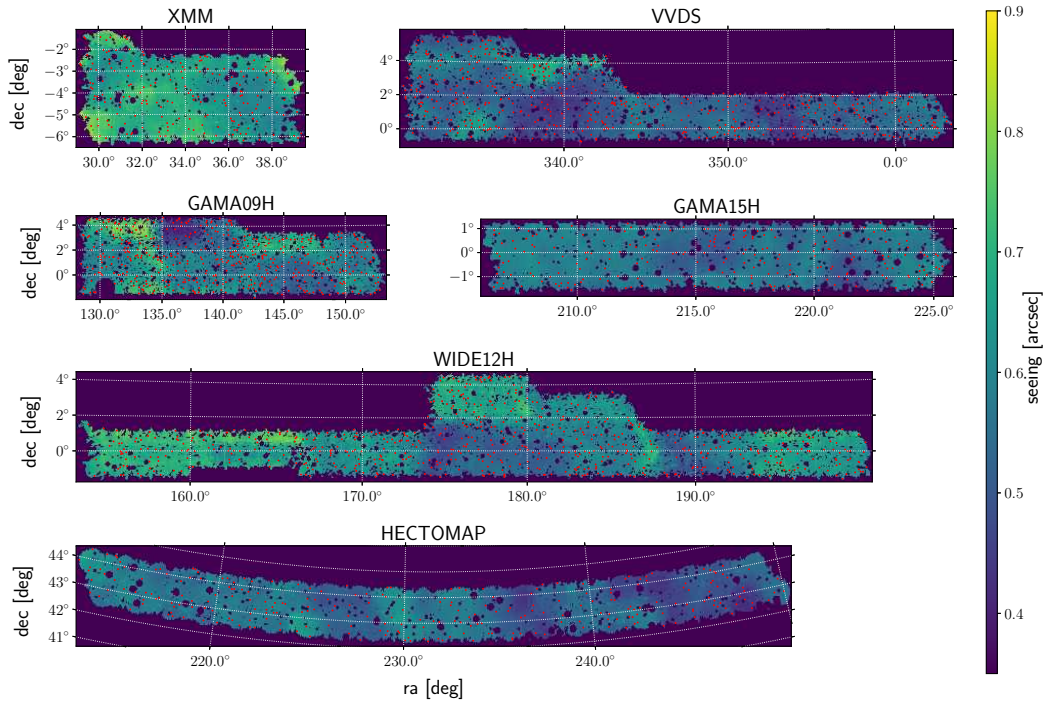}
\end{center}
\caption{
    Map of the $i$-band PSF FWHM across each field. The red dots are the
    sampling positions for PSFs and noise properties that will be used in the
    HSC-like image simulation in Section~\ref{subsec:Sim}. The mean seeing over
    all of the fields is $0\farcs59$. The circular region centered near
    (RA=$130\fdg43$, DEC=$-1\fdg02$) of the GAMA09H field is masked out due to
    the tracking error on the exposure visit 104934.
    }
    \label{fig:fwhmMap}
\end{figure*}

\subsection{Bright star mask}

In this section we describe how bright star masks are applied to the weak
lensing shear catalog. Those who are interested in more details of the bright
star mask construction, please refer to the PDR3 paper \citep{HSC3-data}. The
S19A bright star masks are created using the Gaia second data release
\citep{GAIA-DR2} as a reference catalog in which Gaia magnitudes are converted
to HSC magnitudes. The star masks are defined for stars brighter than 18th
magnitude and for different types of artifacts; halo, ghost, blooming, scratch,
and dip.
\xlrv{
    The scratch mask is designed to mask vertical stripes around bright stars
    in long-wavelength bands (e.g., $y$-band and NB1010-band) due to the
    channel-stop, if the CCD is optically thin with respect to the wavelength
    \citep[for more details, see][]{HSC3-data}. Since the shear catalog is
    based on $i$-band images, the scratch mask is not considered for the shear
    catalog.
}
The dip mask is for masking over-subtracted region in the vicinity of
a star due to the local background subtraction. The over-subtraction affects
the number count of source galaxies but does not have significant influence on
shape estimation. In addition, applying the dip mask reduces the area
significantly. Therefore, the dip mask is not considered for the shear catalog.
The shear estimation near stars is tested in
Section~\ref{subsec:nullTest_posGal}.

For the weak lensing shear catalog, we adopt the star masks for halo, ghost,
and blooming. The flags used for selection are summarized in
Table~\ref{tab:bstarcut}. The halo mask masks an extended smooth halo around a
star whose size depends on the brightness of a star. To define the halo mask, a
median radial profile was computed for stars within a magnitude bin, and the
mask was defined up to the scale where the profiles goes down to the background
level. The size of halo mask decreases as a function of magnitude. The ghost
mask is defined using the median radial profile and a cross-correlation with
objects around bright stars where ghost edges induce spurious detection of
objects. The radius of ghost mask is \hmrv{$350\arcsec$} for stars brighter
than 7th magnitude and $160\arcsec$ for stars between $7$th and $9$th magnitude.
The exact size and shape of ghost depends on the telescope boresight and a
bright star, and fake objects outside the mask are found in some cases. To deal
with such cases, we adopt the ghost mask with 50\% larger than the standard
size defined above. The blooming appears parallel to the channel-stop of a CCD,
which is always horizontal in the image because rotational dithers are not
performed in the SSP survey. The scale of the blooming feature depends on the
star brightness and positions on the CCD inputs, the maximum of which is
\hmrv{$\sim10\arcmin$}. To define the blooming mask, the cross-correlation
measurement was performed along the horizontal and vertical directions, and a
detection excess along the horizontal direction was considered a blooming. The
blooming mask is defined as a function of stellar magnitude.


\begin{table}
\caption{
    Flags of bright star masks considered in our shear catalog. Objects flagged
    as \protect\texttt{True} by any one of the masks are removed.
    }
\begin{center}
\begin{tabular}{ll} \hline
Mask Flag & Meaning \\ \hline
\texttt{i$\_$mask$\_$brightstar$\_$ghost15} & Ghost\\
\hline
\texttt{i$\_$mask$\_$brightstar$\_$halo}    & Halo\\
\hline
\texttt{i$\_$mask$\_$brightstar$\_$blooming}& Blooming\\
\hline
\end{tabular}
\end{center}
\label{tab:bstarcut}
\end{table}


\subsection{Observing strategy}
\label{subsec:obs_strat}

The observing strategy underwent a couple of changes in order to increase the
effective survey completion speed. Firstly, the number of dithers per pointing
in the $i$-, $z$-, and $y$-bands were reduced from $6$ to $5$ since November
2018.  This change results in a survey depth that is shallower by $0.1$
magnitudes on average. The nominal $5\sigma$ depth for point sources in
$i$-band was $26.2$ for PDR2 based on S18A \citep[see Table~2 of][]{HSC2-data}.
Our shear catalog only contains galaxies with $i$-band magnitudes brighter than
$24.5$, and thus the change in depth is not expected to significantly affect
the statistical properties of the shear catalog.

The original requirement on the seeing conditions for procuring $i$-band images
was also relaxed from $0\farcs7$ to $0\farcs9$; this requirement is imposed
using the on-site quick-look software \citep{onSiteSys-Furusawa2018}, which
monitors the data quality with a lag of only a few minutes. Despite the fact
that the requirement was relaxed, the mean $i$-band seeing for the entire three
year data set used in this paper is $0\farcs59$, similar to that of the
first-year HSC shear catalog \citep{HSC1-shape}. We look into the PSF model
errors in the regions observed with 6 dithers and with 5 dithers in
Section~\ref{sec:psftest}, the results of which do not show significant
difference in the PSF model errors between the two observational strategies.

\subsection{Full depth and full color cut}

We restrict ourselves to regions that reach the approximate full depth of the
survey in all five broadband filters ($grizy$), in order to achieve better
uniformity of the shear estimation and photometric redshift quality across the
survey as was also done in \citet{HSC1-shape}. This cut is imposed by requiring
the average number of visits\footnote{Each exposure of the CCD array is termed
a visit.} contributing to the coadds within \texttt{HEALPix} pixels (with
$\rm{NSIDE}=1024$) to be $(g,r,i,z,y) \geq (4,4,5,5,5)$. Note that this is
different from the requirement in the first-year shear catalog that was
$(g,r,i,z,y) \geq (4,4,4,6,6)$. In the first-year shear catalog, some of the
$i$-band visits with the ``very best seeing'' were removed because of the
inability to model the PSFs, and thus the minimum number of $i$-band exposure
was set to $4$ \citep{HSC1-shape}. However, since the PSF determination in the
HSC pipeline was improved as described in Section~\ref{subsec:PSF_improvement},
such exposures are added back to the coadds. In addition, the $5$-dithering
strategy was adopted in November 2018. We thus set the requirement on the
minimum numbers of average input visits for $i$-band to 5. For the $z$- and
$y$-bands, we set the requirement to 5 as well, following the change in
dithering strategy.

As will be discussed in Section~\ref{subsec:psfRegCut}, we also remove a few
regions with large average PSF size modelling errors.  This PSF size modelling
error cut reduces the survey area by $\sim2.2\%$.

\begin{figure*}[ht!]
\begin{center}
\includegraphics[width=0.85\textwidth]{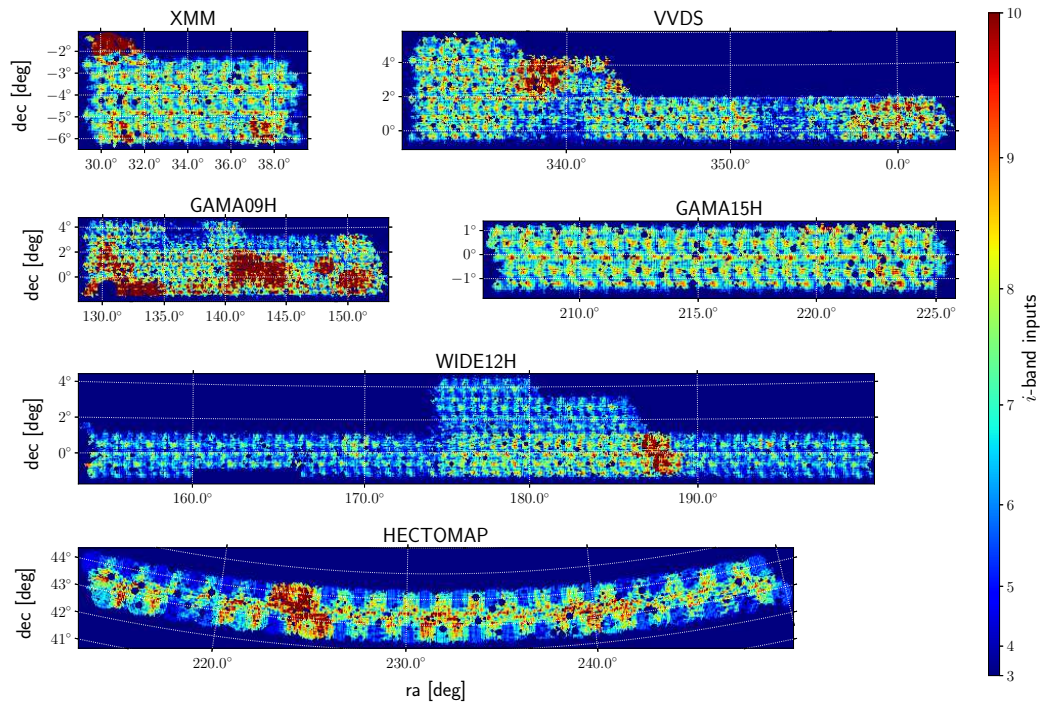}
\end{center}
\caption{
    Number of input visits contributing to the coadds in the $i$-band across
    each field. The mean number of input visits is $6.95$ over all of the
    fields. The way the visits are tiled across each survey area results in the
    repeated pattern of overlap regions with number of inputs more than the
    typical value (see \citet{HSC-SSP2018} for the tiling strategy).
    }
    \label{fig:inputMap}
\end{figure*}

After these cuts, the total area of the catalog is $433.48~\mathrm{deg}^2$.
The footprint of the galaxy catalog is divided into six observational fields,
i.e., XMM, GAMA09H, WIDE12H, GAMA15H, VVDS, HECTOMAP, the areas of which are
33.17~$\mathrm{deg}^2$, 98.85~$\mathrm{deg}^2$, 121.32~$\mathrm{deg}^2$,
40.87~$\mathrm{deg}^2$, 96.18~$\mathrm{deg}^2$, and 43.09~$\mathrm{deg}^2$.
Fig.~\ref{fig:fwhmMap} shows the $i$-band seeing map.  Fig.~\ref{fig:inputMap}
shows the map of the number of $i$-band visits contributing to the coadd.
Fig.~\ref{fig:fwhmHist} shows the seeing histograms, and Fig.~\ref{fig:noiHist}
shows the noise variance histograms.

\section{Shear Catalog}
\label{sec:catalog}
In this section, we introduce the shear catalog measured from the HSC S19A
$i$-band coadded images. We first review the shear estimation process in
Section~\ref{subsec:shearEst}.
In Section~\ref{subsec:Sim}, we present the \xlrv{$i$-band} image simulations
used for the calibration of shear measurements. The selection criteria for the
weak lensing shear catalog are presented in Section~\ref{subsec:WLSample}. We
subsequently determine the intrinsic shape dispersion and the optimal weight
for shear estimation in Section~\ref{subsec:optimal}, calibrate the bias in the
shear estimation in Section~\ref{subsec:calib}, and quantify the amplitude of
the calibration bias residuals in Section~\ref{subsec:Dcalib}. Selection bias
is estimated and calibrated in Section~\ref{subsec:selBias}. Finally, the shear
catalog is characterized in Section~\ref{subsec:charact} and our blinding
strategy to avoid confirmation bias in weak lensing analyses is presented in
Section~\ref{subsec:blind}.

\subsection{Shear estimation}
\label{subsec:shearEst}

\subsubsection{Detection, deblending and source replacement}

\xlrv{
In this subsection, we briefly summarize the processes of source detection,
deblending and source replacement after coadding single exposures and
background subtraction based on $\hscPipe~{\rm v}7$.

The HSC pipeline \citep{HSC1-pipeline} performs a maximum-likelihood source
detection with a $5\sigma$ threshold from the coadded images. Every peak
detected is identified as a source and the connected nearby region above the
threshold is identified as the footprint of the source detection.

For the case that a footprint contains multiple sources, these
sources are taken as blended, and the HSC pipeline apportions the flux to these
blended sources using the SDSS deblending algorithm
\citep{SDSS-pipeline-Lupton2001}. This deblending algorithm takes each peak as a
`child’ source of the `parent’ detection. A template for each `child’ is
constructed with the assumption that each source has $180~\deg$ rotational
symmetry around its detected peak. Then a scaling parameter is determined for
each source by jointly fitting the templates to the blended image.

After deblending, the HSC pipeline performs source measurement (e.g., flux,
size, and shape) on each source. During the deblending and measurement of one
detection, the pipeline replaces the footprints of other sources with
uncorrelated Gaussian noise.
}

\subsubsection{Re-Gaussianization}

Galaxy shapes are estimated with the $\galsim$ \citep{Galsim} implementation of
the re-Gaussianization ($\reGauss$) PSF correction method
\citep{Regaussianization}. This moments-based method has been developed and
used extensively using data from the Sloan Digital Sky Survey
\citep[SDSS;][]{SDSS-shape-Mandelbaum2005,SDSS-gglensDR7-Mandelbaum2013}. The
outputs of the $\reGauss$ estimator are the two components of the ellipticity of
each galaxy:
\begin{equation}\label{eq:e}
    (e_1,e_2)=\frac{1-(b/a)^2}{1+(b/a)^2} (\cos 2\phi,\sin 2\phi),
\end{equation}
where $b/a$ is the axis ratio and $\phi$ is the position angle of the major
axis with respect to sky coordinates (with north being $+y$ and east being
$+x$). Another important output of the pipeline is the resolution factor $R_2$,
which is defined for each galaxy using the trace of the second moments of the
PSF ($T_{{\rm PSF}}$) and those of the observed galaxy image ($T_{{\rm gal}}$):
\begin{equation}
    R_2=1-\frac{T_{{\rm PSF}}}{T_{{\rm gal}}}\,.
\end{equation}
The resolution factor is used to quantify the extent to which the galaxy is
resolved compared to the PSF.

For \xlrv{an isotropically-orientated} galaxy ensemble distorted by a constant
shear, the shear can be estimated with a weighted average of the ellipticity of
all galaxies:
\begin{equation}\label{eq:g}
    \hat{g}_\alpha = \frac{1}{2\res}\left\langle e_\alpha \right\rangle,
\end{equation}
where the shear responsivity ($\res$) is the response of the average galaxy
ellipticity to a small shear distortion
\citep{KSB-Kaiser1995,shearEst-Bernstein2002}, and $\alpha=1,2$ are the indices
for the two components of the ellipticity. The inverse variance weights to be
used while performing the ensemble average are the galaxy shape weights ($w_i$)
defined as
\begin{equation}\label{eq:optimalW}
    w_i=\frac{1}{\sigma_{e;i}^2+e_{{\rm RMS};i}^2},
\end{equation}
\xlrv{where $i$ is an index over galaxies}, $\sigma_e$ is the per-component
$1\sigma$ uncertainty of the shape estimation error due to photon noise, and
$e_{\rm{RMS}}$ denotes the per-component root-mean-square ($\texttt{RMS}$) of
the galaxy intrinsic ellipticity. The parameters $e_{\rm{RMS}}$ and $\sigma_e$
are modeled and estimated for each galaxy using image simulations, as will be
discussed in Section~\ref{subsec:optimal}. The responsivity for the source
galaxy population is estimated as
\begin{equation}\label{eq:response}
    \res=1-\frac{\sum_i w_i e^2_{{\rm RMS};i}}{\sum_i w_i}\,.
\end{equation}
As the PSFs are nearly round, the responsivity for PSFs is approximately one,
and the shear distortion for a PSF image is defined as $g_{{\rm
PSF},\alpha}=e_{{\rm PSF},\alpha}/2$, where $e_{{\rm PSF},\alpha}$ are the two
components ($\alpha=1,2$) of PSF ellipticity defined with the second moments of
the PSF. We refer the reader to Section~\ref{sec:psftest} for tests on
PSF-related systematics.

\subsubsection{Shear estimation bias}

Since the $\reGauss$ algorithm is subject to certain forms of shear estimation
bias (e.g., model bias, noise bias, and selection bias), in this section, we
define the calibration parameters that will encapsulate those biases and review
the calibrated form of the $\reGauss$ shear estimator. The relation between the
estimated shear and the true shear \xlrv{at the individual galaxy level} is
quantified by
\begin{equation}\label{eq:fitmc}
    \hat{g}_{\alpha;i}=(1+m_i)g_{\alpha;i}+a_i e_{\mathrm{PSF},\alpha;i} \,,
\end{equation}
where $m_i$ is the multiplicative bias and $a_i$ is the fractional additive
bias quantifying the fraction of the PSF anisotropy (ellipticity) that leaks
into the shear estimation. \xlrv{Terms involving spin-$4$ quantities, which
average to zero when averaging $\hat{g}_{\alpha;i}$ over all galaxies in the
sample, are neglected.} The two components of the additive bias are thus given
by $c_{\alpha}\equiv a e_{\mathrm{PSF}, \alpha}$. \xlrv{Here we neglect the
additive bias that is independent of PSF anisotropy since, using the image
simulation that will be introduced in Section~\ref{subsec:Sim}, we find that
the amplitude of that term is about $8\times10^{-5}$, which is within the the
HSC three-year science requirements in Section~\ref{sec:requirement}. We also
conduct null tests that are sensitive to the PSF-independent additive bias
within the final shear catalog in Section~\ref{subsec:nullTest_meanShear}}.
Even though shear estimation algorithms can show slightly different biases for
the two different shear components ($g_{1,2}$), we do not distinguish between
the two in this paper. In addition, the value of multiplicative bias is blinded
in this paper to avoid confirmation bias in cosmological analyses.

We will estimate and model the multiplicative bias and the fractional additive
bias for each galaxy as a function of its properties (such as the SNR, $R_2$,
and galaxy redshift) in Section~\ref{subsec:calib}.

The multiplicative bias and the additive bias for the galaxy ensemble are:
\begin{equation}
\begin{split}
    \hat{m}&=\frac{\sum_i w_i m_i}{\sum_i w_i},\\
    \hat{c}_{\alpha}&=\frac{\sum_i w_i a_i e_{{\rm PSF},\alpha;i}}{\sum_i w_i},
\end{split}
\end{equation}
respectively. The calibrated shear estimator is defined as
\begin{equation}\label{eq:calibEstimator}
    \hat{g}_{\alpha}=\frac{\sum_i w_i e_{\alpha;i}}{2\res(1+\hat{m})\sum_i w_i}
    -\frac{\hat{c}_{\alpha}}{1+\hat{m}}\,.
\end{equation}

\xlrv{
Note, here we neglect the selection bias due to the anisotropic selection of
the galaxy ensemble. The shear estimation bias will be estimated using
HSC-like image simulations in Section~\ref{subsec:calib}. The details of the
simulation will be introduced in Section~\ref{subsec:Sim}.
}

\subsubsection{Selection bias}
\xlrv{
Selection bias refers to a multiplicative or  additive bias induced by a
selection criterion that correlates with the true lensing shear and/or the PSF
anisotropy. As a result of the anisotropic selection, the selected galaxies
that are sufficiently close to the edge of the selection coherently align in a
direction that  correlates with the lensing shear and/or the PSF anisotropy.

Here we denote the multiplicative bias and the fractional additive bias caused
by a selection as $\hat{m}^\mathrm{sel}$ and $\hat{a}^\mathrm{sel}$,
respectively. They will be estimated for the galaxy ensemble using the HSC-like
image simulation in Section~\ref{subsec:selBias}. The final shear estimator is
\begin{equation}
    \hat{g}^\mathrm{final}_\alpha =
    \frac{\hat{g}_\alpha - \hat{c}^\mathrm{sel}_\alpha}{1+\hat{m}^\mathrm{sel}},
\end{equation}
where
\begin{equation}
\hat{c}^\mathrm{sel}_\alpha =
\frac{\hat{a}_\alpha^\mathrm{sel} \sum_{i}
w_i e_{\mathrm{PSF},\alpha;i}}
{\sum_i w_i}
\end{equation}
is the estimated additive selection bias.
}

\subsection{Image simulations}
\label{subsec:Sim}

In this section, we introduce the galaxy image simulations used to calibrate
the galaxy shapes output by $\reGauss$ \xlrv{on the HSC $i$-band} coadded
images. Our simulations are divided into $2500$ subfields and each subfield
contains $10^4$ postage stamps each of which is composed of $64\times 64$
pixels. The pixel scale is set to $0.168\arcsec$ to match the pixel scale of
HSC.

\subsubsection{Input noise and PSF}
\label{subsubsec:Sim-noiPsf}

The noise properties (including variance and spatial correlations) and PSF
models are the same in each subfield while they vary between different
subfields in the simulations. We sample $2500$ noise variance values, noise
correlation functions, and PSF models from a set of random positions on the
$i$-band coadded images \xlrv{on which the \reGauss{} shapes are measured}. The
randomly sampled positions are shown as red points in Fig.~\ref{fig:fwhmMap}.

Noise on the coadded images has a spatial correlation between neighboring
pixels, since the fifth-order Lanczos kernel used to warp CCD images during the
coaddition process \citep{HSC1-pipeline} results in correlated noise. We sample
the noise correlations from the blank pixels (where no galaxy is detected) near
the sampled random positions. Subsequently, the sampled noise correlations,
which are noisy on the individual level, are randomly divided into eight
groups, and stacked in each group to create eight different well-measured noise
correlation functions.

We first use the sampled noise variance of each subfield as the input noise
variance for our preliminary simulations. After populating galaxy images into
each subfield, we measure the noise variance from blank (undetected) pixels on
the preliminary simulations. The measured noise variances are in general
greater than the input noise variances due to the light from neighboring
detected sources and undetected sources underlying the blank pixels. We record
the ratio between the measured noise variance and the input noise variance for
each subfield, the average value of which is $1.25$ across all subfields. Then
we divide the sampled noise variance by this ratio for each subfield, and the
rescaled variances are used as the inputs of our fiducial simulations.
By rescaling the sampled noise variances, we match the noise variances measured
from the simulations to those measured from the HSC data in a {\em consistent}
manner. In contrast, we did not perform such a rescaling in the first-year
HSC-like image simulations \citep{HSC1-GREAT3Sim}, but rather {\em inconsistently}
matched the input noise variances in the simulation to the
measured noise variances in the S16A HSC data, which results in a larger noise
variance in image simulations compared to reality.

\begin{figure*}[ht!]
\begin{center}
\includegraphics[width=0.85\textwidth]{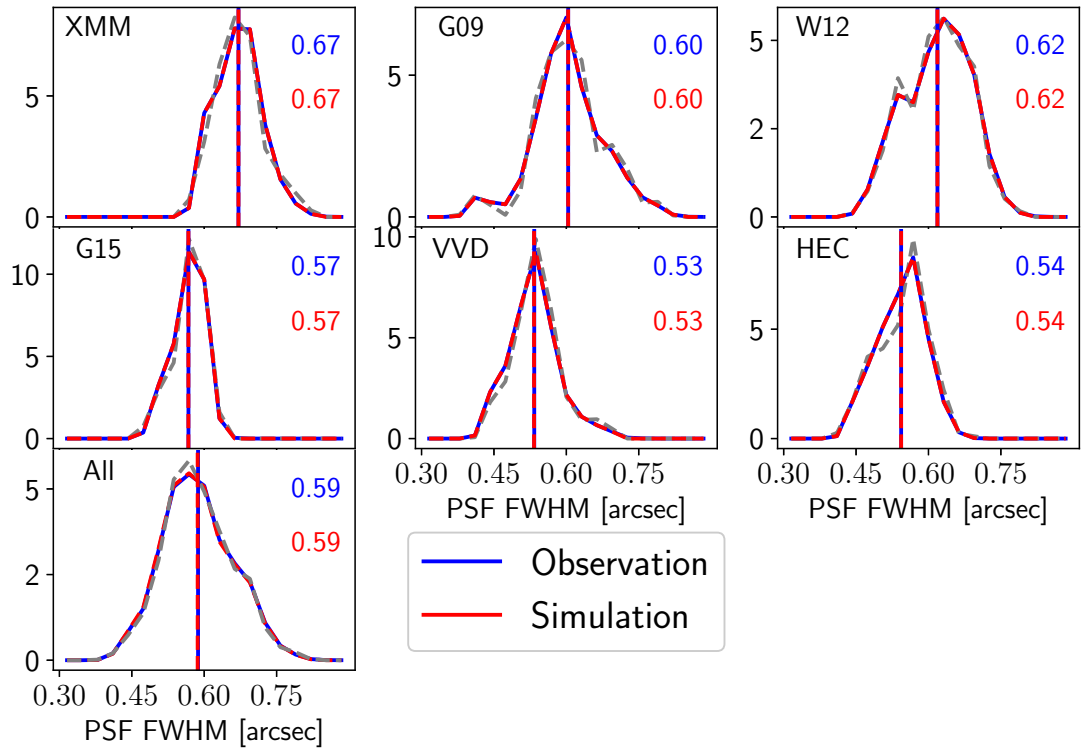}
\end{center}
\caption{
    The first six panels show the normalized number histograms of PSF FWHM for
    the galaxies in the HSC observational fields. The last panel is the
    histogram for galaxies in all fields. The blue solid (red dashed) lines are
    for the HSC data (simulation). The blue (red) text and vertical lines
    indicate the mean averages of the HSC data (simulation). The simulations
    are reweighted to mitigate the difference to the data due to the finite
    sampling for each field. The gray lines show the histograms before the
    reweighting.
    }
    \label{fig:fwhmHist}
\end{figure*}

\begin{figure*}[ht!]
\begin{center}
\includegraphics[width=0.85\textwidth]{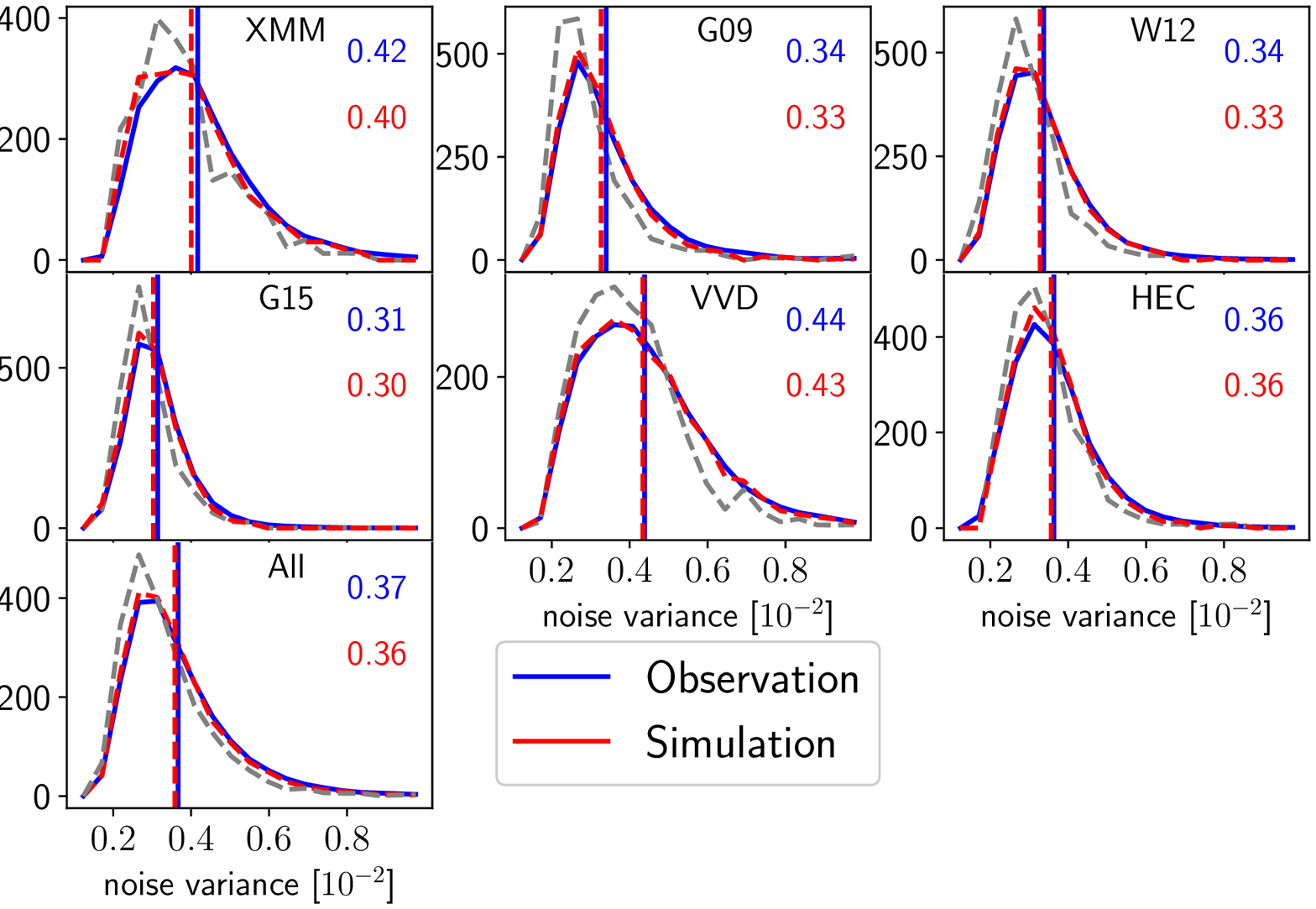}
\end{center}
\caption{
    Same as Fig.~\ref{fig:fwhmHist}, but for noise variance.
    }
    \label{fig:noiHist}
\end{figure*}

To mitigate the differences between the simulations and the HSC data due to the
finite sampling of noise and PSF, we reweight each subfield in the simulations
such that the seeing and noise variance closely histograms match the real data.
Note that we do not reweight the simulations according to any properties of the
input galaxies. The reweighting is conducted separately for each HSC
observational field. The seeing (PSF FWHM) histograms and noise variance
histograms for the observations and the simulations are shown in
Figs.~\ref{fig:fwhmHist} and \ref{fig:noiHist}, respectively.

Note that the input PSF models do not include PSF model errors; that is, the
PSF is assumed to be known perfectly. In addition, we assume the sky
subtraction is perfect, and the residuals of the sky background are not
included in the simulations. As these observational conditions are obtained
from coadded images, the systematics related to the coaddition process can not
be tested with the simulations.

\subsubsection{Input galaxy}
\label{subsubsec:Sim-GalIn}

\citet{HSC1-GREAT3Sim} selected galaxy training samples with CModel magnitudes
less than $25.2$ from the HSC Wide-depth catalogs detected from three stacks of
the HSC Deep/Ultradeep images with typical seeings of $0\farcs5$, $0\farcs7$,
and $1\farcs0$, respectively, in the COSMOS region \citep{HSC1-data}.
\citet{HSC1-GREAT3Sim} determined the centroids of these galaxies on the
exposures of the COSMOS HST Advanced Camera for Surveys (ACS) field
\citep{HST-ACSpipe} in the F814W band. Square postage stamps centered at the
galaxy centroids \xlrv{with width$=10\farcs752$ ($64$ HSC pixels)} were cut out
from the HST exposures. The details of the training samples are described in
\citet{HSC1-GREAT3Sim}. In this paper, we use the training sample selected from
the stack with the best seeing ($0\farcs5$) since it should be the deepest
sample among the three thanks to its best seeing.

\xlrv{
    Note, we do not inject parametric galaxies into images as in, for example,
    \citet{DESY3-BlendshearCalib-MacCrann2021}. Instead, we directly cut out
    postage stamps from the HST F814W images. Since we do not perform any
    deblending or masking on the input HST images before shearing and
    transforming the noise property, all of the neighboring sources are kept on
    the postage stamp to reproduce the effects of both recognized and un-recognized blends. We do
    not input star images into the simulation. Stars could appear on galaxy
    outskirts (not at the centers of the postage stamp) if they happens to
    reside in close proximity to the simulated central galaxy. We will further
    test the influence of stellar contamination in our shear catalog in
    Section~\ref{subsubsec:GalSel}.
}

$\galsim$ \citep{Galsim}, which is an open-source package for galaxy image
simulations, is used to simulate HSC-like images using the COSMOS HST images in
our simulations. The original HST PSF is deconvolved from each input HST
postage stamp and then the image is rotated with a random angle, sheared by a
known input shear distortion, convolved with a collected HSC PSF model, sampled
at the HSC pixel scale, and downgraded to an HSC noise level. The noises and
PSFs used in the simulations are those introduced in
Section~\ref{subsubsec:Sim-noiPsf}.

Each subfield is designed to specifically include $90\degree$ rotated
(intrinsically orthogonal) pairs of galaxies that can be used to nearly cancel
out shape noise \citep{galsim-STEP2}. By keeping track of the members of each
orthogonal pairs, the analysis framework provides options to apply this
cancellation or not. The orthogonal pairs will also be used to derive shape
measurement error, weight bias, and selection bias in the shear estimation
following \citet{HSC1-GREAT3Sim}.

\subsection{Weak lensing galaxy sample}
\label{subsec:WLSample}

\subsubsection{Galaxy selection}
\label{subsubsec:GalSel}

We run $\hscPipe~{\rm v}7$, the pipeline used to process the S19A internal data
release along with the same configuration options, on the simulations for
source detection and deblending.  Subsequently, $\hscPipe~{\rm v}7$ is used to
perform magnitude, size and shape measurements on the deblended sources.  For
all of the analyses shown in this paper based on our image simulations, a basic
set of flag cuts in the ``Basic flag cuts'' section of Table~\ref{tab:icut} are
imposed. Since our simulations do not include image artifacts, only the
following flags actually influence the source selection in the simulations:
\texttt{i$\_$detect$\_$isprimary}, \texttt{i$\_$sdsscentroid$\_$flag}, and
\texttt{i$\_$extendedness$\_$value}.

Following \citet{HSC1-GREAT3Sim}, we only keep the detected source nearest to
the postage stamp center for each postage stamp. In addition, we require the
nearest source to have a centroid that is a maximum of $5$ pixels from the
postage stamp center to eliminate stamps where the detection nearest to the
center was not the intended central object.

Since the input galaxy sample has an $i$-band magnitude limit of $\sim 25.2$,
our simulations are not complete, especially at the very faint end. However,
HSC reaches the 26'th magnitude depth thanks to its longer exposure time than
HST. As a result, the input training sample is not representative to the HSC
galaxies at very faint end. We also note that our simulations do not include
realistic large-scale background light. However, the remaining residuals after
background subtraction are likely to influence the galaxy measurements,
especially on faint galaxies -- and the residuals can also lead to fake
detections that cannot be reproduced in our simulations.

To mitigate the difference between the simulations and the real data due to the
incompleteness of the input HST galaxy training sample and the absence of
realistic background light residuals, a set of cuts on galaxy properties
measured with the pipeline are applied in both the HSC data and the simulations
to define a high-SNR, well-resolved galaxy sample for the weak lensing science.
These cuts serve to remove faint galaxies that are beyond the magnitude limit
of the HST galaxy sample. In addition, such weak lensing cuts are useful
to remove fake detections cased by background light residuals that is not
included in our simulations. The $i$-band cuts, applied to both the
observations and the simulations, are summarized in Table~\ref{tab:icut}.

The cut on \texttt{i$\_$extendedness$\_$value} is applied to reduce stellar
contamination in the weak lensing galaxy catalog. We estimate the stellar
contamination fraction, the number fraction of misclassified stars in our weak
lensing galaxy sample even after this cut, using as a reference the galaxy-star
classification performed on HST COSMOS data by
\citet{HST-shapeCatalog-Alexie2007}. Since the HST images have a much higher
resolution and lower noise level than the HSC images, we regard the HST
galaxy-star separation as the ground truth.

\begin{figure}
\begin{center}
\includegraphics[width=0.45\textwidth]{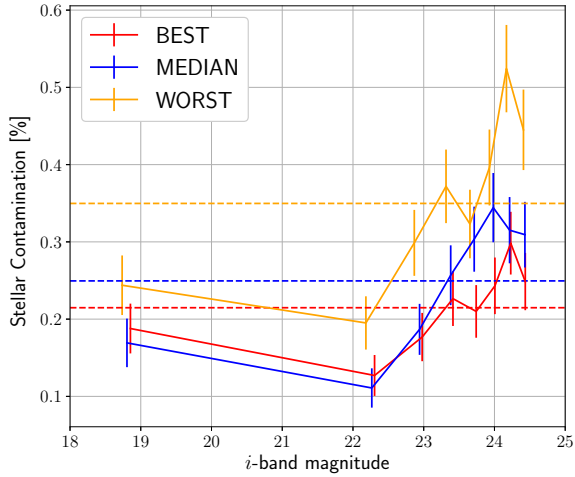}
\end{center}
\caption{
    The stellar contamination fraction due to the incorrect classification by
    $\hscPipe~{\rm v}7$, estimated after application of the weak lensing cuts
    in Table~\ref{tab:icut}. We show the stellar contamination fraction as a
    function of $i$-band CModel magnitude for three different seeing conditions
    (i.e., BEST, MEDIAN, and WORST) estimated with reference to COSMOS HST
    star-galaxy classifications used as an estimate of ground truth. Errorbars
    show the Poisson uncertainties. Dashed lines show the stellar contamination
    fractions for all magnitude bins in the corresponding seeing samples.
    }
    \label{fig:stellarcontam}
\end{figure}

Fig.~\ref{fig:stellarcontam} shows the stellar contamination fractions as a
function of magnitude for the catalogs selected using the weak lensing cuts in
the COSMOS region. For this purpose, we utilize the Deep/Ultradeep data which
consists of multiple exposures in the COSMOS region. We have constructed three
different Wide-depth stacks of the HSC S19A images. These stacks correspond to
the exposures with the best, median, and worst seeing, respectively, with
typical seeing values of $0\farcs5$, $0\farcs7$, and $1\farcs0$, respectively
\citep{HSC1-data}. Even in the worst seeing conditions, the stellar
contamination fraction is below $0.2\%$ for galaxies with $i$-band magnitudes
brighter than $22$, increasing to $0.5\%$ at the faintest end of the shear
catalog with $i$-band magnitude close to $24.5$. Hence we conclude that the
shear estimation biases from the misclassification of stars as galaxies is
negligible, since the fraction of misclassified stars is less than $0.5\%$.

\xlrv{
We do not apply any cuts to remove the potential contamination from binary
stars as in  \citet{KiDS450-CS}. Even though we do find that objects in the
weak-lensing sample with extremely large ellipticity $\abs{e}>0.8$ and $i$-band
determinant radius $r_\mathrm{det}<10^{-0.1r +1.8}$ arcsec show a
characteristic stellar locus in the ($g$-$r$, $r$-$i$) color-color histogram,
its number fraction is only $\sim 0.61 \%$ of the weak-lensing galaxy sample,
which is not likely to cause biases beyond the weak-lensing requirements. We
will remove these potential binary stars from our sample in the three-year
cosmological analysis.
}

In addition to the $i$-band cuts, we follow \citet{HSC1-shape} and apply a
multi-band detection cut to ensure that we have enough color information to
compute photometric redshifts. The multi-band color cut requires at least two
other bands (out of $grzy$-bands) to have at least a $5\sigma$ CModel detection
significance (i.e., SNR$>5$). The multi-band detection cut is applied only on
the HSC data but not on the image simulations since, unfortunately, we do not
have multi-band image simulations. This multi-band detection cut removes a very
small fraction ($<1\%$) of galaxies that pass other selection thresholds.
Therefore, the multi-band cut is not likely to cause significant selection bias
on the shear estimation. On the other hand, this multi-band cut helps remove
junk detections and artefacts \citep{cosmicShear-KiDS2017}.

Compared with the S16A data, the S19A data is processed with a global
background subtraction scheme as summarized in Section~\ref{subsec:bgSub}. The
under-subtraction of sky background in this scheme increases the CModel flux
estimation near bright objects, which makes cuts on CModel flux inefficient at
removing the galaxies beyond the HST magnitude limit and the fake detections
caused by background light residuals in the observations. We find a mismatch in
the SNR-$R_2$ $2$D histograms between the S19A HSC data and the simulations at
the faint end when simply using the first-year $i$-band cuts summarized in
Table~4 of \citet{HSC1-shape}. There are more extended faint detections that
are very likely to be fake detections in the HSC data than in the simulations.
Therefore, we apply an additional cut on $i$-band $1\arcsec$-diameter-aperture
magnitudes (mag$_A$) at $25.5$ to remove the fake detections that cannot be
reproduced in the simulations. The additional aperture magnitude cut removes
$3.9\%$ of the galaxies that pass other selection cuts. The selection bias due
to the cuts is quantified in Section~\ref{subsec:selBias}.

\begin{table*}
\caption{
    Weak lensing cuts: The $i$-band selection criteria that are applied to both
    the simulations and the HSC data. We note that the
    ``\texttt{i$\_$pixelflags$\_$clipped} $==$ False''
    (``\texttt{i$\_$pixelflags$\_$edge} $==$ False'') flag, which identifies
    detections close to the artifacts (edges) resulting in unreliable PSFs, was
    not properly set in the first-year shear catalog. As described in
    Section~4.9 of \citet{HSC2-data}, the flags are correctly set for the
    current data.
    }
\begin{center}
\begin{tabular}{ll} \hline
    Cut & Meaning \\ \hline
\multicolumn{2}{c}{Basic flag cuts} \\ \hline
\texttt{i$\_$detect$\_$isprimary} $==$ True     & Identify unique detections only   \\ \hline
\texttt{i$\_$deblend$\_$skipped} $==$ False     & Deblender skipped this group of objects \\ \hline
\texttt{i$\_$sdsscentroid$\_$flag} $==$ False   & Centroid  measurement failed      \\ \hline
\texttt{i$\_$pixelflags$\_$interpolatedcenter}  $==$ False
                                                & A pixel flagged as interpolated is
                                                close to object center \\ \hline
\texttt{i$\_$pixelflags$\_$saturatedcenter}     $==$ False
                                                & A pixel flagged as saturated is
                                                close to object center \\ \hline
\texttt{i$\_$pixelflags$\_$crcenter}$==$ False  & A pixel flagged as a cosmic ray hit is
                                                close to object center \\ \hline
\texttt{i$\_$pixelflags$\_$bad}$==$ False       & A pixel flagged as otherwise bad is
                                                close to object center \\ \hline
\texttt{i$\_$pixelflags$\_$suspectcenter} $==$ False
                                                & A pixel flagged as near saturation is
                                                close to object center \\ \hline
\texttt{i$\_$pixelflags$\_$clipped} $==$ False  & Source footprint includes
                                                clipped pixels          \\ \hline
\texttt{i$\_$pixelflags$\_$edge} $==$ False     & Object too close to image
                                                boundary for reliable measurements  \\ \hline
\texttt{i$\_$hsmshaperegauss$\_$flag} $==$ False& Error code returned by shape
                                                measurement code        \\ \hline
\texttt{i$\_$hsmshaperegauss$\_$sigma} $!=$ NaN & Shape measurement uncertainty
                                                should not be NaN       \\ \hline
\texttt{i$\_$extendedness$\_$value} $!= 0$      & Extended object       \\ \hline
\multicolumn{2}{c}{Galaxy property cuts} \\ \hline
\texttt{i$\_$cmodel$\_$flux}/\texttt{i$\_$cmodel$\_$fluxerr}
$\ge 10$                                              & Galaxy has high enough $S/N$ in $i$-band     \\ \hline
\texttt{i$\_$hsmshaperegauss$\_$resolution} $\ge 0.3$ & Galaxy is sufficiently resolved             \\ \hline
(\texttt{i$\_$hsmshaperegauss$\_$e1}$^2 + $\texttt{i$\_$hsmshaperegauss$\_$e2}$^2)^{1/2} < 2$
                                                      & Cut on the amplitude of galaxy ellipticity  \\ \hline
$0\le $\texttt{i$\_$hsmshaperegauss$\_$sigma}  $\le 0.4$
                                                      & Estimated shape measurement error is reasonable \\ \hline
\texttt{i$\_$cmodel$\_$mag} $-$ \texttt{a$\_$i} $\le 24.5$
                                                      & CModel Magnitude cut \\ \hline
\texttt{i$\_$apertureflux$\_$10$\_$mag} $\le 25.5$    & Aperture ($1\arcsec$ diameter) magnitude cut \\ \hline
\texttt{i$\_$blendedness$\_$abs} $< 10^{-0.38}$       & Avoid spurious detections and those
                                                        contaminated by blends \\ \hline
\end{tabular}
\end{center}
\label{tab:icut}
\end{table*}


\begin{figure*}[ht!]
\begin{center}
\includegraphics[width=0.85\textwidth]{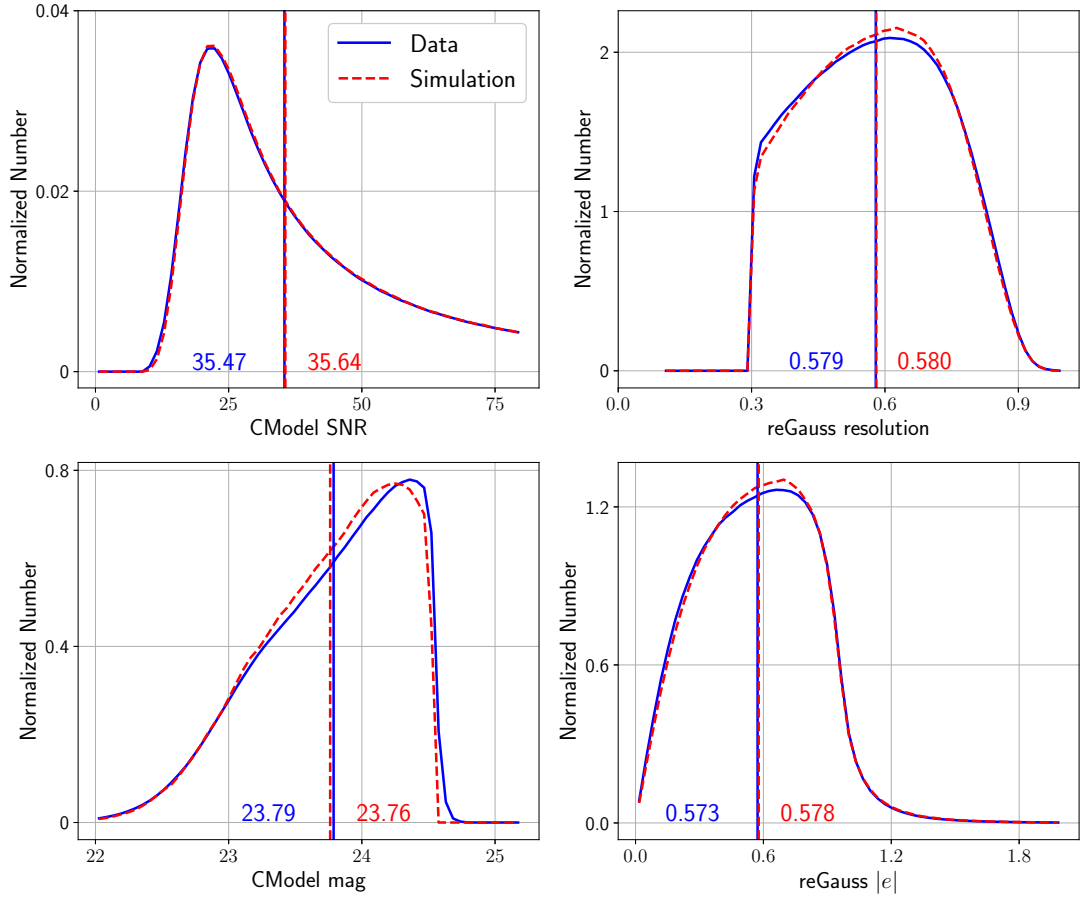}
\end{center}
\caption{
    The normalized number histograms of $i$-band properties including CModel
    SNR (upper left), $\reGauss$ resolution (upper right), CModel magnitude
    (lower left), and $\reGauss$ ellipticity magnitude (lower right), for
    galaxies in all fields combined. The blue solid (red dashed) lines are for
    the HSC data (simulation). The blue (red) text and vertical lines indicate
    the mean averages of the HSC data (simulation).
    }
    \label{fig:galHist}
\end{figure*}

\begin{figure*}[ht!]
\begin{center}
\includegraphics[width=0.85\textwidth]{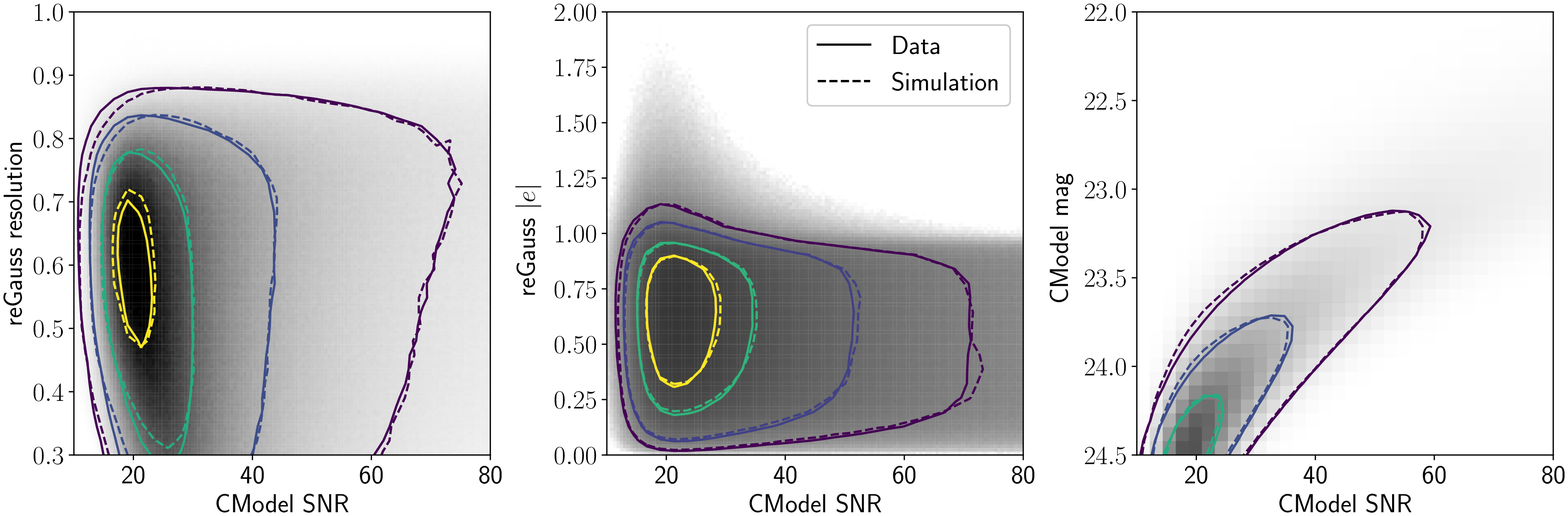}
\end{center}
\caption{
    The color maps are the $2$D histograms for the HSC data. The panels from
    left to right show the (SNR, $R_2$), (SNR, $|e|$) and (SNR, CModel
    magnitude) histograms, respectively. The solid (dashed) lines show the
    contours for the HSC data (simulation). The contours in panels from left to
    right are defined at (0.90, 0.60, 0.30, 0.12), (0.76, 0.54, 0.26, 0.14),
    and (0.68, 0.34, 0.14) of the maximums of the corresponding histograms.
    }
    \label{fig:2DHist}
\end{figure*}

To study the influence of the selection function of $\hscPipe~{\rm v7}$ source
detection on our galaxy sample, in cases where no object is detected within $5$
pixels from the center of a simulated postage stamp, we artificially force one
detection with its peak at the center of the stamp. Flux, size and shape
measurements are conducted on the artificially forced detections. We find that
the number of these forced detections that enter the weak lensing sample after
the weak lensing cuts are applied is far less than $0.1$\% of the total galaxy
number in the weak lensing sample, which indicates that the selection function
of the source detector has a negligible influence on the weak lensing sample;
therefore, the selection bias from the source detector is negligible. This is
aligned with our expectations, since the $5\sigma$ detection limit for point
sources is $26.2$~mag in $i$-band, and our weak lensing galaxy sample is
selected with an $i$-band magnitude cut at $24.5$, far brighter than the
detection limit.
We note that one limitation of our simulations is that several defects from
real data (e.g., sky background residuals, optical ghosts, very bright stars,
etc.) that can affect the object detection are not included.

\subsubsection{Galaxy properties}

The $1$D normalized number histograms for $i$-band galaxy properties (i.e.,
CModel SNR, $\reGauss$ resolution, CModel magnitude, $\reGauss$ ellipticity
magnitude defined as $|e|= \sqrt{e_1^2+e_2^2}$), in the HSC observations and
the simulations are shown in Fig.~\ref{fig:galHist}. When plotting the
histograms, we adopt the same upper limit on the $i$-band CModel SNR (SNR$<80$)
as \citet{HSC1-GREAT3Sim} to compare our results with those shown in the HSC
first-year image simulation paper. We do not find significant differences in
the shapes of the number histograms between the HSC data and the simulations.
The relative difference of the mean values averaged across all of the fields
for these properties between the data and the simulations are $0.5\%$ (CModel
SNR), $0.2\%$ ($\reGauss$ resolution) $0.1\%$ (CModel mag) and $0.8\%$ ($|e|$),
all of which are less than $1\%$. Finally, we show the $2$D joint histograms
of these galaxy properties in Fig.~\ref{fig:2DHist}.

Compared to the first-year HSC-like image simulations
\citep[see][Fig.~8]{HSC1-GREAT3Sim}, the three-year HSC-like simulations have a
better match to the HSC data in the SNR histogram. The average SNR over all
fields was relatively less than the observed SNR by $\sim5\%$ in
\citet{HSC1-GREAT3Sim}, while the discrepancy decreases to $\sim0.5\%$ for the
three-year HSC-like image simulations presented in this paper. The match in SNR
distribution improves because we rescale the sampled noise variance for a
consistent match between the measured noise variances from the HSC data and
those from the simulations as discussed in Section~\ref{subsubsec:Sim-noiPsf}.
Furthermore, the matches between the $2$D histograms are visually better than
those of the first-year HSC simulations shown in Fig.~9 of
\citet{HSC1-GREAT3Sim}, primarily due to the improvement in the match between
the SNR histograms.

In addition, compared to the state-of-art image simulations in other weak
lensing surveys, e.g., Fig.~3 in \citet{DESY3-BlendshearCalib-MacCrann2021}
from the DES survey and Fig.~9 in \citet{KIDS-ImgSim-Kannawadi2019} from the
KiDS survey, our simulations generally have better matches to the observations
in the histograms of galaxy brightness, size and shape.

\subsection{Optimal weighting}
\label{subsec:optimal}

In this section, we estimate and model the statistical uncertainties from
photon noise (shape measurement error) and shape noise (intrinsic shape
dispersion) as functions of galaxy properties, and determine the optimal weight
for the shear estimation.

We first use the simulations to estimate the $1\sigma$  per-component shape
uncertainty due to photon noise ($\sigma_e$) and model it as a function of
galaxy properties (i.e., SNR and $R_2$) following the formalism given in
Appendix~A of \citet{HSC1-GREAT3Sim}. In the estimation, we use the orthogonal
galaxy pairs to nearly cancel out shape noise and measure the statistical error
due to photon noise.

\begin{figure*}[ht!]
\begin{center}
\includegraphics[width=0.324\textwidth]{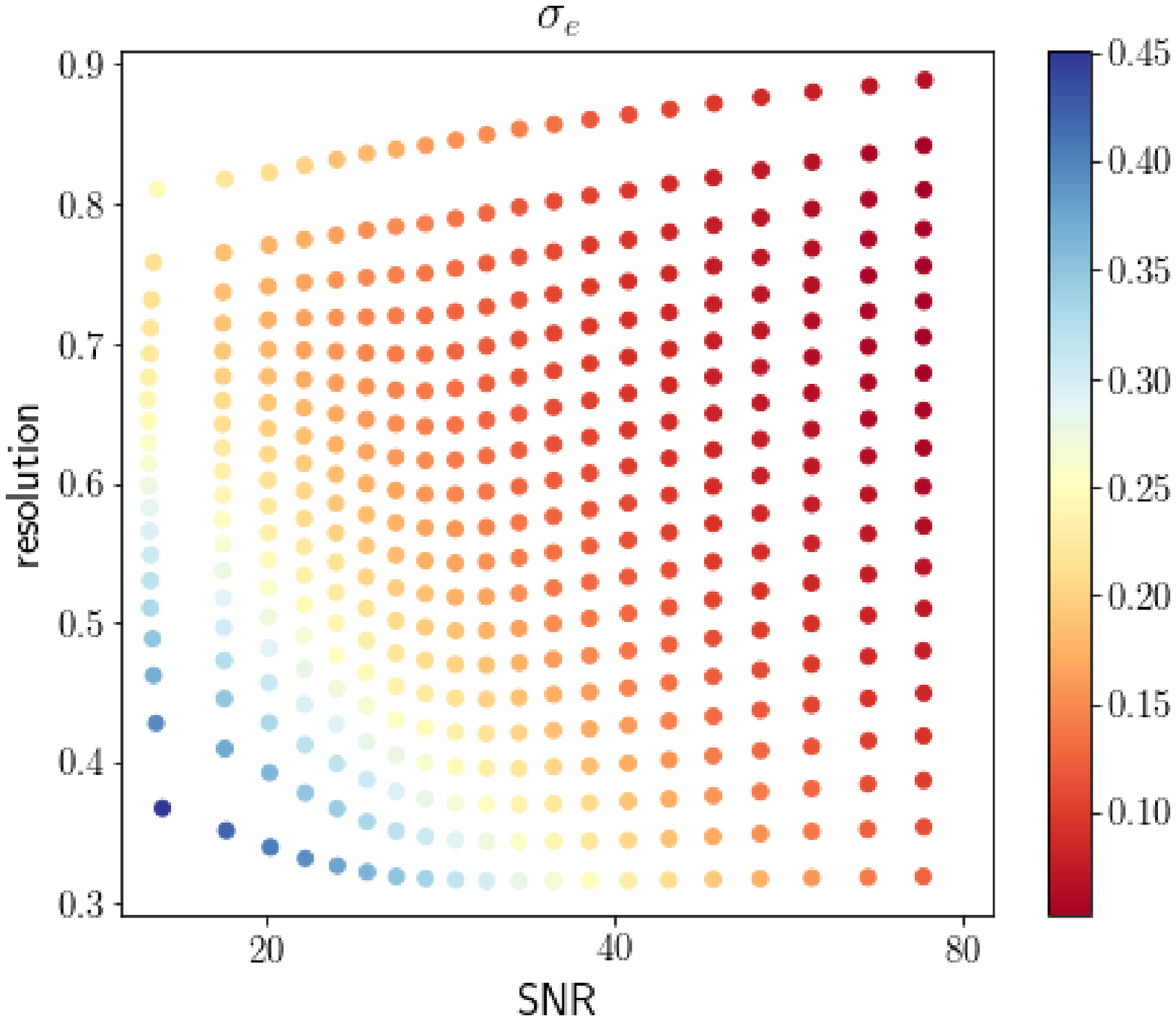}
\includegraphics[width=0.324\textwidth]{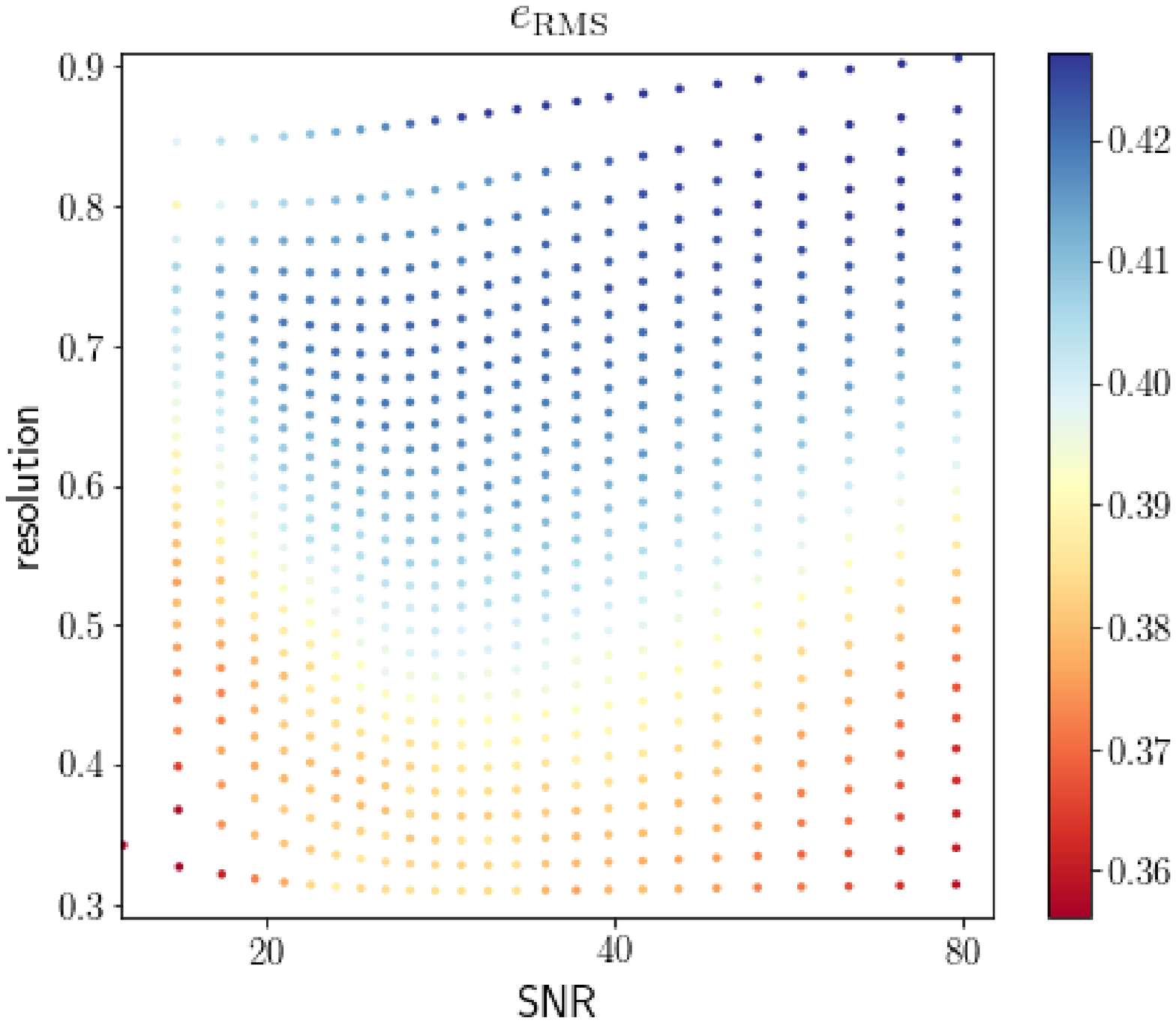}
\includegraphics[width=0.324\textwidth]{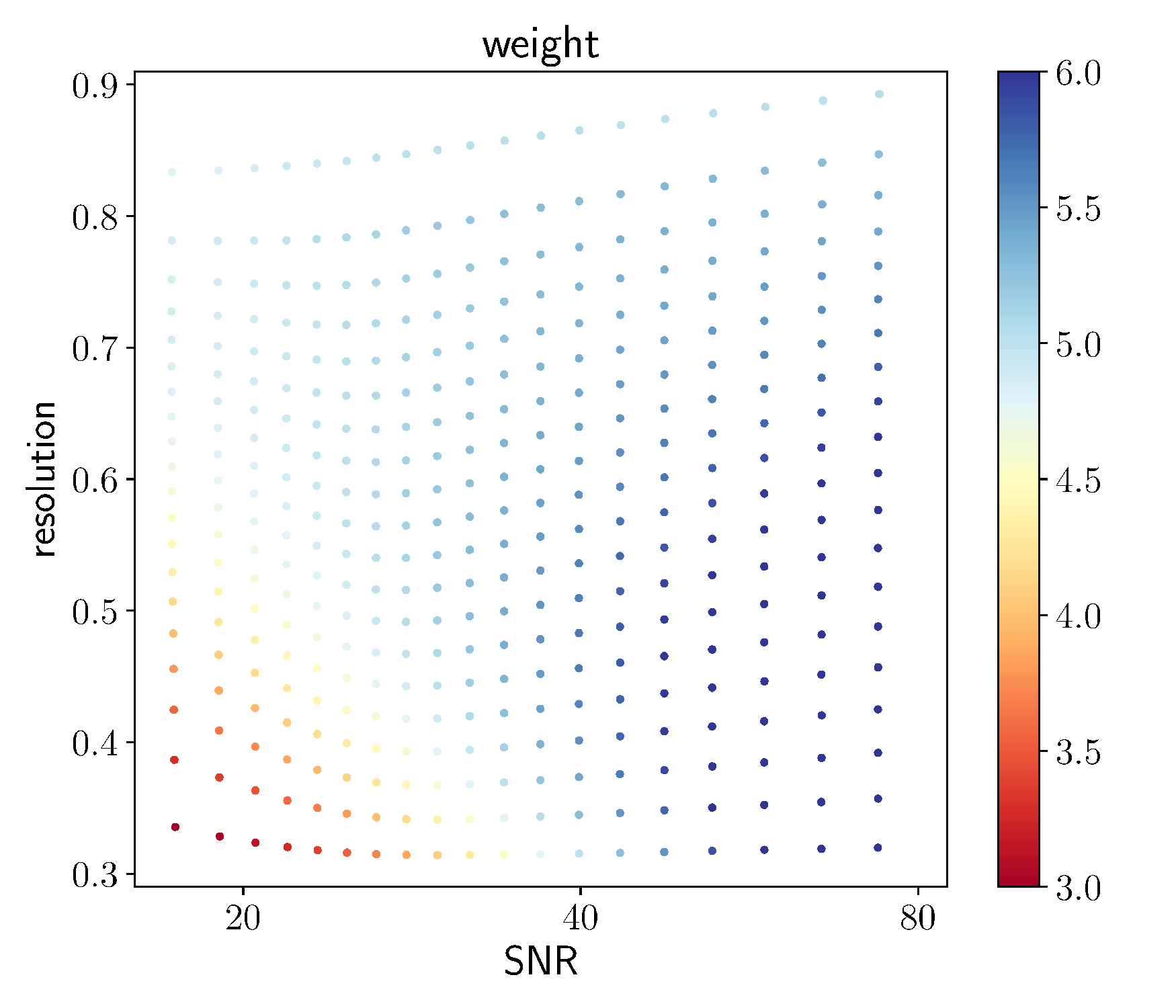}
\end{center}
\caption{
    The left panel shows the $1\sigma$ per-component shape measurement
    uncertainty ($\sigma_e$) estimated with the simulations in different (SNR,
    $R_2$) bins. The middle panel is for the estimated per-component intrinsic
    shape dispersion ($e_{\rm{RMS}}$) following Eq.~\eqref{eq:eRMSEst}. \xlrv{
    The right panel is for the estimated optimal weight.}
    }
    \label{fig:optimal}
\end{figure*}

We define a sliding window in the (SNR, $R_2$) plane with an equal-number
binning scheme and estimate $\sigma_e$ in each bin. The results of this process
are shown in the left panel of Fig.~\ref{fig:optimal}. In order to estimate
$\sigma_e$ for each galaxy in the catalog, we fit a power-law $\sigma_e({\rm
SNR},R_2)$ to the estimated $\sigma_e$, such that
\begin{equation}
    \sigma_e = 0.268 \left(\frac{{\rm SNR}}{20}\right)^{-0.942}
    \left(\frac{R_2}{0.5}\right)^{-0.954},
\end{equation}
\xlrv{
and linearly interpolate the ratio of the estimated values to the fitted
power-law based on the $\log_{10}({\rm SNR})$ and $R_2$ values. For SNR and
$R_2$ outside the bounds of the sliding window, the nearest point within the
sliding window is used for the interpolation of this ratio. As shown, the shape
measurement error from photon noise is a decreasing function in the SNR
direction and the $R_2$ direction since noise has less influence on bright,
large galaxies.
}

Using galaxies in the real HSC shear catalog, we estimate the per-component
intrinsic shape dispersion ($e_{\rm{RMS}}$) by subtracting off (in quadrature)
the shape measurement error from the shape dispersion such that
\begin{equation}
    \label{eq:eRMSEst}
    e_{{\rm RMS}}=\sqrt{\frac{\sum_i \left( e_{1;i}^2+e_{2;i}^2
    -2\sigma_e^2({\rm SNR}_i,R_{2;i})\right)}
    {2 N_\text{gal}}},
\end{equation}
where $i$ is the galaxy index and $N_\text{gal}$ refers to the number of
galaxies in the galaxy ensemble. This estimate is computed in each sliding
window, and the estimated intrinsic shape dispersion as a function of the
position in the (SNR, $R_2$) plane is shown in the middle panel of
Fig.~\ref{fig:optimal}. \xlrv{As shown, the intrinsic shape is a relatively
flat function on the $2$D plane, with a value around 0.4 for most of
parameter space. The corresponding optimal weight defined in
Eq.~\eqref{eq:optimalW} is shown in the right panel of Fig.~\ref{fig:optimal}.}
The shape dispersion is relatively flat with a value around $0.4$; therefore,
we linearly interpolate the function in the $2$D plane to model $e_{{\rm RMS}}$
on the individual galaxy level. The optimal weight is determined with
$\sigma_e$ and $e_{{\rm RMS}}$ following Eq.~\eqref{eq:optimalW}. The
responsivity is determined following Eq.~\eqref{eq:response}.

\subsection{Calibration}
\label{subsec:calib}

In this section, we estimate, model, and remove the shear calibration bias,
except for selection bias, which will be quantified and removed in
Section~\ref{subsec:selBias}. The formalism we applied here generally follows
that introduced in Section~4.5 of \citet{HSC1-GREAT3Sim} but with several
subtle differences that we explicitly flag. \xlrv{ We refer readers to
Section~\ref{subsec:cosshearreq1} and \ref{subsec:additive-ss} for the HSC
three-year weak-lensing science requirements on the residual multiplicative bias
($\abs{\delta m}<9.3\times 10^{-3}$) and the fractional additive bias
($\abs{\delta a}<9.7\times 10^{-3}$), respectively.
}

\subsubsection{Baseline calibration}
\label{subsubsec:baseline}

In order to determine the baseline shear calibration bias in the absence of
selection bias, we keep both galaxies in each $90 \degree$ rotated pair by
imposing the weak lensing cuts on only one randomly chosen galaxy in the pair.
In addition, we force both galaxies in each pair to use the same shape weight
of the randomly chosen galaxy, to avoid weight bias due to the correlation of
shape weight with shear. By doing so, we ensure that both our selection and
weighting processes do not correlate with the input shear, since we wish to
separately quantify and remove those effects.

\begin{figure*}[ht!]
\begin{center}
\includegraphics[width=0.45\textwidth]{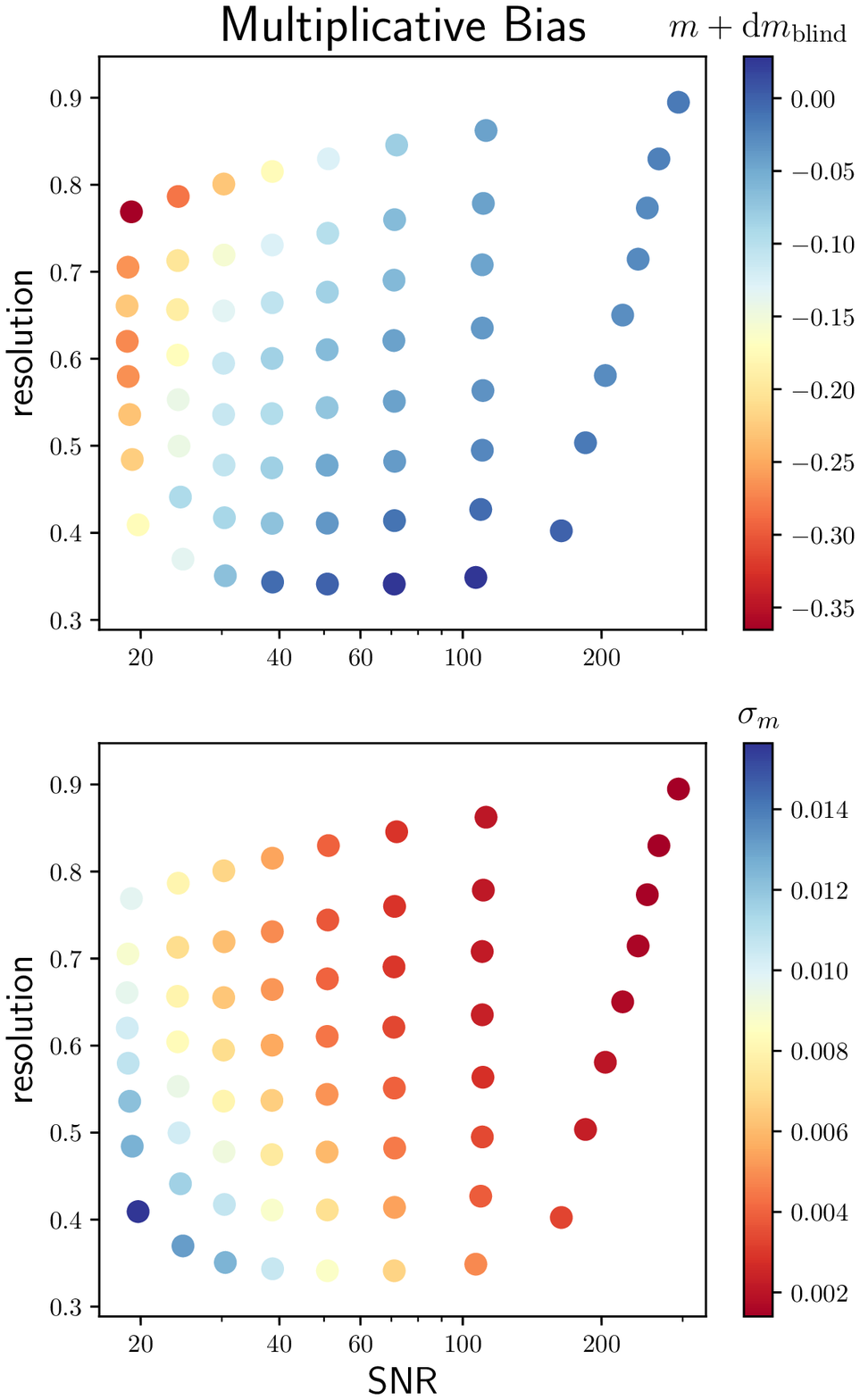}
\includegraphics[width=0.45\textwidth]{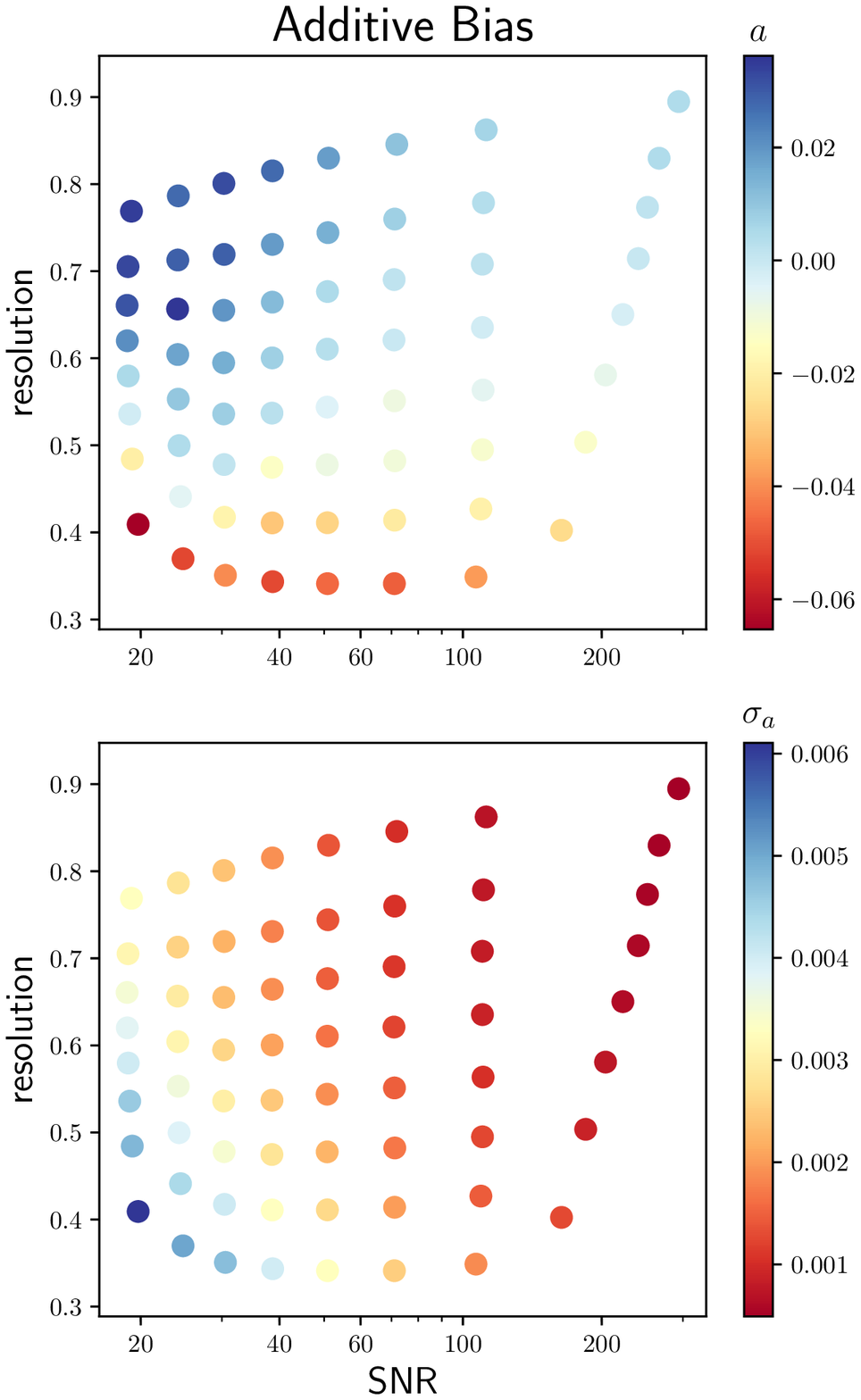}
\end{center}
\caption{
    The left panels show the multiplicative bias (upper left) and its standard
    deviation (lower left) estimated in the (SNR,$R_2$) plane using the image
    simulations. The right panels are for the fractional additive bias. Note
    that the multiplicative bias is blinded by adding a shift $dm_{\rm blind}$.
    }
    \label{fig:baseline_bias}
\end{figure*}

In the upper left panel of Fig.~\ref{fig:baseline_bias}, we show the baseline
multiplicative bias as a function of position in the (SNR, $R_2$) plane with an
equal-number binning scheme for the overall simulation. When making the figure,
an unspecified constant value is added to the multiplicative bias to blind our
shear analysis. For reference, the lower left panel shows the standard
deviation of the multiplicative bias estimation in the upper left panel.
Similarly to what was done to model the shape measurement error in
Section~\ref{subsec:optimal}, we fit $m({\rm SNR},R_2)$ to a power-law in both
parameters plus a constant offset. The best-fit power-law is shown as follows:
\begin{equation}
    m({\rm SNR},R_2)+ {\rm const.} \propto
    \left(\frac{R_2}{0.5}\right)^{1.66}\left(\frac{{\rm SNR}}{20}\right)^{-1.24}\,.
\end{equation}
\xlrv{
We then interpolate a correction to the power-law based on the ratio between
the multiplicative bias estimation and the power-law, and the interpolation
scheme is the same as that for the shape measurement error due to photon noise.
}

In the upper right panel of Fig.~\ref{fig:baseline_bias}, we show the baseline
fractional additive bias as a function of position in the (SNR, $R_2$) plane
with an equal-number binning scheme for the overall simulation. For reference,
the lower right panel shows the standard deviation of the additive bias
estimation in the upper left panel. Similarly to the modelling of the baseline
multiplicative bias, we fit the estimated baseline fractional additive bias to
the model proposed in \citet{HSC1-GREAT3Sim}. The best-fit model is shown as
follows:
\begin{equation}
    a({\rm SNR},R_2) \propto
        (R_2-0.61)\left(\frac{{\rm SNR}}{20}\right)^{-0.94}\,.
\end{equation}
Subsequently, we interpolate a correction to the model based on the difference
between the fractional additive bias estimation and the model.

\subsubsection{Weight bias}
\label{subsubsec:weightBias}

\begin{figure*}[ht!]
\begin{center}
\includegraphics[width=0.45\textwidth]{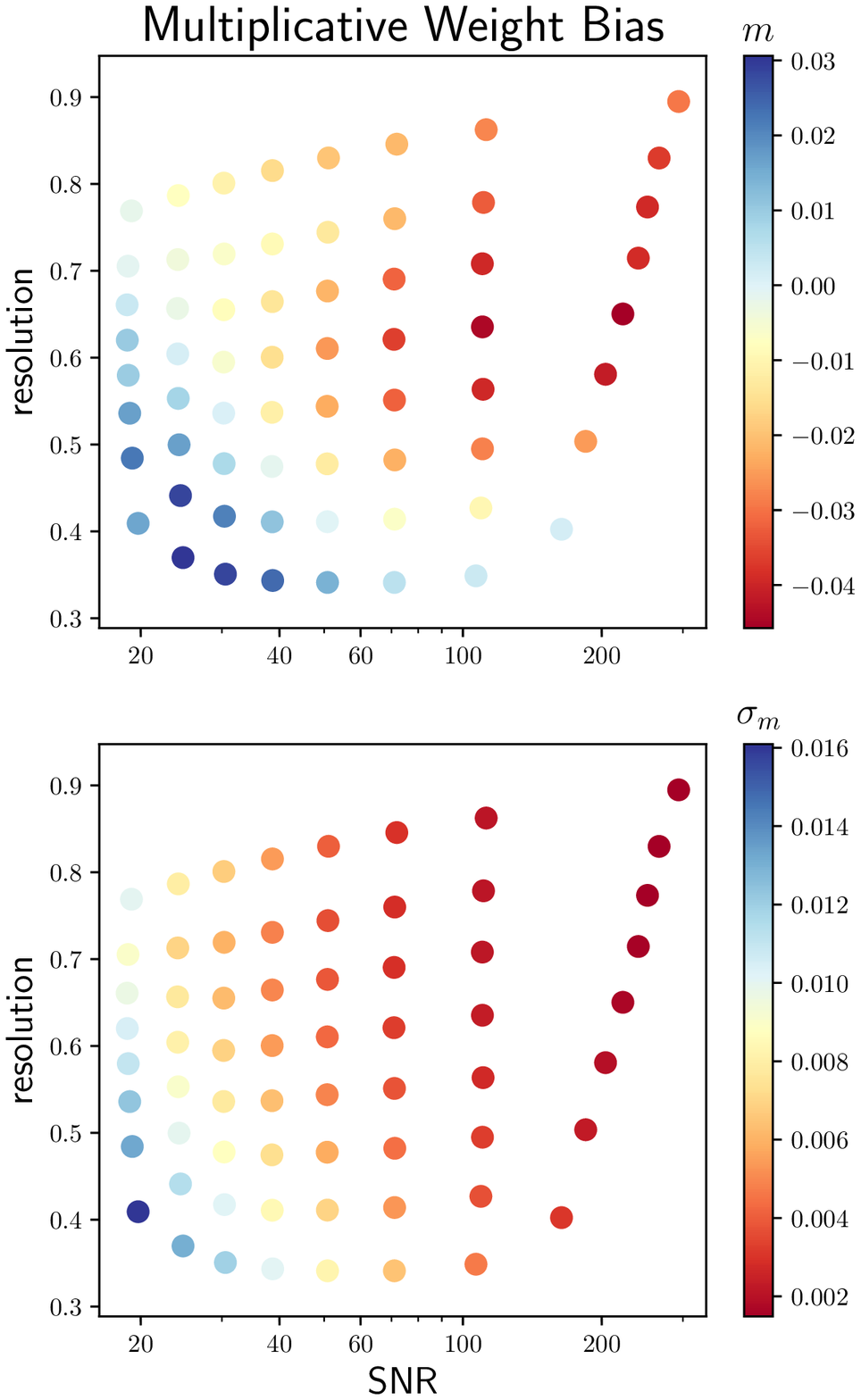}
\includegraphics[width=0.45\textwidth]{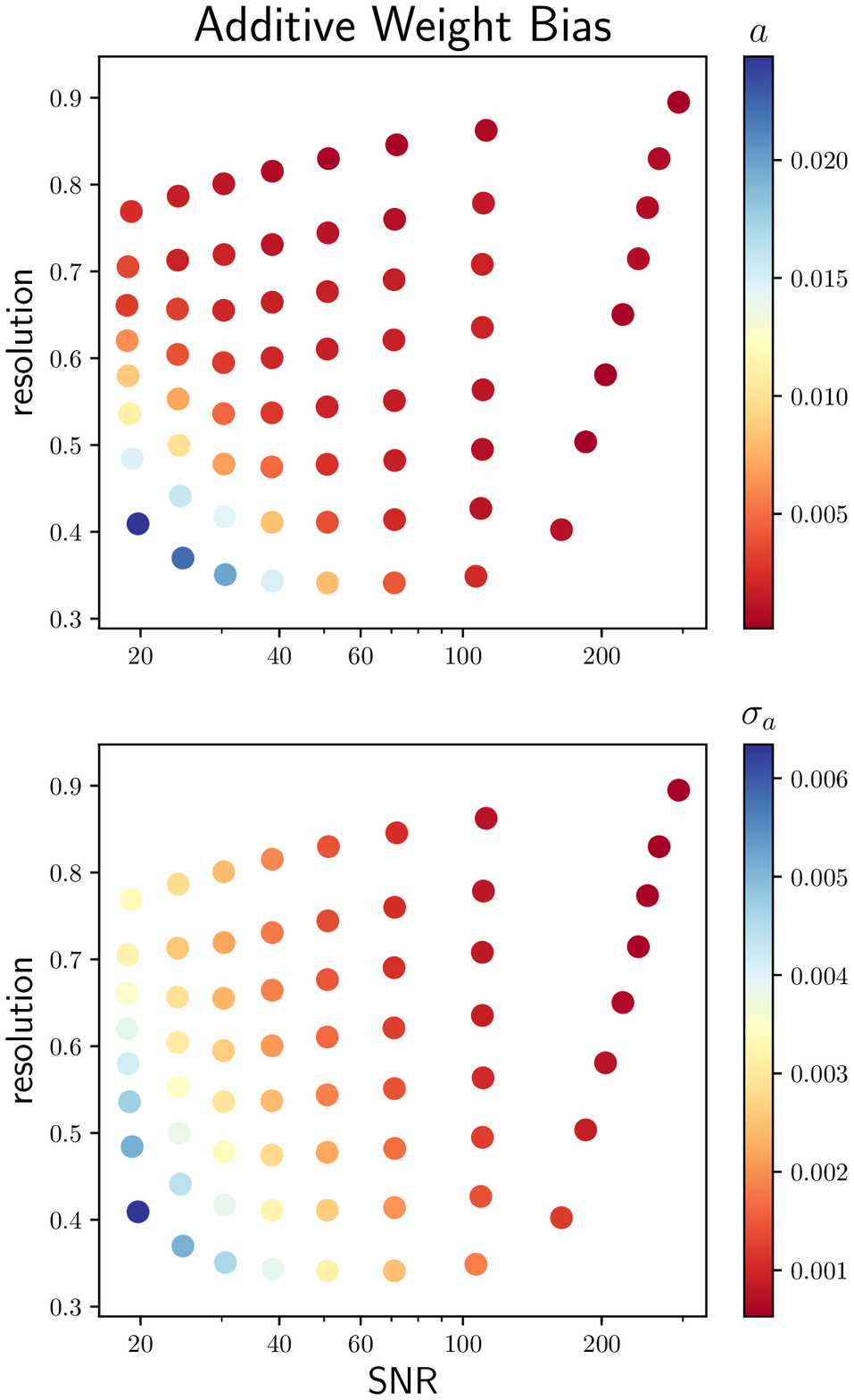}
\end{center}
\caption{
    The left panels show the multiplicative weight bias (upper left) and the
    standard deviation (lower left) of the multiplicative bias estimated in the
    (SNR,$R_2$) plane using the simulation. The right panels are for the
    fractional additive weight bias.
    }
    \label{fig:weight_bias}
\end{figure*}

Weight bias refers to the bias in estimated shear due to a correlation between
the adopted shape weight and the true lensing shear. It can also be regarded as
the bias from a shear-dependent smooth selection, since weighting is
effectively a smooth selection \citep{HSC1-GREAT3Sim}. Weight bias can be
corrected analytically if the response of the weight to the shear distortion is
known \citep[e.g.,][]{FPFS-Li2018}. On the other hand, weight bias can also be
estimated using image simulations containing $90\degree$ rotated pairs
\citep{HSC1-GREAT3Sim}.

Here we follow the scheme of \citet{HSC1-GREAT3Sim} to estimate weight bias
using image simulations by comparing the shear bias estimation with and without
enforcing the same shape weight for each galaxy in an orthogonal galaxy pair.
In Fig.~\ref{fig:weight_bias}, we show the multiplicative weight bias (left
panel) and the fractional additive weight bias (right panel). The binning
scheme here is the same as used in Section~\ref{subsubsec:baseline}. We find a
statistically significant multiplicative weight bias that depends on galaxy
properties. As shown, this bias is negative and reaches a maximum amplitude of
$-0.045$ at high SNR and $R_2$, while it is positive and reaches a maximum
amplitude of $0.03$ at low SNR and $R_2$. We also find a small additive weight
bias with $\lesssim 5\sigma$ significance. The additive weight bias reaches its
maximum of $0.025$ at low SNR and $R_2$, and it decreases as SNR and $R_2$
increase.

Considering that the weight biases are dependent on the location in the $2$D
plane, we use the same process as in Section~\ref{subsubsec:baseline} to model
and interpolate the weight biases as functions of position in the $2$D plane.

\subsubsection{Redshift dependence}
\label{subsubsec:zBias}

\begin{figure*}[ht!]
\begin{center}
\includegraphics[width=0.85\textwidth]{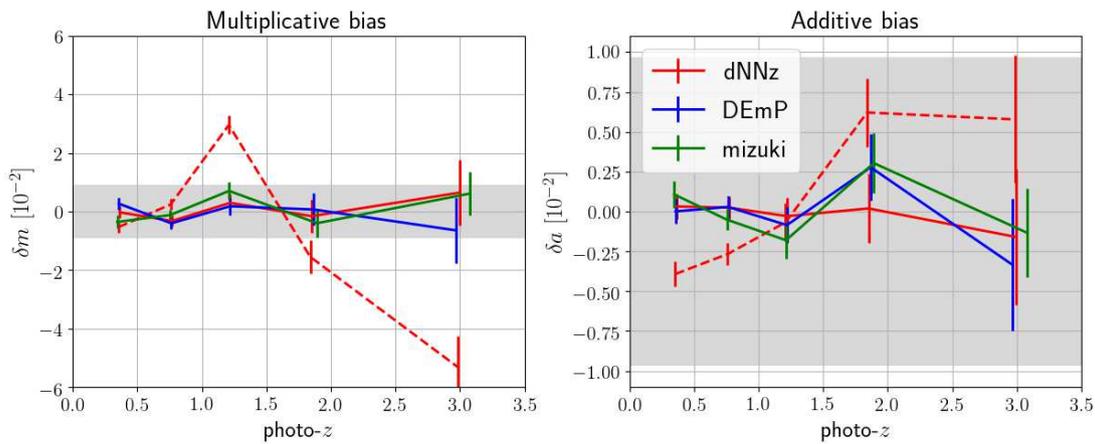}
\end{center}
\caption{
    The left (right) panel shows the redshift-dependent multiplicative
    (fractional additive) bias. The red lines are for $\dNNz$ photo-$z$, the
    blue lines are for $\DEmP$ photo-$z$, and the green lines are for $\mizuki$
    photo-$z$. The dashed lines are the results before removing the
    redshift-dependent bias, whereas the solid lines are the results after
    modelling and calibrating the redshift-dependent bias with $\dNNz$
    photo-$z$. The gray regions indicate the requirements on calibration
    residuals that will be defined in Section~\ref{sec:requirement}.
    }
    \label{fig:redshift_bias}
\end{figure*}

Since weak lensing analyses often divide the galaxy sample into different
photometric redshift (photo-$z$) bins
\citep[e.g.,][]{cosmicShear_HSC1_Chiaki2019,HSC1-cs-real}, or use photometric
redshift-dependent weights
\citep[e.g.,][]{massCalib-CAMIRA-Murata2019,massCalib-actpolCluster-Miyatake2019},
quantifying and correcting the redshift-dependent shear calibration biases are
crucially important. We note that some redshift-dependent biases are already
partially accounted for by the calibrations in
Sections~\ref{subsubsec:baseline} and \ref{subsubsec:weightBias}, which model
the calibration biases as functions of $R_2$ and SNR. In this section, we look
into the remaining redshift dependence of the shear estimation biases after
those effects are already accounted for.

Currently, we only have realistic simulations for $i$-band images since our
input galaxy sample are from the single-band F814W HST exposures. Therefore,
photometric redshifts cannot be directly derived from our simulated images. We
will follow \citet{FPFSHSC1-Li2020}, and use the photo-$z$ estimates of the
input galaxies as a proxy of the measured redshift in the simulations to study
the redshift-dependent shear estimation biases.

In particular, we match the input COSMOS galaxies to the HSC S19A photo-$z$
catalog in the Wide layer according to the angular position of the input
galaxies, and assign each galaxy in the simulations the estimated redshift
of the matched galaxy in the HSC photo-$z$ catalog.

For cross validation, we use three different HSC photo-$z$ estimates: the Deep
Neural Net Photometric Redshift \citep[$\dNNz$;][]{dnnz-Nishizawa2021}, Direct
Empirical Photometric code \citep[$\DEmP$;][]{DEmPZ-Hsieh2014}, and Mizuki
photometric redshift \citep[$\mizuki$;][]{MizukiZ-Tanaka2015}, which are based
on neural network, empirical polynomial fitting, and Bayesian template fitting,
respectively. To be specific, we estimate and remove calibration bias as a
function of the $\dNNz$ photo-$z$. Then we use the $\DEmP$ photo-$z$ and the
$\mizuki$ photo-$z$ for cross-validation tests. The details of the $\DEmP$ and
the $\mizuki$ photo-$z$ catalogs are summarized in \citet{HSC2-photoz}, and the
$\dNNz$ photo-$z$ catalog is described in \citet{dnnz-Nishizawa2021}.

We divide the simulations into $\dNNz$ photo-$z$ bins of equal-numbers of
galaxies with selection bias cancellation by enforcing that orthogonal galaxy
pairs are in the same bin. The multiplicative and additive bias are estimated
for each bin. Then we compare the estimated biases with the predicted biases
using the calibration model derived in Sections~\ref{subsubsec:baseline} and
\ref{subsubsec:weightBias}. Here, we force the shape noise cancellation by
using orthogonal galaxy pairs to cancel out selection bias due to galaxy cuts,
while we do not force the galaxy pairs to have the same shape weight to cancel
weight bias, because weight bias has already been estimated and included in the
calibration parameters (see Section~\ref{subsubsec:weightBias}).

The dashed red lines in Fig.~\ref{fig:redshift_bias} show the residuals of
multiplicative bias (left panel) and additive bias (right panel) as a function
of $\dNNz$ redshift. We model the redshift-dependent biases by linearly
interpolating the bias residuals across the redshift bins.

We note that the first-year HSC shear calibration paper find a redshift
dependence of the per-component intrinsic shape dispersion ($e_{\rm{RMS}}$)
using a training sample of parametric galaxies fitted to the COSMOS HST
galaxies with redshift ranging from $0$ to $1.5$.  \citet{HSC1-GREAT3Sim}
estimated the multiplicative bias caused by such redshift dependence, and
reported a multiplicative biases of $-1\%$ and $3\%$ for galaxies in the
photo-$z$ range $[0,1]$ and $[1,1.5]$, respectively. Our estimation of
redshift-dependent multiplicative bias has the same trend as that in
\citet{HSC1-GREAT3Sim} in the redshift range $[0,1.5]$. In contrast, our
estimation covers the redshift range $[0,4]$ and includes all sources of
redshift-dependent shear measurement bias. The redshift-dependent additive bias
is shown in the right panel of Fig.~\ref{fig:redshift_bias}; even prior to
correction, it is within the three-year systematic error requirements that will
be defined in Section~\ref{sec:requirement}.

\subsubsection{Combined estimates of calibration bias}

The final multiplicative bias and additive bias estimates for each galaxy in
the catalog are the sum of the baseline bias modeled in
Section~\ref{subsubsec:baseline}, the weight bias modeled in
Section~\ref{subsubsec:weightBias}, and the residual redshift-dependent bias
modeled in Section~\ref{subsubsec:zBias}. The outputs of the calibration are
summarized in Table~\ref{tab:simout}.

\begin{table*}
\caption{
    The outputs from the analyses based on the image simulations. The first
    three outputs are derived to optimize the shear estimation as described in
    Section~\ref{subsec:optimal}. The last three outputs are derived to
    calibrate the shear estimation as described in
    Section~\ref{subsubsec:baseline}, Section ~\ref{subsubsec:weightBias}, and
    Section~\ref{subsubsec:zBias}.
    }
\begin{center}
\begin{tabular}{ll} \hline
    Output properties & Meaning \\ \hline
\multicolumn{2}{c}{Optimization} \\ \hline
\texttt{i$\_$hsmshaperegauss$\_$derived$\_$sigma$\_$e}& Measurement error from photon noise \\ \hline
\texttt{i$\_$hsmshaperegauss$\_$derived$\_$rms$\_$e}  & Shape noise dispersion\\ \hline
\texttt{i$\_$hsmshaperegauss$\_$derived$\_$weight}    & Weak lensing shape weight\\ \hline
\multicolumn{2}{c}{Calibration} \\ \hline
\texttt{i$\_$hsmshaperegauss$\_$derived$\_$shear$\_$bias$\_$m} & Multiplicative bias\\ \hline
\texttt{i$\_$hsmshaperegauss$\_$derived$\_$shear$\_$bias$\_$c1}& The first component of additive bias\\ \hline
\texttt{i$\_$hsmshaperegauss$\_$derived$\_$shear$\_$bias$\_$c2}& The second component of additive bias\\ \hline
\hline
\end{tabular}
\end{center}
\label{tab:simout}
\end{table*}

\subsection{Ensemble calibration uncertainties}
\label{subsec:Dcalib}

\begin{figure*}[ht!]
\begin{center}
\includegraphics[width=0.85\textwidth]{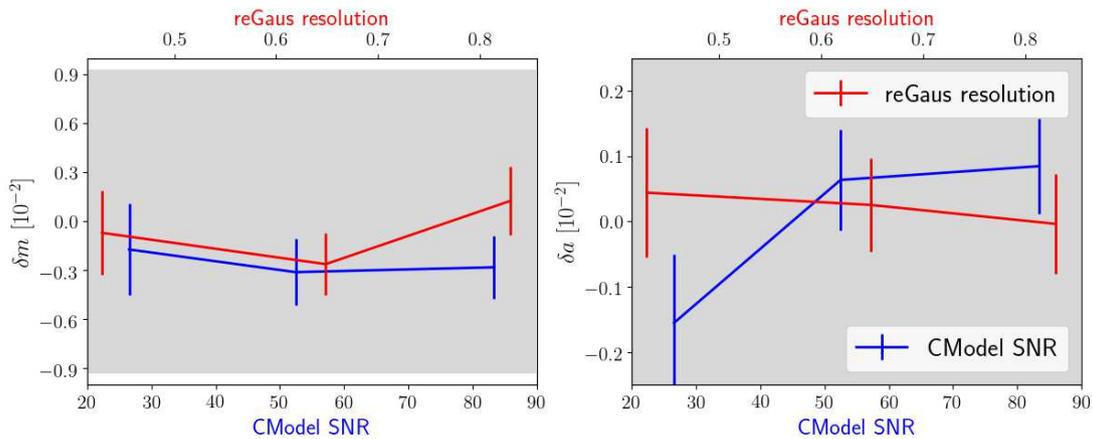}
\end{center}
\caption{
    The calibration residuals for subsamples binned by two modeled galaxy
    properties ,i.e., $R_2$ (red) and SNR (blue). The left (right) panel shows
    the multiplicative (fractional additive) bias. The gray regions indicate
    the requirements on calibration residuals that will be defined in
    Section~\ref{sec:requirement}.
    }
\label{fig:bias_resModel}
\end{figure*}

This section serves to demonstrate the validity and robustness of the
calibration of the shear biases (i.e., multiplicative bias and additive bias)
derived in Section~\ref{subsec:calib}, and assign a systematic uncertainty to
the calibration at the ensemble level. We focus on the systematic calibration
residuals for multiplicative bias ($\delta m$) and fractional additive bias
($\delta a$), which are the remaining bias after the shear calibration of
Section~\ref{subsec:calib}. The selection bias is not taken into account here,
and we force the shape noise cancellation by using orthogonal galaxy pairs to
cancel out selection bias due to galaxy cuts as in Section~\ref{subsec:calib}.

First, we divide the simulations into several subsamples following an
equal-number binning scheme by the galaxy properties including those used for
modeling shear biases (i.e., CModel SNR, $\reGauss$ resolution, and $\dNNz$
photo-$z$) and those that are marginalized, that is, not explicitly taken into
account in the bias modelling (i.e., CModel magnitude, seeing, $\DEmP$ and
$\mizuki$ photo-$z$). Shear is subsequently estimated for each subsample in
each subfield using the calibrated shear estimator. Finally, we determine the
bias residuals for each property-binned subsample using Eq.~\eqref{eq:fitmc}.

\begin{figure*}[ht!]
\begin{center}
\includegraphics[width=0.85\textwidth]{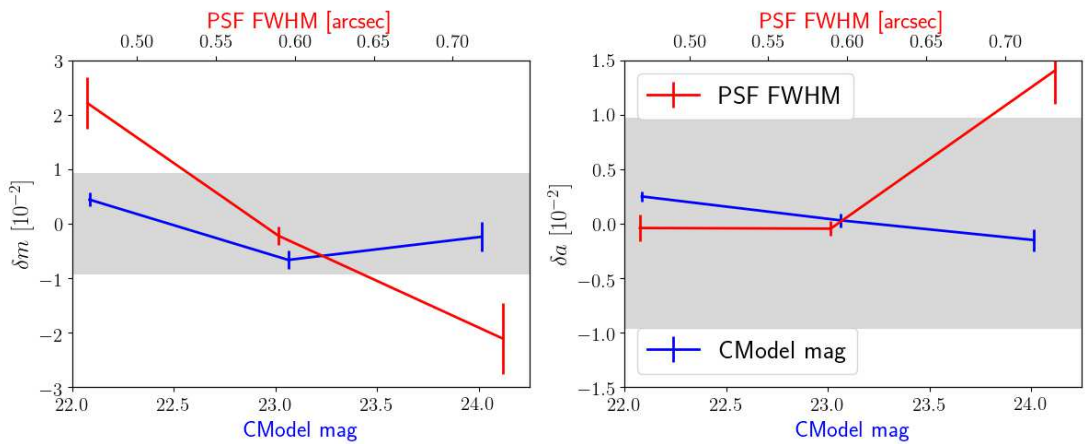}
\end{center}
\caption{
    The calibration residuals for subsamples binned by two marginalized galaxy
    properties, i.e., seeing size (red) and CModel magnitude (blue). The left
    (right) panel shows the multiplicative (fractional additive) bias. The gray
    regions indicate the requirements on calibration residuals defined in
    Section~\ref{sec:requirement}.
    }
\label{fig:bias_resMargin}
\end{figure*}

The red solid lines in Fig.~\ref{fig:redshift_bias} show the bias residuals
with $\dNNz$ photo-$z$ binning. Fig.~\ref{fig:bias_resModel} shows the
calibration bias residuals when binning the simulations with SNR or $R_2$. The
results demonstrate that the amplitude of the multiplicative bias residual
($\delta m$) is less than $0.5\%$, the fractional additive bias residual
($\delta a$) is less than $0.5\%$, both of which are within the systematic
error requirements that will be defined in Section~\ref{sec:requirement}. These
bias residuals are expected to be consistent with zero since these galaxy
properties were used to model the calibration bias calibration.

Finally we test the dependence of the bias residuals on the marginalized
properties. We demonstrate the bias residuals when binning galaxies by $\DEmP$
and $\mizuki$ photo-$z$ in Fig.~\ref{fig:redshift_bias}.
Fig.~\ref{fig:bias_resMargin} shows the bias residuals when binning the
simulations with CModel magnitude and seeing size. We do not find calibration
bias residuals beyond the requirement limits for the cases of $\DEmP$
photo-$z$, $\mizuki$ photo-$z$, and CModel magnitude.

\begin{figure}[ht!]
\begin{center}
\includegraphics[width=0.48\textwidth]{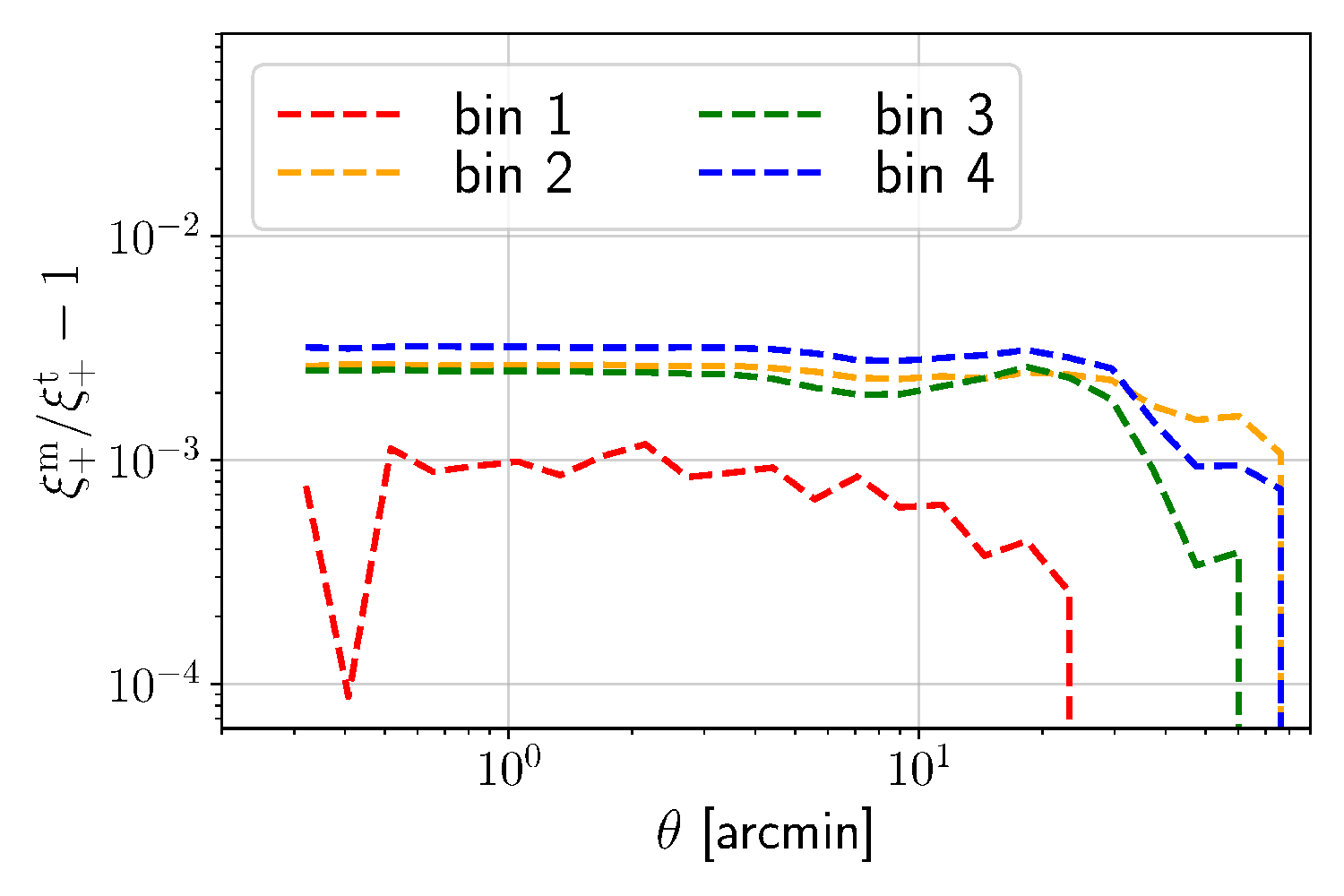}
\end{center}
\caption{
    \xlrv{
    The relative bias on the shear-shear correlation function as a function of
    separation angle caused by the seeing-dependent calibration bias residual.
    Lines with different colors refer to the shear-shear auto-correlations in
    four different redshift bins.
    }
    }
    \label{fig:csSeeing}
\end{figure}

However, when binning by seeing size, the residuals of the multiplicative bias
exceed our requirements for the best and worst seeing bins, and the residuals
of the fractional additive bias slightly exceed the requirements for the worst
seeing bin, which is consistent with \citet{HSC1-GREAT3Sim}. The binning by
seeing size corresponds to an extreme case of splitting up the survey based on
regions with specific properties. Our finding implies that weak lensing
analyses with strict area cuts should evaluate the seeing distribution after
the cuts, and then evaluate whether additional shear calibration biases are
required to be removed for such an area. For a weak lensing analysis that
neither weights galaxies by the seeing size nor divides galaxies into seeing
bins, the calibration bias residuals shown by the red lines in
Figure~\ref{fig:bias_resMargin} will not bias the analysis, since the
calibration bias residuals averaging over seeing sizes are within the
requirement limits. It suggests that this result is not relevant to the cosmic
shear analysis and the galaxy-galaxy lensing analysis using lens samples
covering the entire HSC survey area \citep[e.g., CMASS galaxy
sample;][]{LOWZ_CMASS-Reid2016}.

\xlrv{
    However, we test the impact of this seeing-dependent calibration bias more
    rigorously with a realization of the three-year HSC mock catalog
    \citep{HSC3-mock} constructed using the full-sky lensing simulation of
    \citet{FullSkySim-Takahashi2017}. The mock catalog uses the angular
    coordinates, seeing sizes, galaxy fluxes, photo-$z$ estimation etc.\ of the
    HSC shape catalog, and it samples the true redshift for each galaxy using
    its \dNNz{} photo-$z$ posterior distribution; therefore, the mock has the
    same spatial distribution of the seeing as the data. Lensing shear from the
    full-sky lensing simulation is assigned to each galaxy according to its
    position, and shape noise is not included in this test. We fit the
    calibration residual shown by the red line in the left panel of
    Figure~\ref{fig:bias_resMargin} as a function of seeing FWHM with a linear
    model and use the derived model to assign a multiplicative bias for each
    galaxy in the mock according to its seeing size. Note that when ignoring
    the seeing dependence, the multiplicative bias should give zero spurious
    shear correlations. We subsequently divide galaxies in the mock into four
    redshift tomographic bins from $z=0.3$ to $z=1.5$ with equal separation
    following \citet{HSC1-cs-real} and compute the shear-shear autocorrelation
    function in each redshift bin. In Figure~\ref{fig:csSeeing}, we show the
    relative difference between the results from the mock with the
    seeing-dependent multiplicative bias residual (denoted as $\xi^{\rm m}_+$)
    and the results from the same mock but without multiplicative bias (denoted
    as $\xi^{\rm t}_+$). The results show that the relative difference is less
    than 0.4\%, and the resulting bias on $S_8 = \sigma_8 (\Omega_{\rm
    m0}/0.3)^{0.5}$ should be less than 0.2\%. The method of accounting for
    calibration uncertainties for galaxies in each tomographic bin in the
    cosmic shear analysis will be discussed in detail in the cosmic shear
    paper.
}

\subsection{Selection bias}
\label{subsec:selBias}

\begin{figure*}[ht!]
\begin{center}
\includegraphics[width=0.85\textwidth]{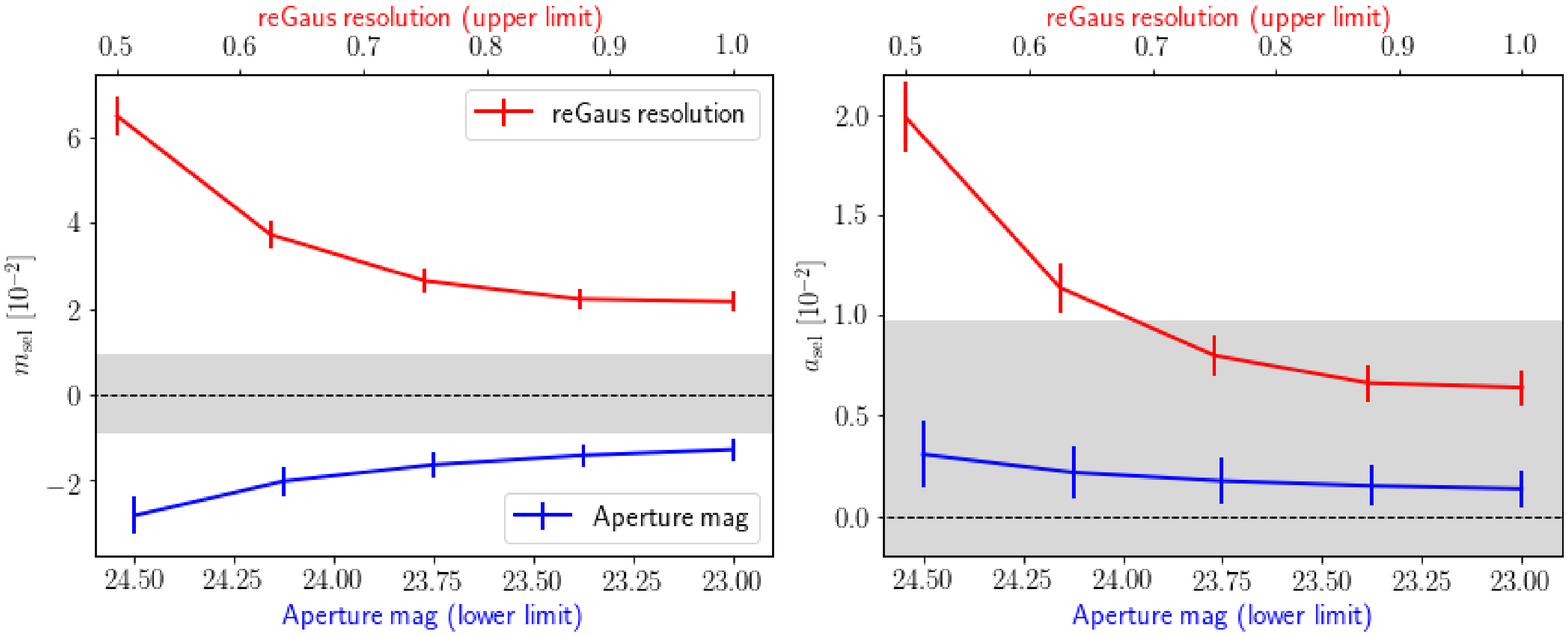}
\includegraphics[width=0.85\textwidth]{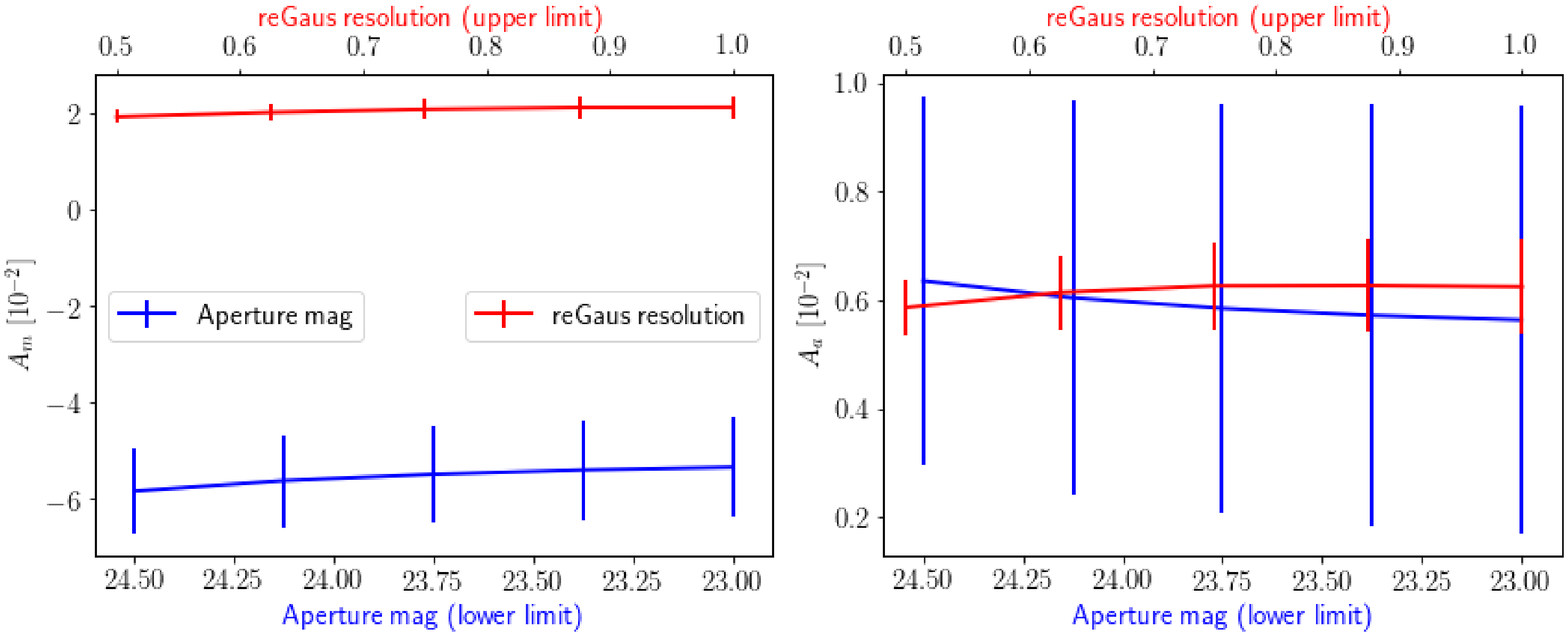}
\end{center}
\caption{
    The upper left (upper right) panel shows the multiplicative (fractional
    additive) bias due to cuts on resolution $R_2$ (red) and aperture magnitude
    mag$_A$ (blue). The gray regions indicate the requirements on calibration
    residuals defined in Section~\ref{sec:requirement}, and the horizontal
    dashed lines are $y=0$. The lower left (lower right) panel shows the
    multiplicative (fractional additive) bias ratio of the cuts on resolution
    $R_2$ and aperture magnitude mag$_A$. For resolution, we fix the lower
    limit and change the upper limit. For aperture magnitude, we fix the upper
    limit and change the lower limit. For each selection cut, errorbars are
    correlated between the points since at least a fraction of the same
    simulated galaxies are used for the calculations.
    }
    \label{fig:selection_bias}
\end{figure*}

Given that the amplitudes of the lensing shear and the PSF anisotropy are
small, the anisotropic selection has little influence on the galaxies that are
far away from the selection edge. The selection bias should be proportional to
the marginal density at the edge \citep[see][for analytical correction of
selection bias]{FPFS-Li2021}. Here we follow \citet{HSC1-GREAT3Sim} to
empirically estimate the selection bias by comparing the shear estimation of
the overall sample with/without forcing the inclusion of $90\degree$ rotated
pairs.

We focus on the correction for the selection bias due to cuts on resolution and
aperture magnitude, as we find that the selection biases for other cuts on
$i$-band galaxy properties (e.g., CModel SNR, CModel Magnitude) are consistent
with zero, and the selection bias for the multi-band detection cut is
negligible since the cut removes less than one percent of the  galaxies from
the parent sample. The upper panels of Fig.~\ref{fig:selection_bias} show the
estimated selection biases for the resolution cut ($R_2>0.3$) and the aperture
magnitude cut (mag$_A<25.5$) listed in Table~\ref{tab:icut}, while changing the
upper (lower) limit of resolution (aperture magnitude) to change the galaxy
ensemble. We find that the multiplicative bias for the resolution (magnitude)
cut is constantly positive (negative), and the fractional additive bias for the
resolution cut is constantly positive. The fractional additive bias for the
aperture magnitude cut is consistent with zero within $2\sigma$ and is within
the three-year HSC science requirement. As we expect, the amplitudes of the
biases decrease when the sizes of the corresponding galaxy ensembles increase
since the fractions of the galaxies that are close enough to the selection
edges and are influenced by the anisotropic selections decrease.

In order to empirically estimate and remove the selection bias for any galaxy
sample due to the two aforementioned cuts, we adopt the method proposed by
\citet{HSC1-GREAT3Sim}. The premise of the method is that, for a galaxy sample,
the ratio between the selection biases, from a cut on galaxy observable ($X$),
versus the marginal galaxy number density at the edge of the cut
($P(X)\mid_{{\rm edge}}$) is approximately constant. The selection bias ratios
are defined as
\begin{equation}
\begin{split}
    A_m(X)=\frac{m^\mathrm{sel} \left(X\right)}{P\left(X\right)\mid_{{\rm edge}}},\\
    A_a(X)=\frac{a^\mathrm{sel} \left(X\right)}{P\left(X\right)\mid_{{\rm edge}}}\,.
\end{split}
\end{equation}
The lower panels of Fig.~\ref{fig:selection_bias} show the selection bias
ratios for $R_2$ and mag$_A$. Here we fix the lower limit of resolution at
$R_2=0.3$ and the upper limit of aperture magnitude at mag$_A=25.5$,
respectively. Then we adjust the upper limit of $R_2$ and the lower limit of
mag$_A$ to change the galaxy sample.

As demonstrated by the lower panels of Fig.~\ref{fig:selection_bias}, the
selection bias ratios vary slowly with the change of the galaxy sample;
therefore, we take the selection bias ratios as constants. The selection bias
ratios are used to estimate selection biases for any galaxy sample by
multiplying them by the marginal galaxy number densities at the edges of the
corresponding selection cuts. The resulting multiplicative and fractional
additive selection biases for $R_2$ and mag$_A$ are shown as follows:
\begin{equation}
\begin{split}
    m^\mathrm{sel}&=-0.05854 P\left(\text{mag}_A=25.5\right)+0.01919 P\left(R_2=0.3\right),\\\notag
    a^\mathrm{sel}&=0.00635 P\left(\text{mag}_A=25.5\right)+0.00627 P\left(R_2=0.3\right),
\end{split}
\end{equation}
respectively. In cosmological analyses, this equation should be used to
estimate the selection biases for specific galaxy ensembles according to the
marginal galaxy number densities. The selection bias should be removed from the
shear estimation if it is beyond the requirement limits.

\subsection{Redshift-dependent blending}

\begin{figure}[ht!]
\begin{center}
\includegraphics[width=0.45\textwidth]{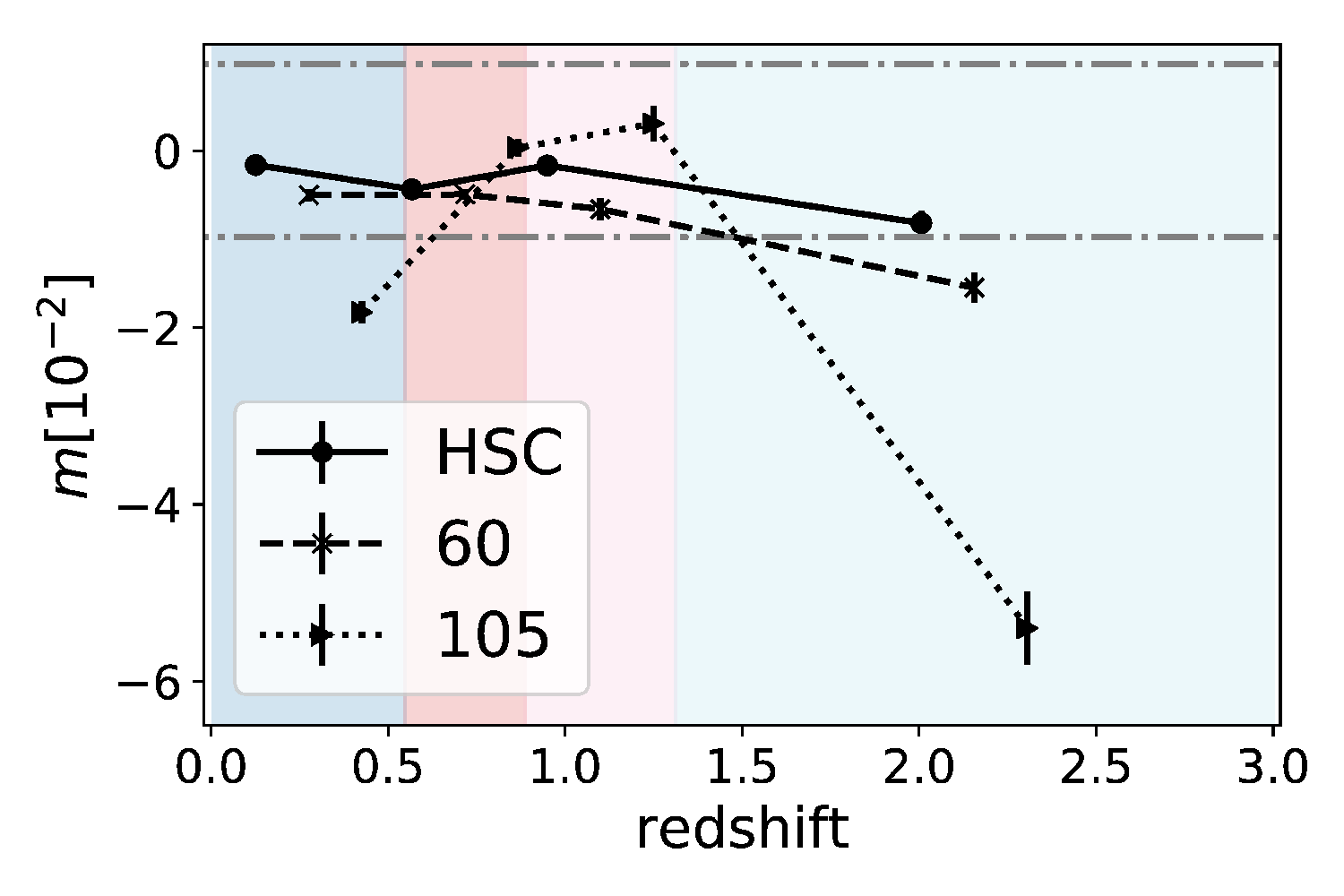}
\end{center}
\caption{
    \xlrv{The excess multiplicative bias in our fiducial calibration due to
    redshift-dependent blending in four redshift bins as indicated by colored
    regions. The solid, dashed and dotted lines are for simulations with an HSC
    PSF with FWHM=$0\farcs58$, Moffat PSF with FWHM$=0\farcs6$, and Moffat PSF with
    FWHM$=1\farcs05$, respectively. The errorbars are estimated using a jacknife
    for different noise realizations. The horizontal dash-dotted lines show
    the three-year HSC requirement.}
    }
    \label{fig:zBlendBias}
\end{figure}

\xlrv{
The DES Y3 analysis in \citet{DESY3-BlendshearCalib-MacCrann2021} used
parametric galaxy models with known redshifts as their image simulation
training sample.  They randomly populated these parametric galaxies with a
detection density matched to the DES observations to simulate multi-band DES
images that were used to test and calibrate \texttt{M{\scriptsize
ETACALIBRATION}} \citep{metacal-Sheldon2017}. They tested for the circumstance
that galaxies at different redshifts were distorted by different shear signals
and compared the results with those from the conventional constant-shear
simulations. According to Fig.~8 of \citet{DESY3-BlendshearCalib-MacCrann2021},
the amplitude of the additional bias due to the redshift-dependent shear is
below $1\%$ for redshift $z<1$, while it reaches $\sim 3$\% for redshifts
$1<z<3$ for the DES observational conditions.

We note that it is impossible to directly apply different shear distortions to
blended galaxies separately in our fiducial simulations since they are
constructed using postage stamps directly cut out from the COSMOS HST
images. Therefore, galaxies in one HST postage stamp can only be distorted by a
single constant shear as a whole, and any bias due to
redshift-dependent blending is not included in our fiducial calibration.  Here
we investigate the multiplicative bias that is not captured by our fiducial
calibration due to the difference between the  constant-shear setup and
a redshift-dependent-shear setup.

We make additional image simulations using parametric galaxy models fitted to
the galaxies in the HST F814W shape catalog
\citep{HST-shapeCatalog-Alexie2007}. We randomly populate these parametric
galaxies into a region with an area of 141~arcmin$^2\,$. The density of the
input galaxies is set to $\sim 88$~arcmin$^{-2}$, which is the same as
\citet{DESY3-BlendshearCalib-MacCrann2021}. The redshifts of the
galaxies are set by matching the HST F814W shape catalog to the COSMOS
photo-$z$ catalog \citep{COSMOSZ-Ilbert2009} using their coordinates. The
galaxies are distorted with different shears and convolved with three different
PSFs (an HSC PSF with FWHM$=0\farcs58$ and Moffat PSFs with input
FWHM$=0\farcs6$ and FWHM$=1\farcs05$). Different realizations of pixel noise
with variance set to the average of the HSC noise variance are added to the image.
We confirm that, after the weak-lensing cuts, the difference between the galaxy
number in the simulation (with the HSC PSF) and that of the real HSC data is
less than 4\%. In addition, the galaxy number histograms over galaxy properties
(e.g.\ CModel SNR, \reGauss{} resolution and CModel magnitude) visually match
those of the real HSC data.

We first determine the calibration bias using a {\it constant-shear} setup
where all of the galaxies in one image are distorted with the same shear. We
simulate two images with shear distortions $\gamma_1=+0.02$ and
$\gamma_1=-0.02$, with the same noise on these two images to reduce the impacts
of shape and pixel noise \citep{preciseSim-Pujol2019}. We repeat the
simulations with different noise realizations. For each image, we detect,
deblend and measure the properties of the sources using the HSC pipeline. The
weak-lensing cut introduced in Section~\ref{subsubsec:GalSel} is then applied
to the detected sources. We match the galaxies selected by the weak-lensing cut
to the input galaxy catalog using their coordinates and assign each selected
galaxy with the redshift of the closest match in the input galaxy catalog. We
measure the difference between the average shears (over noise realization)
measured from the simulations with $\gamma_1=+0.02$ and $\gamma_1=-0.02$, and
divide it by the difference in the input shear distortion ($\Delta
\gamma_1=0.04$) to determine the calibration bias. By dividing the selected
galaxy into four redshift bins as indicated by the four colored regions in
Figure~\ref{fig:zBlendBias}, we estimate the multiplicative bias in each
redshift bin.

Then we apply the multiplicative bias estimated from the {\it constant-shear}
simulation to the {\it redshift-dependent-shear} simulation following
\citet{DESY3-BlendshearCalib-MacCrann2021}. For the redshift-dependent-shear
setup, we select one bin from the four redshift bins and only distort the input
galaxies in the selected redshift bin while leaving the galaxies in the other
three redshift bins undistorted instead of distorting all galaxies with the
same shear. We perform source detection, deblending and measurement using the
HSC pipeline on the simulated images to obtain a galaxy shape catalog. After
that, we apply the weak-lensing cut and estimate the average shear from
galaxies in the selected redshift bin. In order to estimate the additional
multiplicative bias due to the difference between redshift-dependent-shear
setup and constant-shear setup, we use the multiplicative bias obtained from
the constant-shear simulation for the selected redshift bin to calibrate the
average shear measured from the redshift-dependent-shear simulation with
different noise realizations and shear distortions (i.e.\ $\gamma_1=+0.02$ and
$\gamma_1=-0.02$). We carry out this process in four redshift bins to estimate the
excess multiplicative bias in each redshift bin.

Figure~\ref{fig:zBlendBias} shows the additional multiplicative bias due to
redshift-dependent shear in each redshift bin for the three different seeing
setups. It shows that for observations with a larger seeing size, the amplitude
of the excess multiplicative bias due to redshift-dependent blending is
larger. Furthermore, we find that, for the HSC PSF with FWHM close to the HSC
average, the multiplicative bias due to the redshift-dependent blending that is
not captured by our fiducial calibration marginally meets the three-year HSC
requirement. The excess multiplicative bias will be marginalised over during
the cosmological analyses.
}

\subsection{Basic characterization of the catalog}
\label{subsec:charact}

\begin{figure*}[ht!]
\begin{center}
\includegraphics[width=0.85\textwidth]{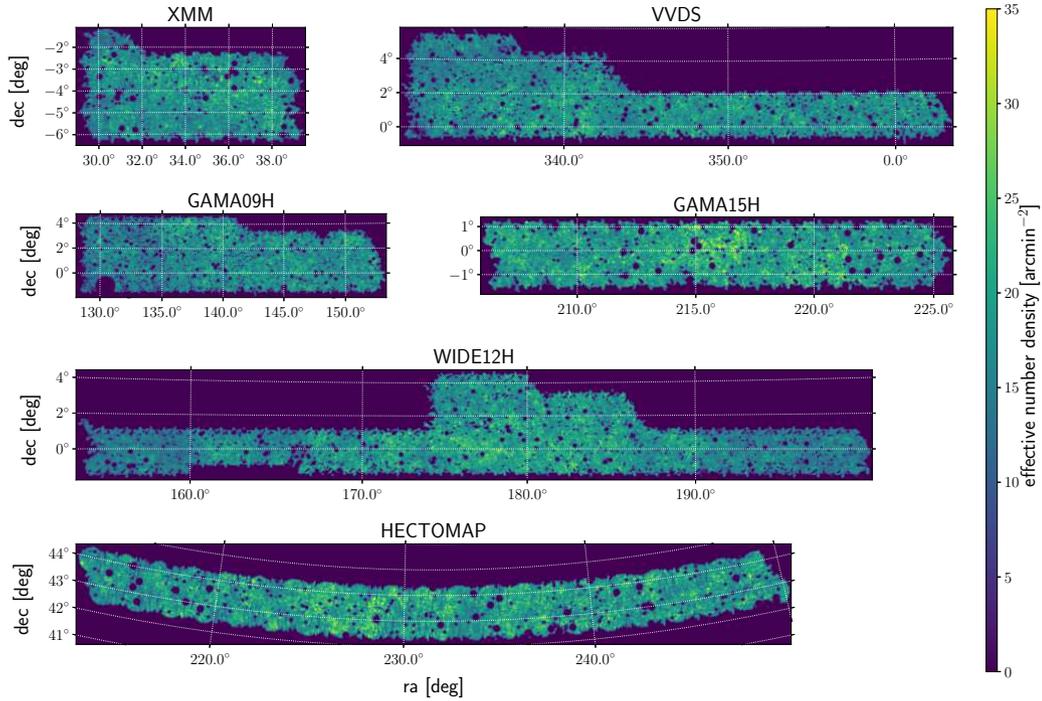}
\end{center}
\caption{
    The effective galaxy number density map of the three-year HSC shape
    catalog. The map is a tangent projection of sky with a regular grid spacing
    of $1\arcmin$ after smoothing with a Gaussian kernel ($\sigma=1\farcs5$).
    The arc feature is masked out by the ``\texttt{i$\_$pixelflags$\_$clipped
    }$==$False'' and ``\texttt{i$\_$pixelflags$\_$edge} $==$ False'' flags
    shown in Table~\ref{tab:icut}, which is only obvious in HECTOMAP field due
    to the difference in the display resolution between the panels.
    }
    \label{fig:neffMap}
\end{figure*}
\begin{figure}[ht!]
\begin{center}
\includegraphics[width=0.5\textwidth]{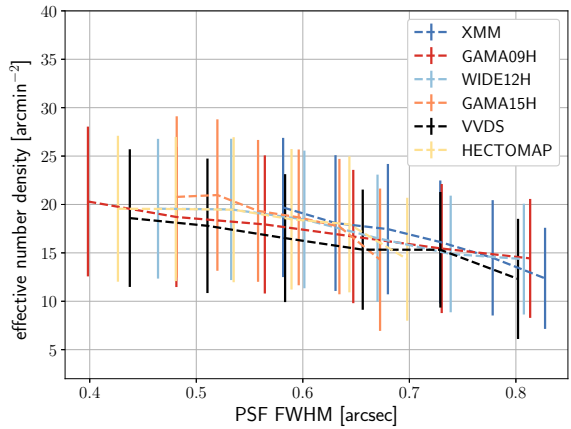}
\end{center}
\caption{
    The mean effective galaxy number density (defined in
    Eq.~\eqref{eq:effectdens}) of the catalog as a function of $i$-band PSF
    FWHM for each field. The errorbars denote the standard deviation of the
    effective number density in each PSF FWHM bin.
    }
\label{fig:neffseeing}
\end{figure}

The catalog, after applying the weak lensing cuts, covers an area of $433.48
~{\rm deg}^2$, split into six fields with an overall mean $i$-band seeing of
$0.59\arcsec$. The shear catalog contains $35,805,482$ galaxies, a number that
is $2.95$ times that of the first-year catalog, primarily due to the increased
area. The raw galaxy source number density for our catalog is $22.9~{\rm
arcmin}^{-2}$, which is comparable with the number density of the first-year
shear catalog. The effective galaxy number density, defined in
\citet{WLsurvey-neffective-Chang2013} as
\begin{equation}\label{eq:effectdens}
    n_{{\rm eff}}=\sum_i \frac{e_{{\rm RMS};i}^2}{\sigma_{e;i}^2+e_{{\rm RMS};i}^2},
\end{equation}
is $19.9~{\rm arcmin}^{-2}$. The effective galaxy number density map for each
field is shown in Fig.~\ref{fig:neffMap}. In Fig.~\ref{fig:neffseeing}, we show
the trend of the average effective number density as a function of the PSF FWHM
for each field. As shown, the effective number density slowly decreases as the
PSF FWHM increases. Since the resolution of a galaxy decreases when the PSF
FWHM increases, the resolution cut ($R_2>0.3$) tends to remove more galaxies in
the regions with larger seeing sizes.


\subsection{Blinding}
\label{subsec:blind}

Multiple cosmological analyses are being conducted by the HSC collaboration
using the three-year HSC shear catalog, each with different analysis PIs. In
order to avoid confirmation bias in cosmological analyses, we blind our catalog
by adding a random additional multiplicative bias with a two-level blinding
scheme \citep[also see][]{cosmicShear_HSC1_Chiaki2019}. The first is a
user-level blinding to prevent an accidental comparison of blinded catalogs
between different analysis teams, while the second is collaboration-level
blinding that is adopted in the cosmological analysis.

For the user-level blinding, we generate a random additional multiplicative
bias $\mathrm{d}m_1$ for each catalog.  The values of $\mathrm{d}m_1$ are
different among different analysis teams, and they are encrypted with the
public keys from the principle investigators (PIs) of the corresponding
analysis teams. This single value of $\mathrm{d}m_1$ should be decrypted by the
PI and subtracted from the multiplicative bias values for each catalog entry to
remove the user-level blinding before the analysis.

For the collaboration-level blinding, we generate three blinded catalogs
with indexes $j=0,1,2$. The additional multiplicative biases $\mathrm{d}m_2^i$
for these three blinded catalogs are randomly selected from the following three
different choices of ($\mathrm{d}m_2^1$, $\mathrm{d}m_2^2$, $\mathrm{d}m_2^3$):
$(-0.1,-0.05,0.)$, $(-0.05,0.,0.05)$, $(0.,0.05,0.1)$. In each case, the
additional multiplicative biases are listed in an ascending order, while the
true catalog has a different index for the three options. The values of
$\mathrm{d}m_2^{1,2,3}$ are encrypted by a public key from one designated
person who will not lead any cosmology analysis.

The final blinded multiplicative bias values for the galaxies in each catalog
are modified as
\begin{equation}
    m_{{\rm blind;i}}^j=
    m_{{\rm true;i}}+\mathrm{d}m_{1}^j+\mathrm{d}m_{2}^j,
\end{equation}
where $i$ is the galaxy index in each blinded catalog.
Each PI receives a separate set of blinded catalogs, and carries out the same
analysis for all three catalogs after decrypting and subtracting the
$\mathrm{d}m_1$ from the multiplicative bias for each catalog.

We provide two types of blinded catalog. The one is the two-level blinding for
cosmology analyses and the other is just user-level blinding for non-cosmology
analysis. As we did for the first year weak lensing science, the additive bias
is not blinded in weak lensing analyses.

\section{Requirements on control of systematic uncertainties}
\label{sec:requirement}
In this section, we set requirements on the control of systematic residuals for
the weak lensing shear catalog defined in Section~\ref{subsec:WLSample}. Note
that the requirements can only be determined after the weak lensing galaxy
sample is defined since the statistical errors that can be obtained from a
cosmological analysis conducted with the shear catalog is the basis for setting
meaningful requirements on the control of systematic residuals.

First, we forecast the statistical errors that are attainable with the shear
catalog defined in Section~\ref{subsec:WLSample}.  Similar to the first-year
shear catalog, we will require the amplitude of each systematic residual
$\delta X_{\text{sys}}$ for an observable denoted as $X$ (e.g., galaxy-shear
cross correlation function or shear-shear correlation function) to contribute
less than one-half of the statistical error on the observable, $\sigma_X$. That
is,
\begin{equation}
    |\delta X_{\text{sys}}|\lesssim 0.5 \sigma_X\,.
\end{equation}
We note that such a requirement is on the systematic residuals after the
removal of known biases that are expected to be calibrated before the use of a
catalog. We will assess the requirements in terms of multiplicative and
fractional additive bias residuals (i.e., $\delta m$ and $\delta a$) in shear
estimation, which are defined in Eq.~\eqref{eq:fitmc}.

Not all weak lensing science cases will require the same level of control of
systematic residuals as we use in this paper, because some analyses will have
lower SNR. The requirements defined here are tuned such that for certain key
cosmological science cases, the statistical error will continue to dominate
over the systematic residuals.  When adding statistical and systematic errors
in quadrature, this threshold would mean that ignoring the systematic residuals
would result in an underestimation of the total uncertainty (statistical +
systematic) by at most $12$ percent.

\xlrv{
In this paper, we will check the magnitude of a number of systematics. If a
particular systematic effect does not meet the requirements we set, the
systematic residuals must be explicitly modeled and marginalized, potentially
contributing a significant portion of the total error budget in the
cosmological constraints. We also model and marginalize some of the systematics
in our cosmological analyses even though they meet the requirements, to avoid a
scenario where multiple residual systematic uncertainties that are each at the
$0.5\sigma$ level combine such that the total systematic uncertainty dominates
the final results.
}

\subsection{Requirement from cosmic shear}
\label{subsec:cosshearSNR}

Cosmic shear and galaxy-galaxy lensing are two of the most important scientific
applications of the HSC shear catalog. They require measurements of the
shear-shear correlations and the correlations between galaxy position and
shear, respectively. The requirements on control of systematic residuals for
galaxy-galaxy lensing are comparable to that for cosmic shear
\footnote{
In \citet{HSC1-shape}, the requirement on PSF model size errors from the
galaxy-galaxy lensing is twice as stringent as that from the cosmic shear.
However, that was due to a mistake in the calculation of the requirement on PSF
size errors from cosmic shear -- the right-hand-side of their Eq.~(19) should
be $0.25\sigma_{\xi+}(\theta)\,$.
}.
Therefore, we will determine the requirements on systematic residuals using the
expected covariance of shear-shear correlation functions. We estimate this by
rescaling the covariance matrix, denoted as $\mathbf{C}$, of the first-year
shear-shear correlations by the inverse square of the galaxy number ratio
between the three-year catalog and the first-year catalog.

The shear-shear correlations are defined as
\begin{equation}
\label{eq:xipm}
\xi_\pm(\theta) = \langle \hat{g}_+ (\vec{r})\hat{g}_+ (\vec{r}+\vec{\theta}) \rangle
    \pm \langle \hat{g}_\times (\vec{r})  \hat{g}_\times (\vec{r}+\vec{\theta}) \rangle\,.
\end{equation}
Here we decompose the per-object shear estimates $\hat{g}$ for pairs of
galaxies into the tangential  ($\hat{g}_+$) and cross component
($\hat{g}_\times$).

The first-year shear catalog paper \citep{HSC1-shape} used a covariance matrix
measured from mock catalogs
\citep{wl-covariance-Shirasaki2017,Shirasakietal:2019} to estimate the SNR for
a cosmic shear measurement without tomographic binning. The estimated SNR over
angular scales $5\arcmin<\theta<285\arcmin$ is $12.6$. Cosmic shear analyses
have been conducted using the first-year HSC shear catalog in both Fourier
space \citep{HSC1-cs-fourier} and configuration space \citep{HSC1-cs-real} with
tomographic binning. The estimated SNRs of $15.6$ and $18.4$ were achieved with
a fiducial multi-pole range $300<l<1900$ and an angular range
$4\arcmin<\theta<50\arcmin$, respectively. The differences between the SNR
measurements are mostly due to the different tomographic setups, angular ranges
and cosmological models adopted by these studies. Here we take their average
and rescale it according to the increase in galaxy number.  This process yields
a rescaled SNR$_{s-s}=27$. Note that even though we consider tomographic SNR
measurements when deriving SNR$_{s-s}$, we adopt a non-tomographic formalism
when deriving the requirements in the following context for simplicity.

We use this SNR to derive the upper limit of the amplitude of systematic
residuals on the cosmic shear as
\begin{equation}\label{eq:requirement0}
    \delta \xi_{\pm,\text{max}}(\theta) =
    \frac{\xi_{\pm}(\theta)}{2\,\text{SNR}_{s-s}},
\end{equation}
which has a statistical significance of $0.5\sigma\,$. In summary, the
requirement on the amplitude of systematic residuals that originate from any
sources on the cosmic shear measurement is given by
\begin{equation}\label{eq:requirement}
    |\delta \xi_{\pm}|<\delta\xi_{\pm,\text{max}}\,.
\end{equation}

\subsection{Multiplicative bias residuals}
\label{subsec:cosshearreq1}

In this section, we place a requirement on the overall residual multiplicative
shear bias ($\delta m$) after the calibration process described in
Section~\ref{subsec:calib}. To focus on multiplicative bias residuals, we
consider the situation that the additive bias is zero. If we neglect the
high-order terms of $\delta m$ with the assumption that $\delta m \ll 1$, the
multiplicative bias residual primarily affects the shear-shear correlations as
\begin{equation}
    \langle\hat{g}^\dagger \hat{g}\rangle \approx
        (1+2\,\delta m) \langle g^\dagger g\rangle,
\end{equation}
where $g^\dagger$ refers to the complex conjugate of $g$. The systematic
residual on the correlation function due to the multiplicative bias residual is
\begin{equation}
    \delta \xi_{+,\delta m}=2 \,\delta m \,\xi_+\,.
\end{equation}
According to Eq.~\eqref{eq:requirement}, the value of $2 |\delta m|$ should be
$\lesssim 0.5/\text{SNR}_\text{s-s}$, or
\begin{equation}
    |\delta m| \lesssim \frac{0.25}{\text{SNR}_\text{s-s}}=9.3\times10^{-3}\,.
\end{equation}
with integrated SNR (SNR$_\text{s-s}=27$) for cosmic shear. This requirement is
$\sim\sqrt{3}$ times as stringent as the first-year requirement on the
multiplicative bias.

\subsection{PSF model size errors}
\label{subsec:psfmodelsize-ss}

The systematic residual on PSF model size is quantified by the fractional PSF
size residual:
\begin{equation}
    f_{\delta\sigma}=\frac{\delta \sigma_\text{PSF}}{\sigma_\text{PSF}},
\end{equation}
where $\sigma_\text{PSF}$ is the determinant radius calculated from the second
moments of the PSF. Such systematics lead to an additive shift in the
shear-shear correlation function $\xi_{+}(\theta)$ \citep{Jarvis2016MNRAS},
which can be written as
\begin{equation}
    \delta \xi_{+,\delta\sigma}
    =4 \langle f_{\delta\sigma_\text{PSF}} \rangle \xi_{+}(\theta),
\end{equation}
if we use the approximation: $\langle T_\text{PSF}/T_\text{gal} \rangle=1$, and
set the fractional PSF model area (trace) error to twice the fractional PSF
determinant radius error \citep{HSC1-shape}.

Therefore, we place a specific requirement on the PSF model size errors -- the
systematics should be less than $0.5|\delta\xi_{\pm}|$ -- as we did in
\citet{HSC1-shape}. With the integrated SNR ($\text{SNR}_\text{s-s}=27$), the
requirement is written as
\begin{equation}
    |\left\langle f_{\delta\sigma}\right\rangle|
    \lesssim \frac{1}{16\,\text{SNR}_\text{s-s}} \approx 2.3\times 10^{-3}\,.
\end{equation}
This requirement is also $\sim\sqrt{3}$ times as stringent as the first-year
requirement on the PSF model size errors.

\subsection{Additive bias residuals}
\label{subsec:additive-ss}

In this section, we place a requirement on the correlation of the overall
additive shear bias residual ($\delta c$), originating from, e.g., an
inadequate removal of PSF anisotropy in the shear estimation or the PSF model
shape errors, etc. Similar to Section~\ref{subsec:cosshearreq1}, we set $\delta
m=0$ to focus on the additive bias residual ($\delta c$). The additive bias
($c$) propagates into an additive term in the correlation function through
\begin{equation}
\langle\hat{g}^\dagger \hat{g}\rangle = \langle g^\dagger g\rangle
    +\langle \delta c^\dagger \delta c \rangle\,.
\end{equation}
Then the systematic residual that originates from additive bias residual is
given by
\begin{equation}
\delta\xi_{+,\delta c}=
    \langle \delta c^\dagger \delta c \rangle\,.
\end{equation}

According to Eq.~\eqref{eq:requirement} and the conservative integrated SNR
(SNR$_\text{s-s}=27$), the requirement on the correlation of fractional
additive bias is
\begin{equation}
\begin{split}
\langle \delta c^\dagger \delta c \rangle
&<\frac{\xi_+(\theta)}{2 \,\text{SNR}_\text{s-s}}\\
    &=\frac{\xi_+(\theta)}{54}.
\end{split}
\end{equation}
Using the relation $\delta c=\delta a \, e_{\text{PSF}}$, we transform the
requirement on $\langle \delta c^\dagger \delta c \rangle $ to the requirement
on $\langle \delta a \,\delta a \rangle$:
\begin{equation}
\label{eq:additive-psf-req}
\langle \delta a \,\delta a \rangle
    <\frac{\xi_+(\theta)}{54 \langle e^{\dagger}_\text{PSF} e_\text{PSF}\rangle(\theta)}\,.
\end{equation}
Note , when using $\delta c=\delta a \, e_{\text{PSF}}$, we neglect the PSF
model shape errors, the requirement of which will be quantified in
Section~\ref{subsec:psfmodele-ss}.

In order to use the cosmic shear signals at scales where baryonic effects are
unimportant, we only consider the $\xi_+$ measurements on scales from
$4\arcmin$ to $50\arcmin$ as in \citet{HSC1-cs-real}. The quantity $\langle
\delta a \,\delta a \rangle$ declines as a function of angular scale, ranging
from $\sim 2.0\times 10^{-4}$ at our minimum scale of $\theta=4\arcmin$ to
$\sim 4.3\times 10^{-5}$ at $\theta=50\arcmin$ since the scale-dependence of
the PSF-PSF shape correlation (denominator) is much flatter than that of the
cosmic shear correlation function (numerator).  As we have already
conservatively used the integrated SNR in this equation, it is not necessary to
use the lowest $\langle\delta a \,\delta a \rangle$ value as well; rather, we
use the geometric mean of these values, requiring $|\delta a|^2<9.4\times
10^{-5}$, or $|\delta a|<9.7\times 10^{-3}$.

\subsection{PSF model shape errors}
\label{subsec:psfmodele-ss}

Systematic correlations in the errors in the shape of PSF model, which are
quantified by the PSF model shape residual ($\delta g_\text{PSF}$), also
produce an additive term in the galaxy shear-shear correlations. Here we place
a requirement on the spatial correlations of PSF model shape errors by
requiring the additive terms induced by PSF shape errors to be half the
statistical error \citep{HSC1-shape}.

The additive terms are expressed in Eq.~(3.17) in \citet{Jarvis2016MNRAS},
which depends on the five $\rho$ statistics, two of which were defined in
\citet{Rowe2010MNRAS} and the last three in \citet{Jarvis2016MNRAS}. The $\rho$
statistics are summarized as follows:
\begin{align}
\rho_1(\theta) &\equiv \left\langle \delta g_\text{PSF}^\dagger(\vec{r})\delta
    g_\text{PSF}(\vec{r}+\vec{\theta})\right\rangle, \\\label{eq:rho1}
\rho_2(\theta) &\equiv \left\langle g_\text{PSF}^\dagger(\vec{r})\delta g_\text{PSF}(\vec{r}+\vec{\theta})\right\rangle,\\
\rho_3(\theta) &\equiv \left\langle \left(g_\text{PSF}^\dagger\frac{\delta
    T_\text{PSF}}{T_\text{PSF}}\right)(\vec{r}) \left(g_\text{PSF}\frac{\delta
    T_\text{PSF}}{T_\text{PSF}}\right)(\vec{r} + \vec{\theta})\right\rangle, \\
\rho_4(\theta) &\equiv \left\langle\delta g_\text{PSF}^\dagger(\vec{r}) \left(g_\text{PSF}\frac{\delta
    T_\text{PSF}}{T_\text{PSF}}\right)(\vec{r} + \vec{\theta})\right\rangle, \\
\rho_5(\theta) &\equiv \left\langle g_\text{PSF}^\dagger(\vec{r}) \left(g_\text{PSF}\frac{\delta
    T_\text{PSF}}{T_\text{PSF}}\right)(\vec{r} + \vec{\theta})\right\rangle.  \label{eq:rho5}
\end{align}
$\rho_1$ is the auto-correlation function of PSF model shape residuals, while
$\rho_2$ is its cross-correlation with the PSF shape itself. The other
statistics, i.e., $\rho_{3,4,5}$, involve $T_\text{PSF}$ -- the trace of the
second moment matrix of the PSF.

We place requirements on the $\rho$ statistics following Eqs.~(33)--(34) of
\citet{HSC1-shape} using the conservative integrated SNR (SNR$_\text{s-s}$) to
avoid a binning-dependence of the requirement. These requirements become
\begin{equation}\label{eq:rho134-req}
\begin{split}
|\rho_{1,3,4}(\theta)| < \frac{\xi_+(\theta)}{2\,\text{SNR}_\text{s-s}}
    = \frac{\xi_+(\theta)}{54},\\
|\rho_{2,5}(\theta)| < \frac{\xi_+(\theta)}{4|\delta a| \,\text{SNR}_\text{s-s} }
    = \frac{\xi_+(\theta)}{108|\delta a|},
\end{split}
\end{equation}
where $\delta a$ is the remaining fractional additive bias due to the leakage
of PSF anisotropy into the shear estimation after the calibration with image
simulations. Here we set $\delta a$ to $0.02$, which is greater than the value
we detect in Section~\ref{subsec:Dcalib}, to ensure that our requirement is
stringent enough for the weak lensing science.

\section{PSF Model Tests}
\label{sec:psftest}
In this section, we carry out tests to ascertain the fidelity of the PSF
modelling for $i$-band coadds. In Section~\ref{subsec-starsamples}, we define
the star samples used for PSF tests. In Section~\ref{subsec:psfRegCut}, we
select regions where PSF adequately modeled. We test the PSF model size and PSF
model shape residuals in Section~\ref{subsec:psfTestsize} and
Section~\ref{subsec:psfTestshape}.

\subsection{Star sample}
\label{subsec-starsamples}

In the HSC pipeline, the PSF modelling is carried out with a modified
version\footnote{The modification fixes the sub-pixel interpolation problem for
the ``very best seeing'' images as shown in Section~4.6 of \citet{HSC2-data}.}
of PSFEx \citep{PSFEx11} at the single exposure level after correction for the
brighter-fatter effect \citep{HSC-BFE-Coulton2018}. The PSFs on coadded images
are reconstructed based on coaddition of the PSF models estimated in each CCD
visit by interpolating star images. The selection of stars used for PSF
modelling is based on the k-means clustering of high-SNR (i.e., SNR$>50$)
objects in size, typically resulting in $\sim 80$ star candidates per CCD chip
with an area of $\sim 60$~arcmin$^2$ (see \citealt{HSC1-pipeline} for
details). At the single exposure CCD processing, $\sim20\%$ of the stars in a
given single exposure are randomly selected and reserved for cross-validation
and are not used for PSF modelling. Since the star sample used in PSF modelling
is derived on individual exposures, different exposures will not necessarily
select the same set of stars. At the coadded image level, stars that were used
by $\geq 20\%$ of the input visits is labelled as having been used in the
modeling, namely ``\texttt{i$\_$calib$\_$psf$\_$used}$==$True''.

\begin{figure}
\begin{center}
\includegraphics[width=0.45\textwidth]{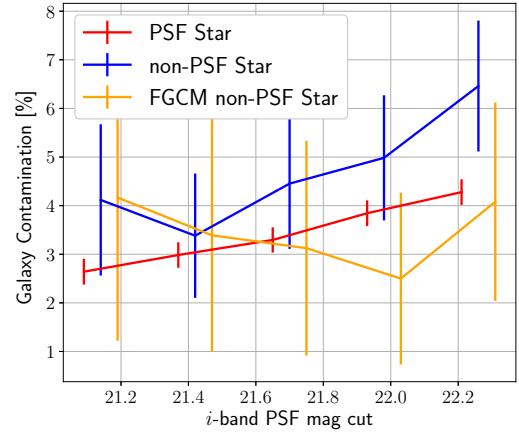}
\end{center}
\caption{
    The galaxy fractions misclassified as stars in three star catalogs (i.e.,
    PSF-star, non-PSF star, and FGCM non-PSF star) as a function of $i$-band
    PSF magnitude cut at the faint end. The results are estimated with a
    reference to the COSMOS HST star-galaxy classifications as ground truth.
    Errorbars show the Poisson uncertainties.
    }
    \label{fig:galcontam}
\end{figure}

The systematic tests are conducted at the coadded image level. The star sample
used in the tests is selected by ``\texttt{i$\_$extendedness$\_$value}$==0$'',
which is a cut indicating whether an object is extended (galaxy) or point-like
(star) as shown in Table~\ref{tab:icut}. Following \citet{HSC1-shape}, a $22.5$
magnitude cut in $i$-band is applied to select a high-SNR star sample. In this
magnitude limited star sample, those flagged by
``\texttt{i$\_$calib$\_$psf$\_$used} $==$ True'' are defined as PSF stars, and
the others are defined as non-PSF stars. We match these star samples in the
COSMOS region to the HST catalog of \citet{HST-shapeCatalog-Alexie2007}, and
use the HST galaxy-star classification as a reference to estimate the galaxy
contamination in our star catalogs. Fig.~\ref{fig:galcontam} shows the
estimated galaxy contamination as a function of $i$-band PSF magnitude limit.
As shown, the fraction of galaxy contamination is smaller by $\sim2\%$ in
PSF-star sample than in non-PSF star sample for the $22.5$th magnitude cut. This
is because ``\texttt{i$\_$calib$\_$psf$\_$used} $==$ True'' is a stricter
selection of stars based on both size and brightness information. In contrast,
``\texttt{i$\_$extendedness$\_$value}$==0$'' selects stars only based on size
information.

In order to further improve the purity of the non-PSF stars and reduce the
contamination by galaxies, we cross-match this star sample to the star catalog
selected from the HSC S20A data release \citep{HSC3-data} for the Forward
Global Calibration Method \citep[FGCM;][]{FGCM-Burke2018} photometry
calibration by their sky coordinates. Hereinafter, we term this cross-matched
star catalog as FGCM non-PSF stars. The number of PSF stars, non-PSF stars and
FGCM non-PSF stars are $2260229$, $186529$, and $87131$, respectively. As shown
in Fig.~\ref{fig:galcontam}, the purity of this FGCM non-PSF stars is better
than non-PSF stars and comparable to PSF stars for the 22.5th magnitude cut.
This is because the FGCM selection requires each star in the catalog to be
identified as point-like sources in at least two observations (not limited to
$i$-band) in the single exposure image processing. In addition, the FGCM star
catalog is downsampled for a homogeneous distribution on the sky. We refer
readers to \citet{HSC3-data} for more detailed description of the FGCM star
catalog.


To quantify the fidelity of the PSF reconstruction we perform tests of the PSF
size and shape residuals. Using the image of a star on coadded images as a proxy
for the true PSF, the fractional PSF model size residual ($f_{\delta\sigma}$)
can be quantified as
\begin{equation}
    \label{eq:fdelta}
    f_{\delta\sigma}= \frac{\sigma_\text{PSF}-\sigma_\text{*}}{\sigma_*},
\end{equation}
and the PSF model shape residual ($\delta g_\text{PSF}$) is quantified as
\begin{equation}
\delta g_\text{PSF}=g_*-g_\text{PSF},
\end{equation}
where $\sigma_*$ ($g_*$) is the size (shape) of the star, and
$\sigma_\text{PSF}$ ($g_\text{PSF}$) is the PSF model size (shape) evaluated at
the position of the star.

\subsection{PSF region cut}
\label{subsec:psfRegCut}

In the first-year shear catalog, significant PSF model size residuals were
identified in the VVDS region, which included some of the ``very best seeing''
data with PSF FWHM less than $0\farcs5$ (see Fig.~9 of \citealt{HSC2-data}). As
described in Section~\ref{sec:dataPipe}, in the current data release, the
improved PSF interpolation for small-size PSFs as detailed in \citet{HSC2-data}
has ameliorated this issue. Therefore, unlike the first-year shear catalog
\citep{HSC1-shape}, we do not remove the ``very best seeing'' regions.

\begin{figure*}[ht!]
\begin{center}
\includegraphics[width=0.9\textwidth]{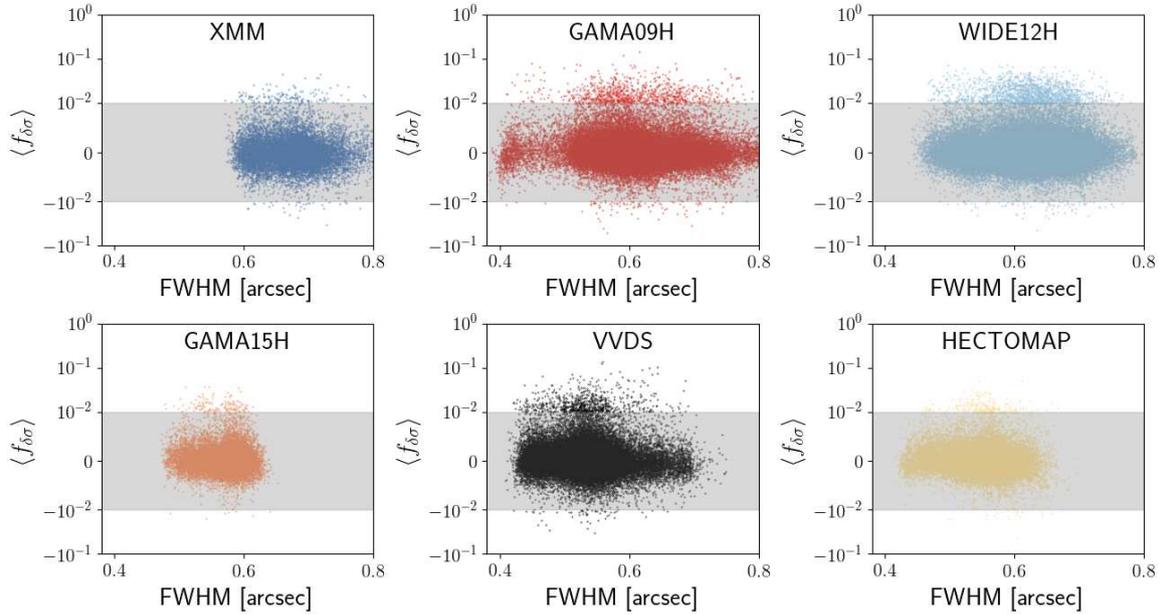}
\end{center}
\caption{
    The average fractional size residual $\langle f_{\delta\sigma}\rangle$
    reconstructed at star positions, averaged over the PSF stars within
    \texttt{HEALPix} pixels with ${\tt NSIDE}=1024$, and shown as a function of
    seeing. A symlog scale is used to allow negative residuals to be shown.
    The gray region indicates the linear part of the symlog scale, with the
    rest being logarithmic. The upper boundary of the gray regions are the
    fractional size cut ($f_{\delta\sigma}<0.01$) we apply to remove the
    \texttt{HEALPix} pixels with large positive amplitude of size residual.
    }
    \label{fig:psf_healpix}
\end{figure*}

However, we do perform a cut since we find the histogram of the fractional size
residuals is skewed toward positive. In addition, the cut is applied to ensure
that the PSF model size is adequately modeled in the selected regions. To
suppress measurement noise, we average the fractional size residuals
($f_{\delta\sigma}$) within each \texttt{HEALPix} pixel with ${\tt NSIDE}=1024$
corresponding to an area of $\sim12\ {\rm arcmin}^2$. The number of PSF stars
in a \texttt{HEALPix} pixel varies from $\sim10$ to $\sim24$. We then plot the
average fractional size residual (Eq.~\eqref{eq:fdelta}) as a function of
average seeing within the corresponding \texttt{HEALPix} pixels in
Fig.~\ref{fig:psf_healpix}. We limit the weak lensing FDFC (WLFDFC) region to
\texttt{HEALPix} pixels with $\langle f_{\delta\sigma}\rangle<0.01$. This cut
reduce the PSF model size residuals on average while only removing $\sim2.2\%$
of the FDFC region.

\subsection{PSF model size}
\label{subsec:psfTestsize}

\begin{figure*}
\begin{center}
\includegraphics[width=0.85\textwidth]{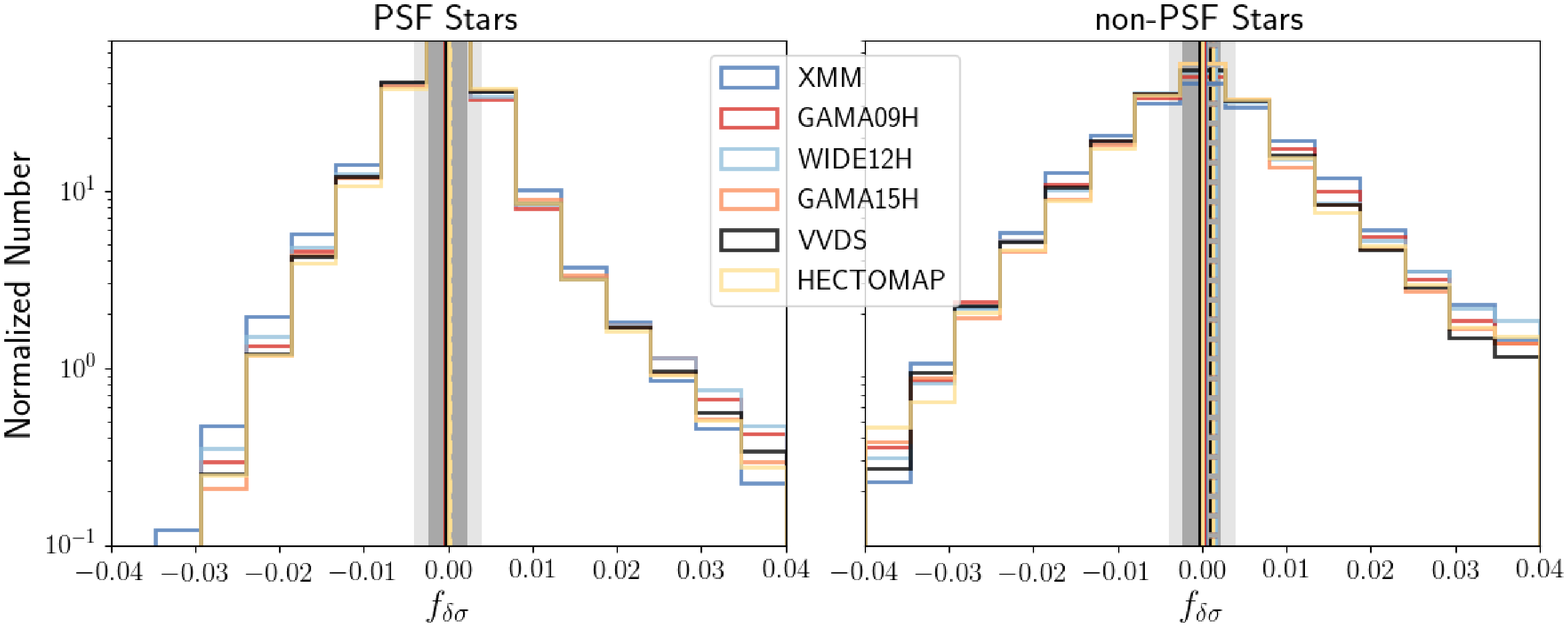}
\includegraphics[width=0.85\textwidth]{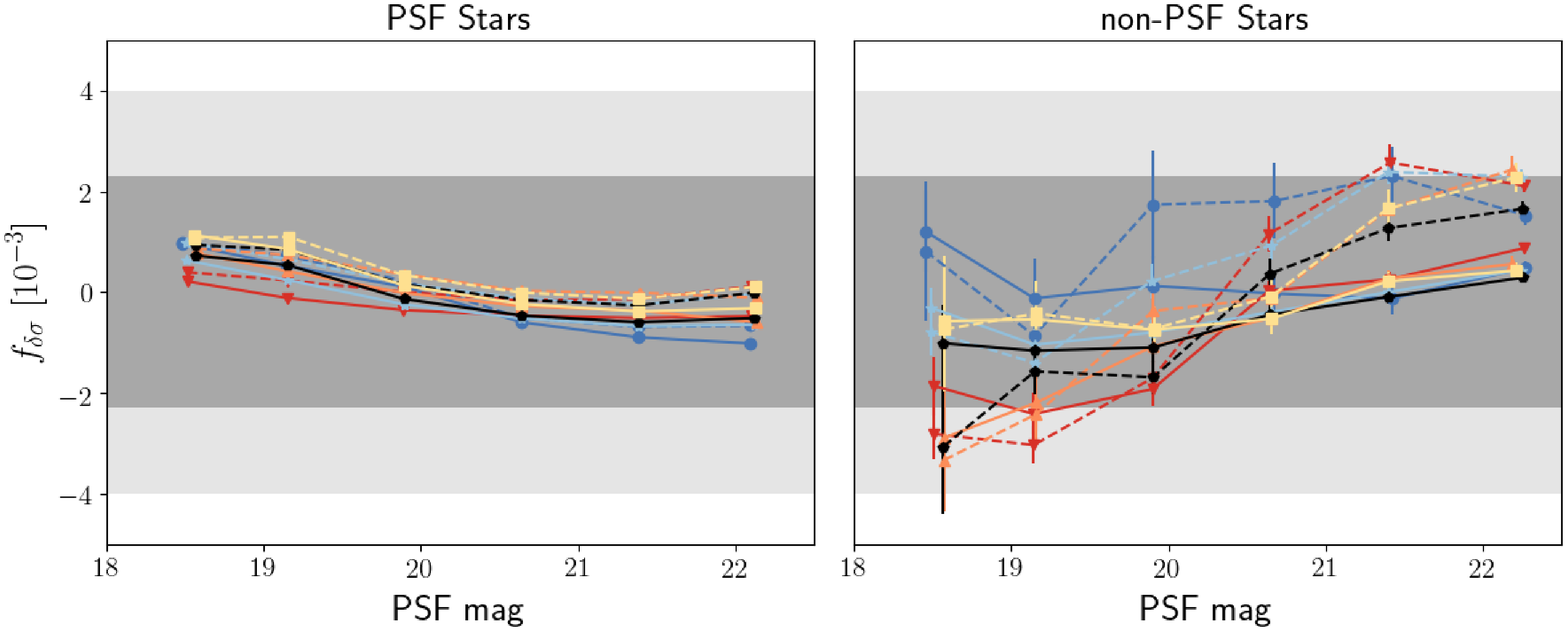}
\end{center}
\caption{
    {\em Upper left:} Distribution of the fractional size residual
    ($f_{\delta\sigma}$) for PSF stars in each field. The dark- (light-)gray
    region indicate the requirements on the residual for the three- (first-)year
    shear catalog. The vertical dashed (solid) lines show the mean (median) of
    $f_{\delta\sigma}$.
    {\em Upper right:} Same as the {\em upper left}, but for FGCM non-PSF stars.
    {\em Lower left:} $f_{\delta\sigma}$ as a function of the $i$-band PSF
    magnitude for PSF stars in each field. The dark- (light-)gray region
    indicates the three- (first-)year requirement. The dashed (solid) lines
    show the mean (median).
    {\em  Lower right:} Same as {\em lower left}, but for FGCM non-PSF stars.
    }
    \label{fig:psf_mag_fdelta}
\end{figure*}

The results of the PSF model size residual tests are shown in
Fig.~\ref{fig:psf_mag_fdelta}. The two plots in the upper panels are the number
distributions of $f_{\delta\sigma}$ for PSF stars (left) and FGCM non-PSF stars
(right). Here we show the mean and median for each field and compare with the
overall three-year requirement. All of the results are well within the
three-year requirement. Compared to the PSF stars, the results (mean and
median) for FGCM non-PSF stars have slightly larger deviations from zero. The
deviation in the mean values is more pronounced compared to that in the median,
as the mean value is more sensitive to the outliers and/or the skewness of the
distribution.

The two lower panels show fractional size residuals as a function of PSF
magnitude. The lower left panel shows the mean (solid lines) and median (dashed
lines) for PSF stars \citep{HSC1-pipeline}. The lower right panel shows the
results of FGCM non-PSF stars. The magnitude dependence of the mean size
residual for FGCM non-PSF stars is slightly different from that of PSF stars.
The results of FGCM non-PSF stars are within our requirements, except perhaps
at the bright and the faint ends of the magnitude bins in the WIDE12H and
GAMA09H fields. However, given the large size of the uncertainties, the
evidence for the difference between the PSF stars and the FGCM non-PSF stars is
not entirely conclusive.

\begin{figure}[ht!]
\begin{center}
\includegraphics[width=0.48\textwidth]{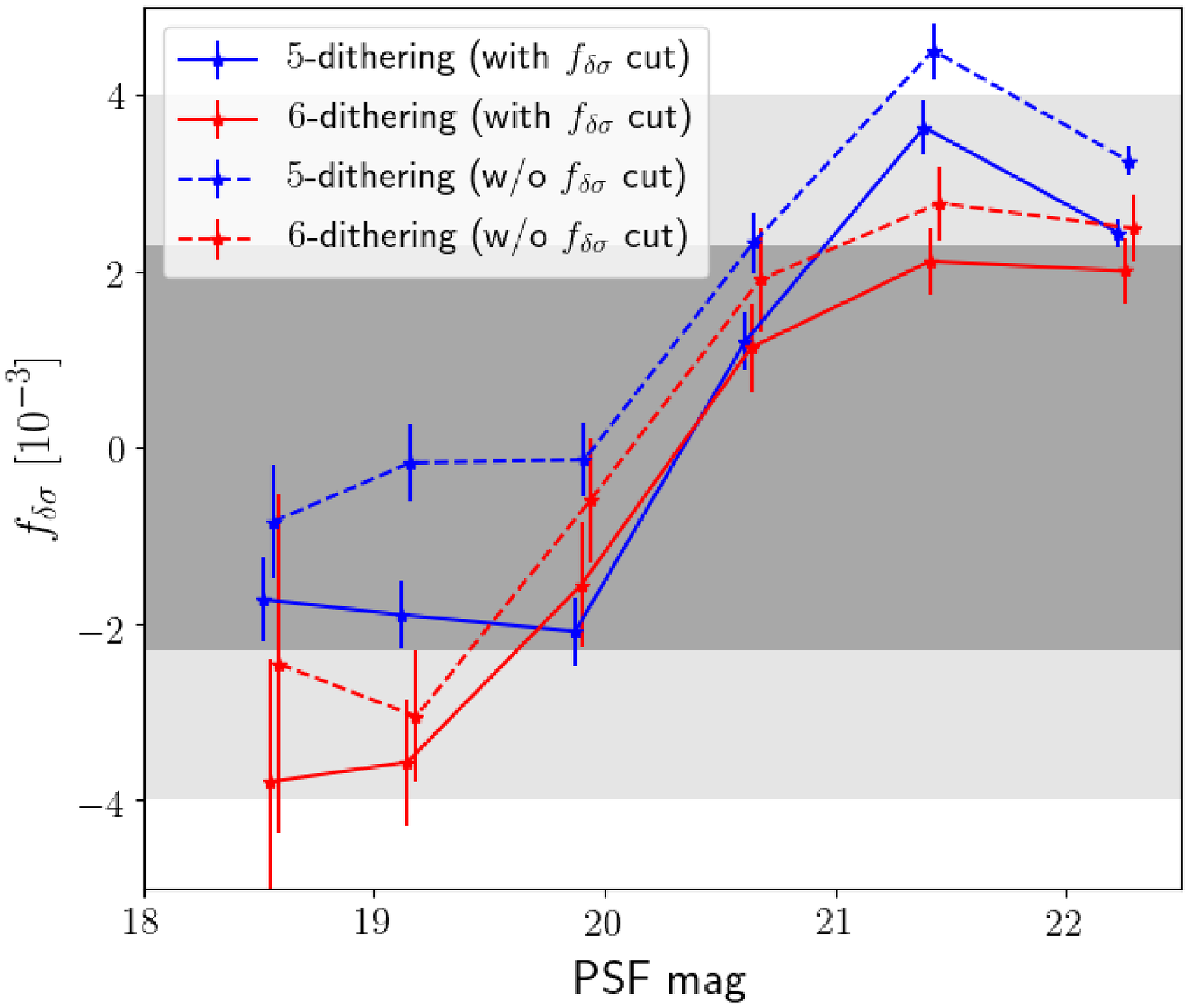}
\end{center}
\caption{
    The fractional size residual of FGCM non-PSF stars from the WIDE12H field
    divided into the $5$-dithering regions (blue lines) and the $6$-dithering
    regions (red lines) as a function of $i$-band PSF magnitude. The dashed
    lines and solid lines represent the results with and without $\langle
    f_{\delta\sigma} \rangle<0.01$, respectively. The gray region indicate the
    requirements on the residual for the three-year shear catalog. Results are
    shifted horizontally for illustrative purposes.
    }
    \label{fig:psf_dither_test}
\end{figure}

As described in Section~\ref{subsec:obs_strat}, the dithering strategy was
changed in S19A from a $6$-dithering pattern to a $5$-dithering pattern in the
$i$-band. Thus the definition of the FDFC region has also been changed as
described in Section~\ref{sec:dataPipe}. To assess the impact of this change in
observing strategy on PSF modelling quality, we have tested the PSF modelling
from regions with different dither patterns. Here we show the results of the
WIDE12H field as an example, where we contrast the PSF model size residuals in
a patch with a $6$-dithering pattern to that observed in patches with a
5-dithering pattern. As illustrated by the color map of numbers of input visits
in Fig.~\ref{fig:inputMap}, the central patch of the WIDE12H field with redder
color corresponds to the region with the $6$-dithering pattern, with
$174\fdg0\leq \text{ra}<190\fdg5$ and $\text{dec}<=1\fdg6$. The other three
patches surrounding the central region correspond to the $5$-dithering pattern.

Fig.~\ref{fig:psf_dither_test} shows $\langle f_{\delta\sigma} \rangle$ as a
function of the star magnitude from the aforementioned $6$-dithering region and
the $5$-dithering region in WIDE12H field with/without the $\langle
f_{\delta\sigma} \rangle <0.01$ cut. WIDE12H field is used to test the
different dithering region because it has much larger $5$-dithering region than
the other fields. We observe small differences between the $6$-dithering region
and the $5$-dithering region. At the bright end, $f_{\delta\sigma}$ is slightly
lower than the lower boundary of the requirement for the $6$-dithering region,
while it is slightly above the upper boundary of the requirement for the
$5$-dithering region. The results of both of the dithering strategies slightly
improve after applying the cut on $f_{\delta\sigma}$.

\begin{figure}[ht!]
\includegraphics[width=0.48\textwidth]{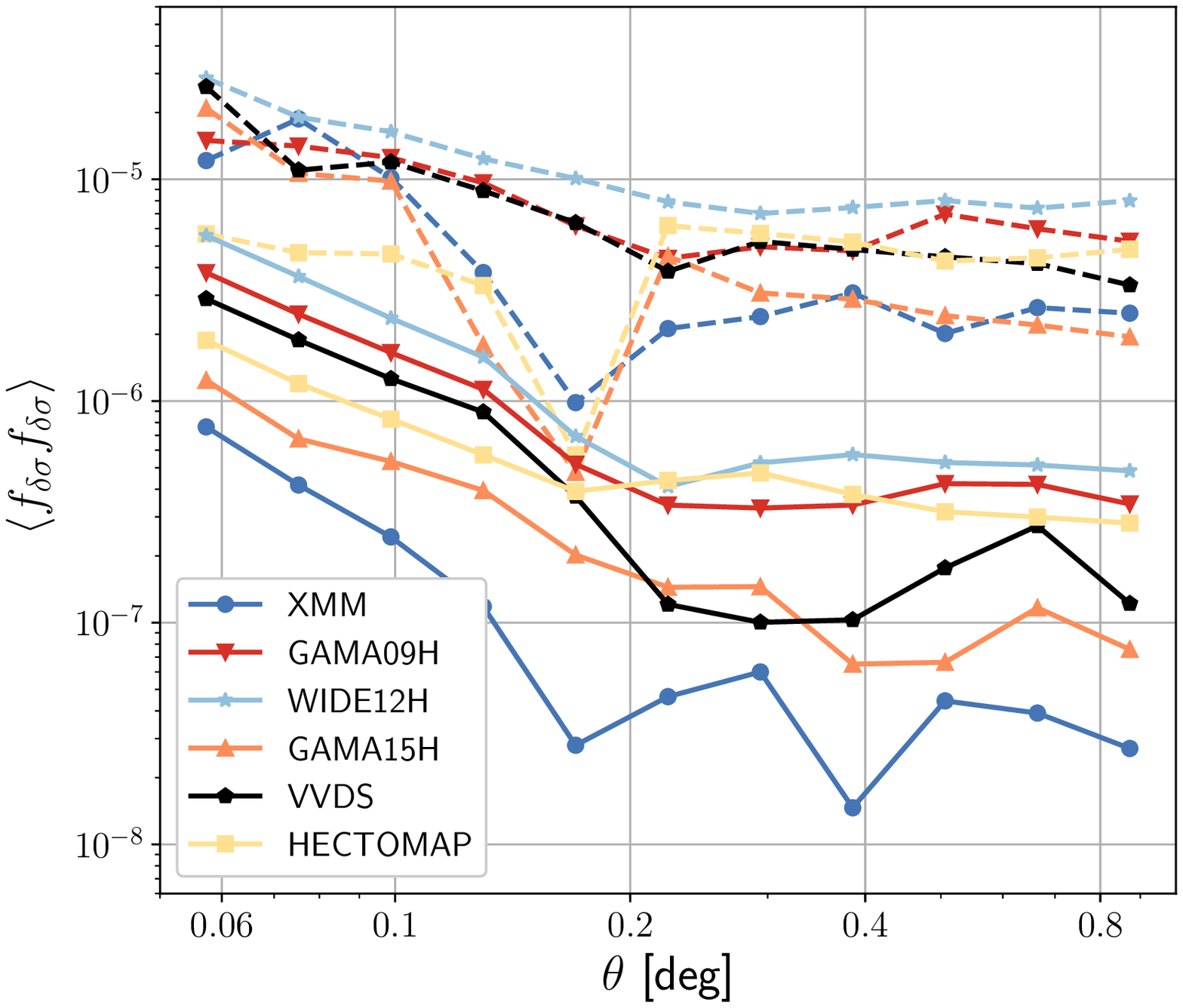}
\caption{
    The star fractional size residual ($f_{\delta\sigma}$) correlation for PSF
    stars (solid lines) and FGCM non-PSF stars (dashed lines) in each field.
    }
    \label{fig:psf_starsize_corr}
\end{figure}

The correlation function of the fractional size residual ($f_{\delta\sigma}$)
for each field is shown in Fig.~\ref{fig:psf_starsize_corr}. Comparing with the
results of the first-year shear catalog shown in the lower right panel of
Fig.~6 in \cite{HSC1-shape}, we conclude that although the three-year results
show scale dependence and the first-year results demonstrate weaker scale
dependence, the amplitudes are much lower. The correlations have decreased by a
factor of $\sim$10 at $1\degree$ scale and a factor of $\sim$2 at $0.1\degree$
scale compared to what was seen in the first-year shear catalog.

\subsection{PSF model shape}
\label{subsec:psfTestshape}

\begin{figure}[ht!]
\includegraphics[width=0.48\textwidth]{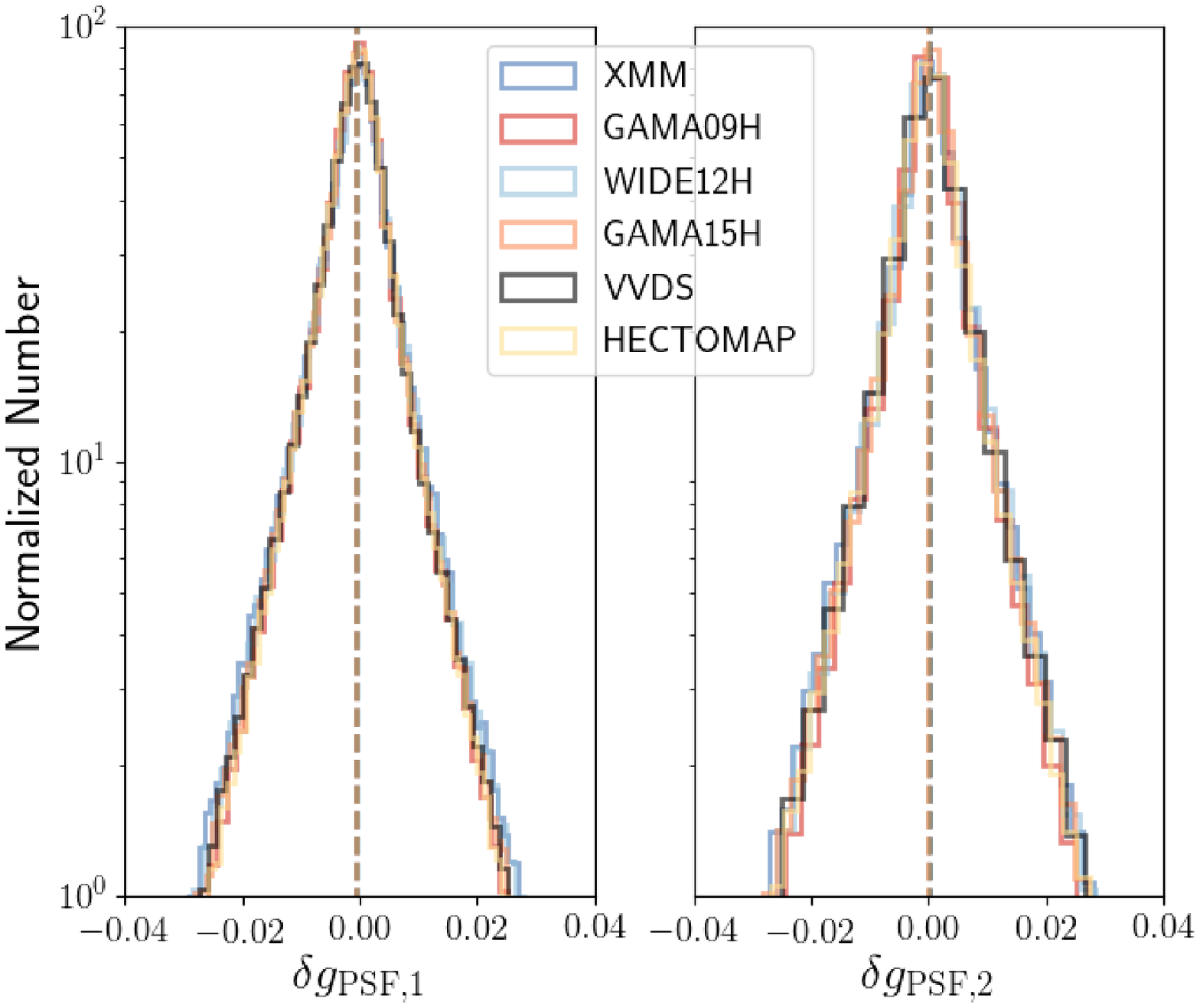}
\caption{
    Distribution of PSF model shape residual ($\delta g_{\text{PSF}}$) for the
    FGCM non-PSF stars in each field. The vertical lines show the average of
    $\delta g_{\text{PSF}}\,$.
    }
    \label{fig:psf_shape_residual}
\end{figure}

The PSF model shape residual distributions for both PSF and FGCM non-PSF stars
are shown Fig.~\ref{fig:psf_shape_residual}. We plot the median and mean of the
distribution of each field. We do not place a requirement on the average of
shape residuals since the average additive bias from PSF model shape errors can
be removed by cross-correlating with a random catalog that has the same area
coverage as the lens sample for galaxy-galaxy lensing measurements, and we
directly place requirements on the correlations of the PSF model shape errors
for cosmic shear measurements.

\begin{figure*}[ht!]
\begin{center}
\includegraphics[width=0.9\textwidth]{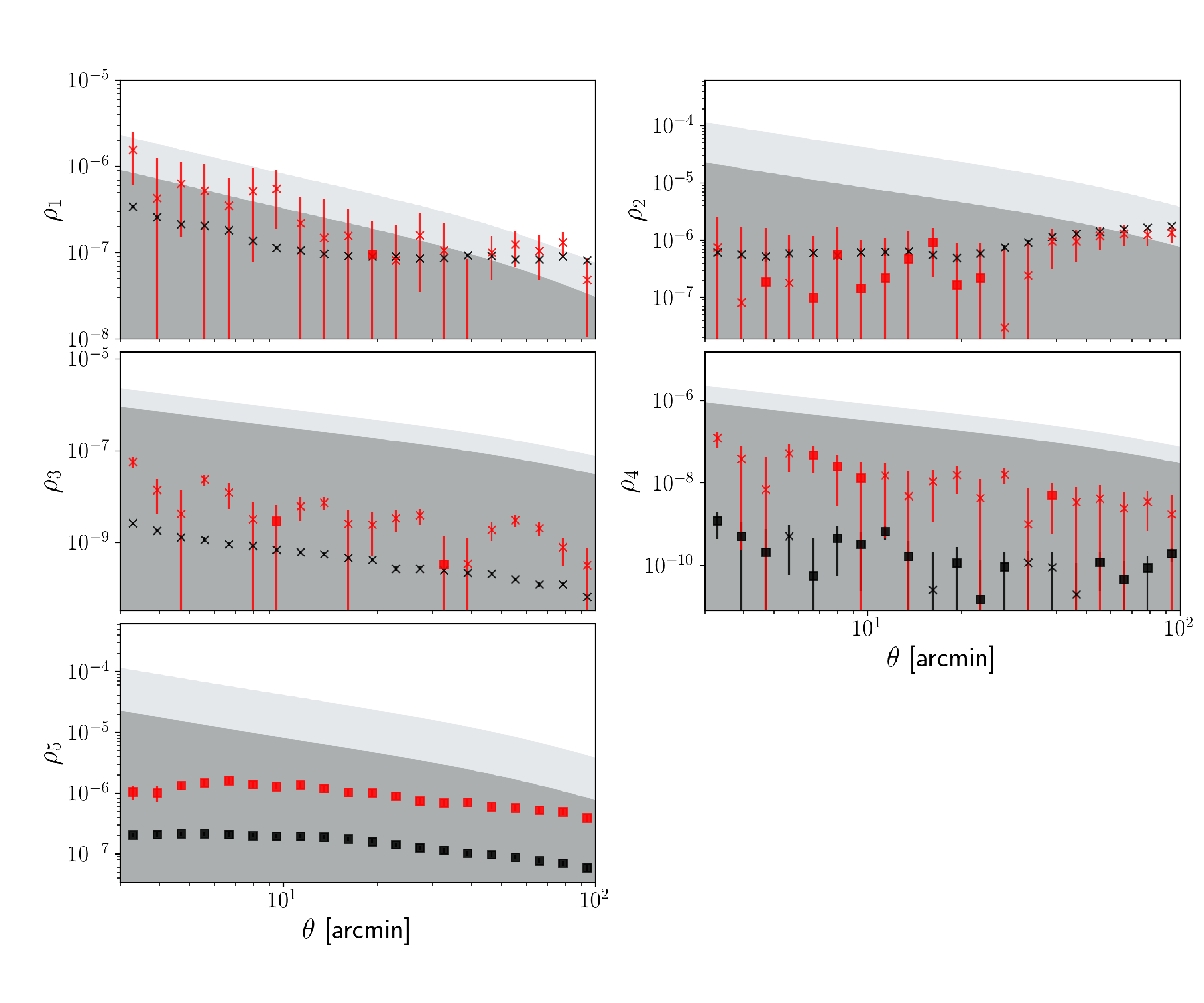}
\end{center}
\caption{
    PSF model shape residual correlations, or $\rho$ statistics, $\rho_1$
    through $\rho_5$ (defined in Section~\ref{subsec:psfmodele-ss}) as a
    function of separation $\theta$ on the sky. Negative values are shown in
    absolute values and denoted by `$\sqcdot$', whereas positive values are
    shown as they are and denoted by `$\times$'. Black (red) points are for PSF
    stars (FGCM non-PSF stars). The regions with dark- (light-)gray background
    are within the three- (first-)year HSC requirement.
    \xlrv{
    The errorbars are estimated by bootstrap resampling.
    }
    }
    \label{fig:psf_rho}
\end{figure*}

In Fig.~\ref{fig:psf_rho}, we show the $\rho$ statistics, defined in
Eqs.~\eqref{eq:rho1}--\eqref{eq:rho5} which are constructed from the spatial
correlation functions of PSF model shape and size residuals. The requirements
on the $\rho$ statistics discussed in Section~\ref{subsec:psfmodele-ss} are
shown by the dark-gray regions in all the panels. On all $\rho$ statistics, we
show the results for PSF stars (black points) and FGCM non-PSF stars (red
points). In order to avoid the potential difference in the $\rho$ statistics
between the PSF stars and the FGCM non-PSF stars due to the population
difference between these two star samples caused by the SNR cut in the PSF
star selection, we perform a reweighting of the FGCM non-PSF stars so that they
match the magnitude distribution of the PSF stars. The $\rho_2$--$\rho_5$
statistics are generally within the three-year requirements at scales
$\theta<1^\circ$, regardless of the noisiness of these statistics for the FGCM
non-PSF stars due to their small numbers. We observe some flattening in the
$\rho_2$ statistics on scales nearing the size of the field of view of our
camera. Similar issue can be seen in the {\em top left panel}, where there is
some evidence that $\rho_1$, which is the auto-correlation of the PSF shape
residuals, marginally exceeds the requirements on scales greater than
$50\arcmin\,$. Another concern is the $\rho_1$ statistic for the FGCM non-PSF
stars. Although it is noisy, the $\rho_1$ statistic is at the very edges of our
three-year requirements. In addition, we have confirmed that whether applying
the aforementioned reweighting processing to match the magnitude distributions
or not does not change the conclusion.

\xlrv{
In the cosmic shear analyses, this particular systematic and its impact on the
cosmological inference needs to be carefully accounted for \citep[see e.g., the
treatment in][]{cosmicShear_HSC1_Chiaki2019, HSC1-cs-real,DESY3-cosmicShearA}.
}
Similarly large values for the $\rho_1$ statistic were seen in the DES year 1
data \citep{Zuntz2018} and are likely related to the use of \textsc{PSFex} for
modeling the PSF. In DES year 3 data, this issue was significantly reduced by
the use of a new PSF extraction software called PIFF\footnote{
https://github.com/rmjarvis/Piff} (PSFs In the Full FOV) introduced in
\citet{Jarvis2021}. The implementation of PIFF in the LSST pipeline (and
subsequently the HSC pipeline) is currently in progress and future releases of
the shape catalog from HSC are likely to include these improvements.

\section{Null Tests}
\label{sec:nullTest}
In this section, we conduct internal null tests related to galaxy and star
shapes within the shear catalog. We first show the mean shear value as
functions of different galaxy properties in
Section~\ref{subsec:nullTest_meanShear}. Then we cross-correlate galaxy shapes
with positions (e.g., stars and random positions) in
Section~\ref{subsec:nullTest_posGal}. Subsequently, we test the systematics
related to mass map reconstruction in Section~\ref{subsec:nullTest_massMap}.
Finally, we cross-correlate galaxy shapes with star shapes to test the
systematics related to PSF revision and PSF model errors in the shear
estimation.

\begin{figure*}[ht!]
\begin{center}
\includegraphics[width=0.85\textwidth]{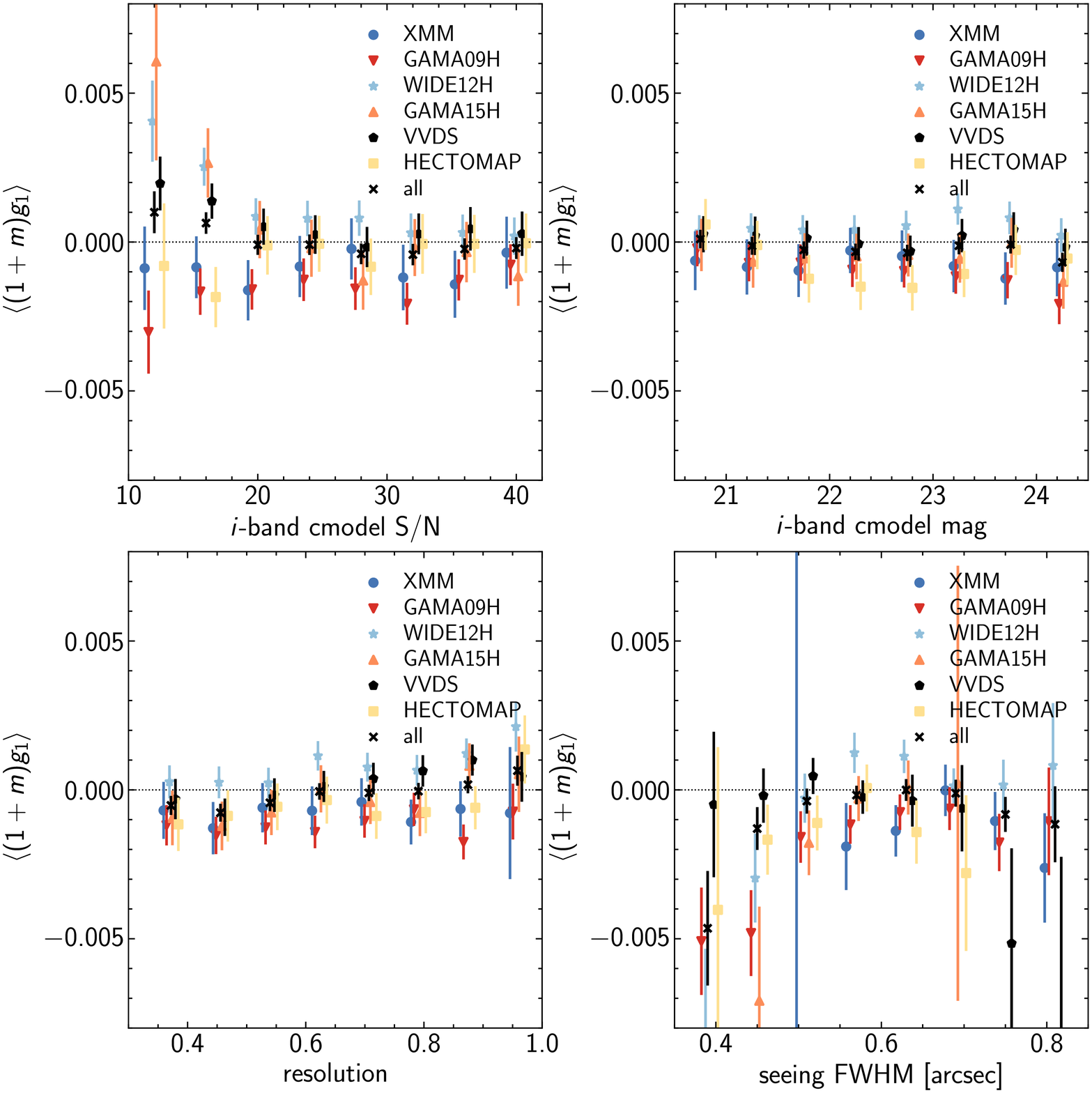}
\end{center}
\caption{
    Weighted mean shear values $\langle g_1\rangle$ as a function of $i$-band
    CModel SNR ({\em top-left}), $i$-band CModel magnitude ({\em top-right}),
    the \reGauss{} resolution factor corresponding to galaxy size ({\em
    bottom-left}), and PSF FWHM ({\em bottom-right}). Errorbars are the
    $1\sigma$ uncertainties estimated from mock shear catalogs and include
    cosmic variance.
    }
    \label{fig:systest_eave_par_e1}
\end{figure*}

\subsection{Mean shear values}
\label{subsec:nullTest_meanShear}

\xlrv{
    We first calculate the mean shear $\langle g_1\rangle$ and $\langle
    g_2\rangle$, which is sensitive to any additive bias residual that is
    independent of the PSF ellipticity.
}
Throughout the paper we derive mean shear values in sky coordinates, which are
quite close to the CCD coordinates in most cases. To check whether the mean
shear values are dominated by systematic errors, we derive uncertainties on
those mean shear values from mock shear catalogs including both shape noise and
cosmic variance from $N$-body simulations.  Specifically, we create $200$
realizations of mock catalogs following the method described in
\citet{HSC1-wlmap} \citep[also see][]{Shirasakietal:2019}, which adopts
ray-tracing results of \citet{FullSkySim-Takahashi2017}. We derive the
$p$-value for a fit to zero signal for the weighted mean value of each shear
component in the six fields.  We find that only one of the 12 $p$-values is
below a nominal threshold of 0.05, with an $p$-value of $0.024\,$. We therefore
conclude that the mean shear values do not exhibit signs of significant
systematic errors.

Following \citet{HSC1-shape}, we also check weighted mean shear values $\langle
g_1\rangle$ as a function of four properties of the $i$-band images: CModel
SNR, CModel magnitude, the \reGauss{} resolution parameter corresponding to
galaxy size, and the PSF FWHM. The error for the PSF FWHM bin at $\sim
0\farcs5$ in XMM field blows up due to the limited galaxy number in this bin as
shown by the number histogram in Fig.~\ref{fig:fwhmHist}. Results for all
fields combined, as shown in Fig.~\ref{fig:systest_eave_par_e1}, indicate that
most of mean shear values are consistent with zero within $2\sigma$. Moreover,
the average shear values do not show strong dependence on these galaxy
properties, although the average shear values for some observational fields are
persistently positive or negative in almost all galaxy property bins this is
very likely due to the bin-to-bin correlations, ranging from $0.3$ to $0.6$
\citep{HSC1-shape}, caused by cosmic variance.

\begin{figure}[ht!]
\begin{center}
\includegraphics[width=0.5\textwidth]{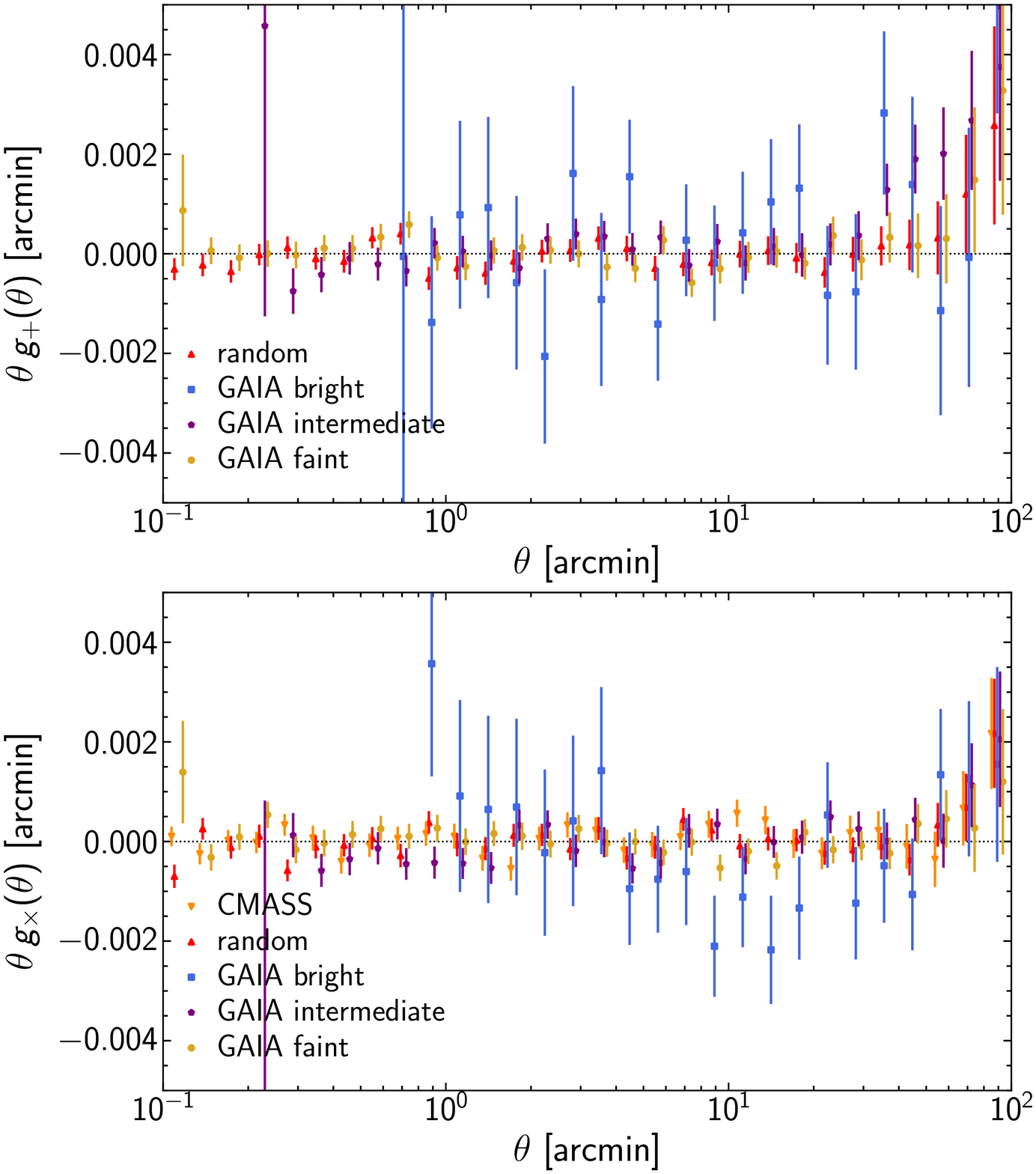}
\end{center}
\caption{
    Stacked tangential ({\em upper}) and cross ({\em lower}) shear profiles,
    averaged over the entire survey, around the CMASS galaxy sample ({\it
    inverted triangles}), random points ({\em triangles}), bright Gaia stars
    with $G<10$ ({\em squares}), intermediate Gaia stars with $13<G<14$ ({\em
    pentagons}), and faint Gaia stars with $18<G<18.2$ ({\em circles}). Only
    cross shear profiles are shown for stacking around CMASS galaxies. Errors
    are estimated from mock shear catalogs including cosmic shear. The $\chi^2$
    and $p$ values are summarized in Tables~\ref{table:stack_e} and
    \ref{table:stack_b} in Appendix~\ref{app:1}.
    }
    \label{fig:systest_stack}
\end{figure}

\subsection{Stacked shear signals}
\label{subsec:nullTest_posGal}

When measuring stacked shear signals for a source sample in annuli around a
sample of massive lenses, the stacked cross (not tangential) shear signals
should be zero due to symmetry \citep[e.g.,][]{Massey2007}. This fact therefore
provides a useful null test. In addition, stacked tangential shear signals
around objects that do not induce any weak lensing signals (e.g., stars and
random positions, etc.) can also be used for null tests. In this subsection we
explore stacked shear profiles around the following objects:

\begin{enumerate}
\item We adopt the CMASS galaxy sample of the SDSS-III Baryon Oscillation
    Spectroscopic Survey Data Release 12 \citep{BOSS-cmass} with an additional
    redshift cut of $0.4<z<0.7$. The galaxy density is about 90~deg$^{-2}$.
    Since the tangential shear profiles around CMASS galaxies have clear
    positive signals, we only use the cross shear profiles for our null tests.

\item We use random points generated in the HSC-SSP footprint with a density of
    100~deg$^{-2}$. In this case, we use both tangential and cross shear
    profiles for our null tests.

\item We use star catalogs generated from the Gaia Data Release 2 data
    \citep{GAIA-DR2}, which is currently used to create bright star masks for
    the HSC-SSP data \citep{HSC2-data}. We use the Gaia bright, intermediate,
    and faint star catalogs consisting of Gaia stars with $G$-band magnitude
    $G<10$, $13<G<14$, and $18<G<18.2$, respectively, and use these three
    catalogs for our null tests. Bright stars are suitable for testing the
    validity of our bright star masks and the impacts of \xlrv{background
    residuals} near bright stars on the shape measurements, whereas faint stars
    are suitable for testing the effect of possible residual systematics of the
    PSF correction in shape measurements.
\end{enumerate}

Fig.~\ref{fig:systest_stack} summarizes results of our null tests from stacked
shear signals averaged over the entire survey area. We find that the shear
profiles are mostly consistent with zero, suggesting no evidence for any
significant detection of systematic effects. To quantify the significance of
any deviations from zero, in Tables~\ref{table:stack_e} and \ref{table:stack_b}
in Appendix~\ref{app:1}, we tabulate $\chi^2$ and $p$ values for the null
hypothesis of the stacked shear profiles for the six individual fields as well
as all fields combined. To do so, we fully account for correlations between
different radial bins, which are caused by e.g., cosmic shear, by deriving the
full covariance matrix of the measurements using the 200 realizations of the
mock shear catalog mentioned above.
\xlrv{
    We include the correction factor ($\sim 0.85-0.9$) of \citet{Cov-Hartlap}
    to correct the covariances estimated from a limited number of
    realisations}\footnote{
    \citet{cov-Percival2021} provides a better motivated correction factor, and
    it will improve the p-value for this test.
}.
We find that only four out of 63 $p$ values fall below a nominal threshold of
0.05, which is consistent with statistical fluctuations.  Therefore we conclude
that stacked shear profile tests show no significant evidence for significant
systematic errors in the shear catalogs.

\begin{figure}[ht!]
\begin{center}
\includegraphics[width=0.5\textwidth]{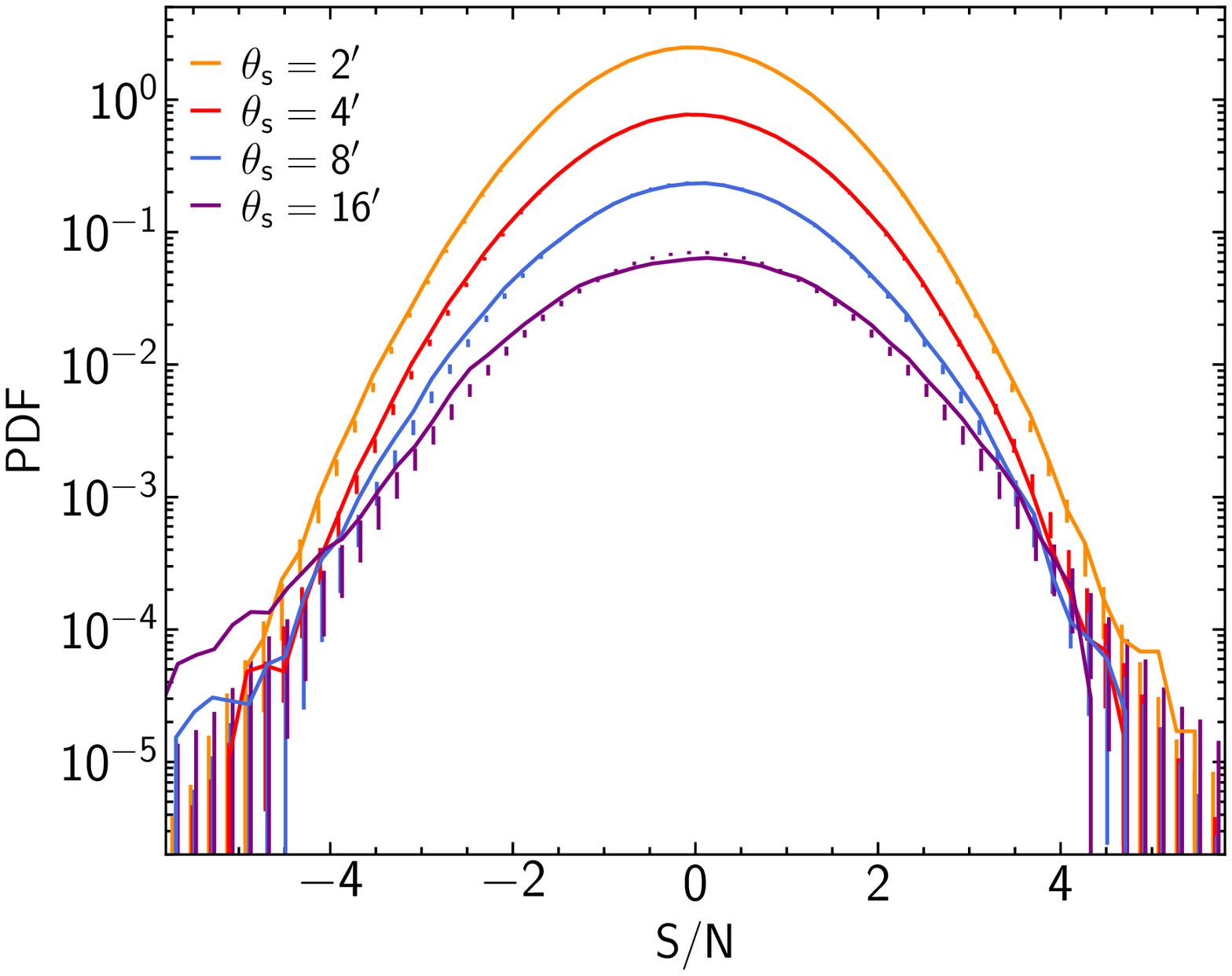}
\end{center}
\caption{
    Observed $B$-mode mass map probability distribution functions (PDFs) for
    different smoothing length $\theta_{\rm s}$, which are shown by solid
    lines, are compared with those from mock shear catalogs with errors shown
    by bars. Results for $\theta_{\rm s}=2'$, $4'$, and $8'$ are shifted upward
    by 1.5~dex, 1~dex, and 0.5~dex, respectively, for illustrative purposes.
    }
    \label{fig:systest_map}
\end{figure}

\begin{figure*}[ht!]
\begin{center}
\includegraphics[width=0.85\textwidth]{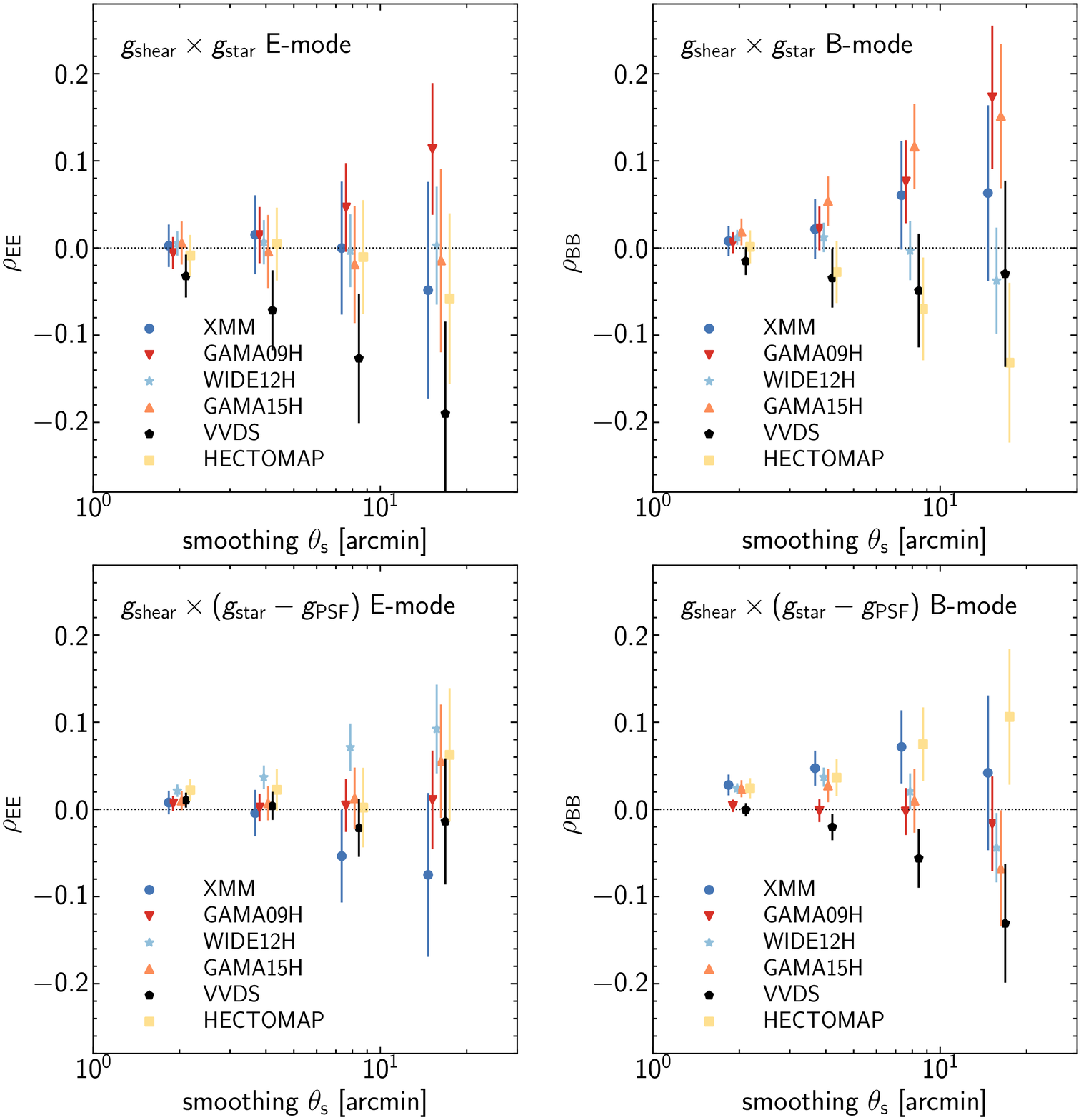}
\end{center}
\caption{
    Pearson cross-correlation coefficients of $E$- ({\em left panels}) and
    $B$-mode ({\em right panels}) mass maps and star mass maps constructed
    using star ellipticities. We consider cases both with ({\em lower panels})
    and without ({\em upper panels}) the PSF correction for star mass maps.
    Pearson cross-correlation coefficients are shown as a function of the
    smoothing length of mass maps. Different symbols show results for different
    observed fields. Errors are estimated from mock shear catalogs including
    cosmic shear.
    }
    \label{fig:systest_map_cross_star}
\end{figure*}

\subsection{Mass maps}
\label{subsec:nullTest_massMap}

The observed shear field can be converted to the projected density field
\citep{KS93}. Since weak lensing produces mostly ``$E$-mode'' convergence
fields, we can use ``$B$-mode'' convergence fields as additional null tests.
For this purpose, we reconstruct Gaussian-smoothed convergence maps adopting
four different smoothing lengths, following the methodology detailed in
\citet{HSC1-wlmap}. Fig.~\ref{fig:systest_map} shows the PDFs of the $B$-mode
mass maps for four different smoothing lengths as compared to the average PDFs
from mock shear catalogs that correctly capture effects of the survey boundary
and masking that mix $E$- and $B$-modes. We find that the $B$-mode mass map
PDFs follow an approximately Gaussian distribution and are roughly consistent
with those from mock shear catalogs. There are small deviations from the mock
results, which were also seen in the HSC-SSP S16A shear catalog
\citep{HSC1-shape} and must originate from PSF leakage or PSF modelling errors
as we will discuss below.

Mass maps can be used for a complementary check for PSF leakage and PSF
modelling errors \citep[see e.g.,][]{HSC1-shape}. Specifically, we derive the
Pearson correlation coefficient $\rho_{\kappa_1\kappa_2}$ between the $E$- or
$B$-mode mass map and the $E$- or $B$-mode star mass map, where the star mass
map refers to the smoothed convergence map created using star ellipticities. We
consider two types of star ellipticities: one uses observed star ellipticities
to check for PSF leakage, and the other uses star ellipticities after PSF
correction to check for PSF modelling errors. In this analysis, we only use
reserved stars that are not used for modelling the PSF.  We show the results in
Fig.~\ref{fig:systest_map_cross_star}. We find that correlations between mass
maps and star mass maps without the PSF correction are consistent with zero
within $\sim 2 \sigma$, indicating that this test shows no sign of PSF leakage.
On the other hand, we see small deviations from zero for the case of star mass
maps with the PSF correction, which suggests that PSF modelling errors may be a
source of small deviations of the $B$-mode mass map PDFs from mock results as
shown in Fig.~\ref{fig:systest_map}. This likely has a similar origin as that
of the issue related to the $\rho_2$ statistic on large scales seen in
Fig.~\ref{fig:psf_rho} in Section~\ref{sec:psftest}. These PSF effects and
their impact need to be carefully evaluated during cosmological analyses as was
done in \citet{HSC1-cs-fourier} and \citet{HSC1-cs-real}.

\subsection{Star-galaxy cross correlation}
\label{subsec:star-galaxy}

\begin{figure}
\begin{center}
\includegraphics[width=0.48\textwidth]{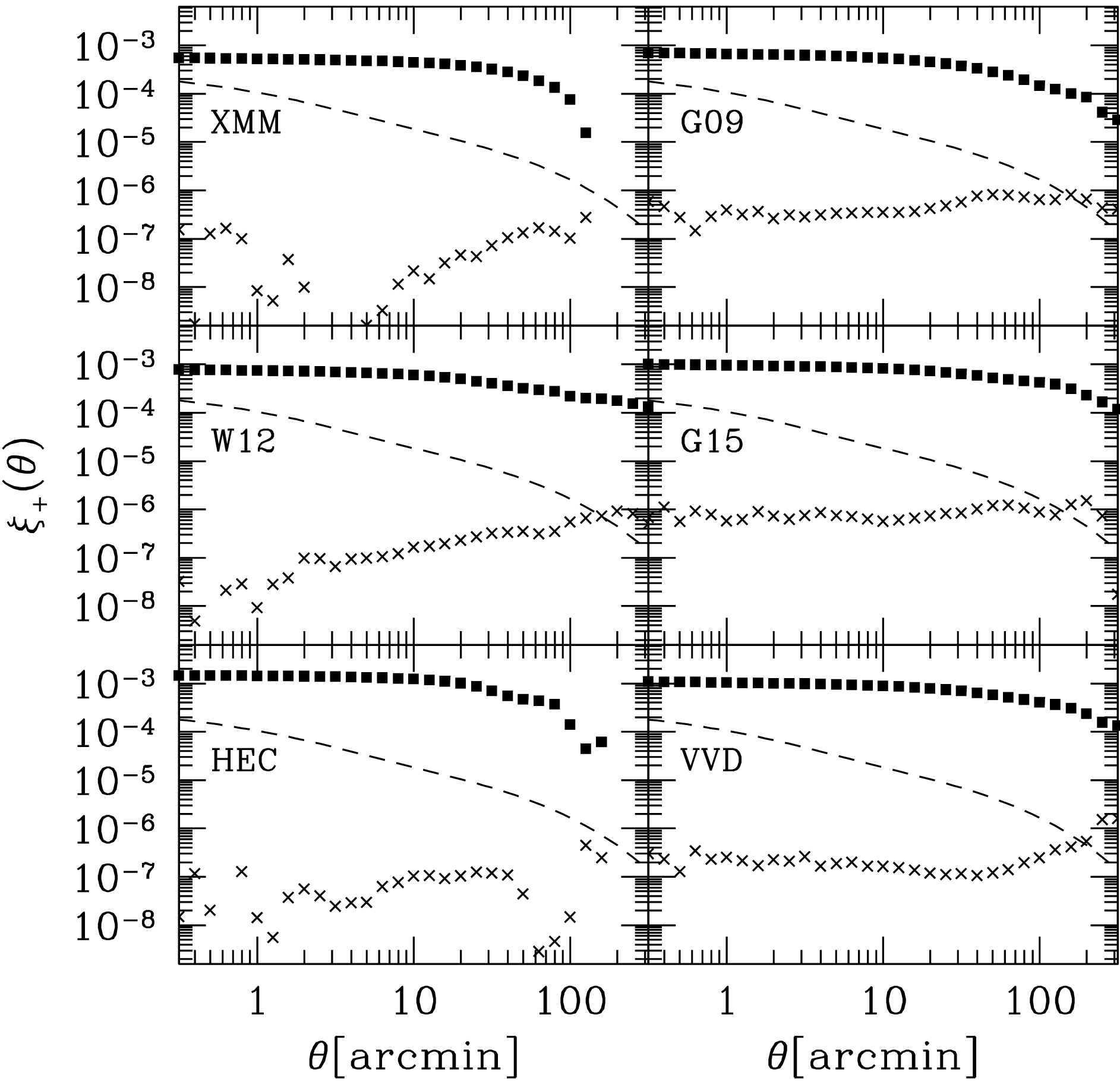}
\end{center}
\caption{
    Separate panels show (for each survey field) the shape–shape correlation
    function $\xi_+(\theta)$ for PSF star shapes as points (the errors are
    smaller than the size of the points); the predicted cosmic shear
    correlation function with a WMAP9 cosmology using the $n(z)$ from HSC
    photometric redshifts without any correction for photo-$z$ errors (which
    illustrates the approximate magnitude of the expected cosmic shear signal)
    as dashed lines; and $\xi_{\rm sys}$, defined in equation
    (\ref{eq:xi_sys}), as crosses.
    }
    \label{fig:xisys_s19a_v2}
\end{figure}

Next, following \citet{HSC1-shape}, we present results of an empirical test for
the possible impact of either PSF modelling errors or residual PSF anisotropy
in galaxy shapes on cosmic shear two-point correlation function measurements.
We calculate the following combination of the star-galaxy cross correlation
function and the star auto correlation function,
\begin{equation}
\label{eq:xi_sys}
\xi_{\rm sys}=
\frac{\langle g^\dagger_\ast \hat{g}_{\rm gal}\rangle^2}
{\langle g^{\dagger}_\ast \hat{g}_\ast\rangle}.
\end{equation}
Adopting the prescription given in Section~\ref{subsec:additive-ss}, one finds
that this combination gives an estimate of a residual correlation caused by PSF
anisotropy leakage to the galaxy-galaxy correlation function ($\langle
g^\dagger g \rangle$), $\Delta \langle g^\dagger g \rangle \sim a^2 \langle
g^\dagger_\ast g_\ast \rangle$. Note that $\xi_{\rm sys}$ can also detect
additive PSF modelling errors that contribute to $\langle g^\dagger g \rangle$.
Fig.~\ref{fig:xisys_s19a_v2} shows $\xi_{\rm sys}$ for each field along with
the standard $\Lambda$CDM prediction of $\xi_+$, the cosmic shear correlation
function. Overall, the amplitudes and shapes of $\xi_{\rm sys}$ are similar to
those of the first-year shear catalog (see Fig.~18 of \citealt{HSC1-shape}).
The amplitude of $\xi_{\rm sys}$ varies among fields and can be comparable to
$\langle g^\dagger g \rangle$ on degree scales. This indicates that a careful
choice of angular scales used in cosmological analyses of cosmic shear
two-point correlation function (or power spectrum) and a correction for the
impact of PSF errors are required, as was done in
\citet{cosmicShear_HSC1_Chiaki2019} and \citet{HSC1-cs-real}.

\section{Summary and Outlook}
\label{sec:summary}

In this paper, we presented the galaxy shear catalog measured from the $i$-band
wide layer of the HSC S19A internal data release. The galaxy shapes were
calibrated with HSC-like image simulations that transfer the galaxy images from
COSMOS HST to the HSC observing conditions. We confirmed that the simulated
galaxy sample has the same distributions of galaxy properties as the real HSC
data. Then we used the simulation to calibrate the galaxy property-dependent
shear estimation bias, including redshift-dependent bias. We tested the
residuals of the shear calibration by applying the calibrated shear estimator
to sub-samples of the simulation divided by several different galaxy
properties. The selection bias was removed empirically from ensemble shear
estimates using the simulation.

In summary, the resulting galaxy shear catalog covers an area of
$433.48~\rm{deg}^2$ of the northern sky, split into six fields, with a mean
$i$-band seeing of $0.59\arcsec$. With conservative galaxy selection criteria,
the raw galaxy number density is $22.9~\rm{arcmin}^{-2}$ and the effective
galaxy number density is $19.9~\rm{arcmin}^{-2}$. The galaxy catalog has a
depth of $24.5$th magnitude.

We defined the requirements for cosmological weak lensing science for this
shear catalog, and quantified potential systematics in the catalog using a
series of internal null tests for problems with point-spread function modelling
and shear estimation.

\subsection{Future improvements}

Here we summarize the areas to improve in our future shear catalog, beyond the
already-highlighted issue of PSF model shape residuals
(Section~\ref{sec:psftest}).

\subsubsection{Shear catalogs from multi-band images}

One limitation of our current shear catalog is that we only have $i$-band image
simulations to validate and calibrate shear estimations obtained from $i$-band.
That is, we are not able to use the galaxy shapes observed from other filter
bands to reduce shape measurement uncertainties and shape noise in weak lensing
science. The potential difficulty that need to be overcome to generate shear
catalogs from multi-band HSC images is that other bands (i.e., $grzy$) are
different in wavelength from F814W band filter, the transmission curve of which
has a much larger overlap with that of the $i$-band filter than other bands.
Therefore, it would be necessary to carefully check whether the input training
samples are still representative of the galaxy images in other bands of HSC.

\subsubsection{Unrecognized blending}

Unrecognized blending refers to the case that multiple blended sources are
identified as one single source by the detector. It has been shown by many
existing works that unrecognized blending has two influences on shear
estimation:
\begin{enumerate}
    \item The possibility of unrecognized blends depends upon the underlying
        shear distortion. Such a shear-dependent blending identification leads
        to an anisotropic selection in the galaxy sample; therefore, it can
        lead to a few percent multiplicative shear bias
        \citep{metaDet-Sheldon2020}.
    \item Shear estimated from a detection containing unrecognized blended
        galaxies is a weighted average of the shear signals at different
        redshifts if the blended galaxies are located at different redshift
        planes. Such effect biases the effective galaxy number density on
        redshift: $n(z)$ \citep{DESY3-BlendshearCalib-MacCrann2021}.
\end{enumerate}

\xlrv{
Since we directly used real images from the HST COSMOS survey in our image
simulations, both recognized and unrecognized blended galaxies with magnitude
brighter than the HST's magnitude limit were fully included in our fiducial
image simulation.  Therefore, the fiducial calibration corrects the biases from
shear-dependent blending identification. As the fiducial image simulation
distorts images in units of postage stamps, it does not includes
redshift-dependent shear. We generated another image simulation that distorts
parametric galaxies with redshift-dependent shear under the HSC-like
observational condition and found that the multiplicative bias, which is not
included in the fiducial calibration due to redshift-dependent shear,
marginally meets the HSC three-year science requirements. This additional
multiplicative bias will be marginalized in our cosmological analysis.
}

\subsubsection{Independent shear estimator}

Several shear estimators have been proven to have sub-percent level accuracy on
isolated galaxy image simulations, e.g., BFD \citep{BFD-Bernstein2016},
\texttt{M{\scriptsize ETACALIBRATION}}
\citep{metacal-Huff2017,metacal-Sheldon2017}, \texttt{Fourier$\_$Quad}
\citep{Z17,FQ-Li2021}, and \texttt{FPFS} \citep{FPFS-Li2018,FPFS-Li2021}.
\citet{metaDet-Sheldon2020} proposed \texttt{M{\scriptsize ETADETECTION}}
algorithm, which is able to estimate shear to sub-percent accuracy even from
blended galaxies if the blended sources are distorted by the same shear after
carefully removing the bias from shear-dependent blending identifications.

Fourier Power Function Shapelets \citep[\texttt{FPFS},][]{FPFS-Li2018,
FPFS-Li2021} shear estimator is one of the shear estimators that can reach
sub-percent accuracy on isolated galaxies, but it relies on a calibration of a
few percent ($\sim -5.7\%$) multiplicative shear bias in the presence of
blending. The \texttt{FPFS} shear estimator has been applied to the S16A HSC
data release after being calibrated with HSC-like image simulations
\citep{FPFSHSC1-Li2020}.

For future shear catalogs, we would benefit from the application of a shear
estimator that has minimum reliance on calibration from external image
simulation and produce independent shear catalogs. Cross-comparisons between
independent catalogs will be valuable given the very different assumptions
behind the shear estimators.

\subsection{Outlook for three-year HSC weak lensing science}

In summary, for the systematics that can be characterized with the image
simulations and null tests, the shear catalogs presented in this paper meet the
requirements for the HSC three-year weak lensing science. Additional papers
will detail the methods used to assess the systematics that were not fully
addressed here including systematics in photometric redshift estimation and
systematics from redshift-dependent shear.

For the three-year HSC weak lensing science, some initial papers will be
presented covering topics such as mass mapping and cluster galaxy lensing.
Furthermore, cosmological analyses (e.g. cosmic shear and galaxy-galaxy
lensing) will come in the following months. We will release this catalog
publicly when the three-year cosmological results are published. Details of
data access will be made public at that time.

\begin{ack}

XL was supported by Global Science Graduate Course (GSGC) program of University
of Tokyo and JSPS KAKENHI (JP19J22222). This work was supported in part by JSPS
KAKENHI Grant Nos. JP18H04350, JP18K13561 JP18K03693, JP19H00677, JP20H00181,
JP20H01932, 20H05850, 20H05855, JP20H05856, and by JST AIP Acceleration
Research Grant Number JP20317829. HM was supported by the Jet Propulsion
Laboratory, California Institute of Technology, under a contract with the
National Aeronautics and Space Administration. RM was supported in part by a
grant from the Simons Foundation (Simons Investigator in Astrophysics, Award ID
620789).

\xlrv{
We thank the anonymous referee for feedback that improved the quality of the
paper.
}

The Hyper Suprime-Cam (HSC) collaboration includes the astronomical communities
of Japan and Taiwan, and Princeton University. The HSC instrumentation and
software were developed by the National Astronomical Observatory of Japan
(NAOJ), the Kavli Institute for the Physics and Mathematics of the Universe
(Kavli IPMU), the University of Tokyo, the High Energy Accelerator Research
Organization (KEK), the Academia Sinica Institute for Astronomy and
Astrophysics in Taiwan (ASIAA), and Princeton University. Funding was
contributed by the FIRST program from Japanese Cabinet Office, the Ministry of
Education, Culture, Sports, Science and Technology (MEXT), the Japan Society
for the Promotion of Science (JSPS), Japan Science and Technology Agency (JST),
the Toray Science Foundation, NAOJ, Kavli IPMU, KEK, ASIAA, and Princeton
University.

This paper is based on data collected at the Subaru Telescope and retrieved
from the HSC data archive system, which is operated by Subaru Telescope and
Astronomy Data Center, National Astronomical Observatory of Japan.

This paper makes use of software developed for the Large Synoptic Survey
Telescope. We thank the LSST Project for making their code available as free
software at \texttt{http://dm.lsst.org}.

We acknowledge the public packages used in this paper: \texttt{TreeCorr}
\citep{TreeCorr-Jarvis2004}, a code
(\url{https://github.com/rmjarvis/TreeCorr/}) for fast correlations
measurements based on a ball tree method (similar to a $k$-d tree), is to
compute the correlation functions; \texttt{smatch}
(\url{https://github.com/esheldon/smatch}), a code for points matching on the
sphere based on binary tree and \texttt{HEALPix}, is used to match sources
between catalogs.
\end{ack}

\bibliographystyle{apj}
\bibliography{citation}

\begin{thebibliography}{}
\expandafter\ifx\csname natexlab\endcsname\relax\def\natexlab#1{#1}\fi

\bibitem[{{Abbott} {et~al.}(2018){Abbott}, {Abdalla}, {Alarcon}, {Aleksi{\'c}},
  {Allam}, {Allen}, {Amara}, {Annis}, {Asorey}, {Avila}, {Bacon}, {Balbinot},
  {Banerji}, {Banik}, {Barkhouse}, {Baumer}, {Baxter}, {Bechtol}, {Becker},
  {Benoit-L{\'e}vy}, {Benson}, {Bernstein}, {Bertin}, {Blazek}, {Bridle},
  {Brooks}, {Brout}, {Buckley-Geer}, {Burke}, {Busha}, {Campos}, {Capozzi},
  {Carnero Rosell}, {Carrasco Kind}, {Carretero}, {Castander}, {Cawthon},
  {Chang}, {Chen}, {Childress}, {Choi}, {Conselice}, {Crittenden}, {Crocce},
  {Cunha}, {D'Andrea}, {da Costa}, {Das}, {Davis}, {Davis}, {De Vicente},
  {DePoy}, {DeRose}, {Desai}, {Diehl}, {Dietrich}, {Dodelson}, {Doel},
  {Drlica-Wagner}, {Eifler}, {Elliott}, {Elsner}, {Elvin-Poole}, {Estrada},
  {Evrard}, {Fang}, {Fernandez}, {Fert{\'e}}, {Finley}, {Flaugher}, {Fosalba},
  {Friedrich}, {Frieman}, {Garc{\'\i}a-Bellido}, {Garcia-Fernandez}, {Gatti},
  {Gaztanaga}, {Gerdes}, {Giannantonio}, {Gill}, {Glazebrook}, {Goldstein},
  {Gruen}, {Gruendl}, {Gschwend}, {Gutierrez}, {Hamilton}, {Hartley}, {Hinton},
  {Honscheid}, {Hoyle}, {Huterer}, {Jain}, {James}, {Jarvis}, {Jeltema},
  {Johnson}, {Johnson}, {Kacprzak}, {Kent}, {Kim}, {King}, {Kirk}, {Kokron},
  {Kovacs}, {Krause}, {Krawiec}, {Kremin}, {Kuehn}, {Kuhlmann}, {Kuropatkin},
  {Lacasa}, {Lahav}, {Li}, {Liddle}, {Lidman}, {Lima}, {Lin}, {MacCrann},
  {Maia}, {Makler}, {Manera}, {March}, {Marshall}, {Martini}, {McMahon},
  {Melchior}, {Menanteau}, {Miquel}, {Miranda}, {Mudd}, {Muir}, {M{\"o}ller},
  {Neilsen}, {Nichol}, {Nord}, {Nugent}, {Ogando}, {Palmese}, {Peacock},
  {Peiris}, {Peoples}, {Percival}, {Petravick}, {Plazas}, {Porredon}, {Prat},
  {Pujol}, {Rau}, {Refregier}, {Ricker}, {Roe}, {Rollins}, {Romer}, {Roodman},
  {Rosenfeld}, {Ross}, {Rozo}, {Rykoff}, {Sako}, {Salvador}, {Samuroff},
  {S{\'a}nchez}, {Sanchez}, {Santiago}, {Scarpine}, {Schindler}, {Scolnic},
  {Secco}, {Serrano}, {Sevilla-Noarbe}, {Sheldon}, {Smith}, {Smith}, {Smith},
  {Soares-Santos}, {Sobreira}, {Suchyta}, {Tarle}, {Thomas}, {Troxel},
  {Tucker}, {Tucker}, {Uddin}, {Varga}, {Vielzeuf}, {Vikram}, {Vivas},
  {Walker}, {Wang}, {Wechsler}, {Weller}, {Wester}, {Wolf}, {Yanny}, {Yuan},
  {Zenteno}, {Zhang}, {Zhang}, {Zuntz}, \& {Dark Energy Survey
  Collaboration}}]{2018PhRvD..98d3526A}
{Abbott}, T.~M.~C., {Abdalla}, F.~B., {Alarcon}, A., {et~al.} 2018, \prd, 98,
  043526

\bibitem[{{Aihara} {et~al.}(2018{\natexlab{a}}){Aihara}, {Armstrong},
  {Bickerton}, {Bosch}, {Coupon}, {Furusawa}, {Hayashi}, {Ikeda}, {Kamata},
  {Karoji}, {Kawanomoto}, {Koike}, {Komiyama}, {Lang}, {Lupton}, {Mineo},
  {Miyatake}, {Miyazaki}, {Morokuma}, {Obuchi}, {Oishi}, {Okura}, {Price},
  {Takata}, {Tanaka}, {Tanaka}, {Tanaka}, {Uchida}, {Uraguchi}, {Utsumi},
  {Wang}, {Yamada}, {Yamanoi}, {Yasuda}, {Arimoto}, {Chiba}, {Finet},
  {Fujimori}, {Fujimoto}, {Furusawa}, {Goto}, {Goulding}, {Gunn}, {Harikane},
  {Hattori}, {Hayashi}, {He{\l}miniak}, {Higuchi}, {Hikage}, {Ho}, {Hsieh},
  {Huang}, {Huang}, {Imanishi}, {Iwata}, {Jaelani}, {Jian}, {Kashikawa},
  {Katayama}, {Kojima}, {Konno}, {Koshida}, {Kusakabe}, {Leauthaud}, {Lee},
  {Lin}, {Lin}, {Mandelbaum}, {Matsuoka}, {Medezinski}, {Miyama}, {Momose},
  {More}, {More}, {Mukae}, {Murata}, {Murayama}, {Nagao}, {Nakata}, {Niida},
  {Niikura}, {Nishizawa}, {Oguri}, {Okabe}, {Ono}, {Onodera}, {Onoue}, {Ouchi},
  {Pyo}, {Shibuya}, {Shimasaku}, {Simet}, {Speagle}, {Spergel}, {Strauss},
  {Sugahara}, {Sugiyama}, {Suto}, {Suzuki}, {Tait}, {Takada}, {Terai}, {Toba},
  {Turner}, {Uchiyama}, {Umetsu}, {Urata}, {Usuda}, {Yeh}, \&
  {Yuma}}]{HSC1-data}
{Aihara}, H., {Armstrong}, R., {Bickerton}, S., {et~al.} 2018{\natexlab{a}},
  \pasj, 70, S8

\bibitem[{{Aihara} {et~al.}(2018{\natexlab{b}}){Aihara}, {Arimoto},
  {Armstrong}, {Arnouts}, {Bahcall}, {Bickerton}, {Bosch}, {Bundy}, {Capak},
  {Chan}, {Chiba}, {Coupon}, {Egami}, {Enoki}, {Finet}, {Fujimori}, {Fujimoto},
  {Furusawa}, {Furusawa}, {Goto}, {Goulding}, {Greco}, {Greene}, {Gunn},
  {Hamana}, {Harikane}, {Hashimoto}, {Hattori}, {Hayashi}, {Hayashi},
  {He{\l}miniak}, {Higuchi}, {Hikage}, {Ho}, {Hsieh}, {Huang}, {Huang},
  {Ikeda}, {Imanishi}, {Inoue}, {Iwasawa}, {Iwata}, {Jaelani}, {Jian},
  {Kamata}, {Karoji}, {Kashikawa}, {Katayama}, {Kawanomoto}, {Kayo}, {Koda},
  {Koike}, {Kojima}, {Komiyama}, {Konno}, {Koshida}, {Koyama}, {Kusakabe},
  {Leauthaud}, {Lee}, {Lin}, {Lin}, {Lupton}, {Mandelbaum}, {Matsuoka},
  {Medezinski}, {Mineo}, {Miyama}, {Miyatake}, {Miyazaki}, {Momose}, {More},
  {More}, {Moritani}, {Moriya}, {Morokuma}, {Mukae}, {Murata}, {Murayama},
  {Nagao}, {Nakata}, {Niida}, {Niikura}, {Nishizawa}, {Obuchi}, {Oguri},
  {Oishi}, {Okabe}, {Okamoto}, {Okura}, {Ono}, {Onodera}, {Onoue}, {Osato},
  {Ouchi}, {Price}, {Pyo}, {Sako}, {Sawicki}, {Shibuya}, {Shimasaku},
  {Shimono}, {Shirasaki}, {Silverman}, {Simet}, {Speagle}, {Spergel},
  {Strauss}, {Sugahara}, {Sugiyama}, {Suto}, {Suyu}, {Suzuki}, {Tait},
  {Takada}, {Takata}, {Tamura}, {Tanaka}, {Tanaka}, {Tanaka}, {Tanaka},
  {Terai}, {Terashima}, {Toba}, {Tominaga}, {Toshikawa}, {Turner}, {Uchida},
  {Uchiyama}, {Umetsu}, {Uraguchi}, {Urata}, {Usuda}, {Utsumi}, {Wang}, {Wang},
  {Wong}, {Yabe}, {Yamada}, {Yamanoi}, {Yasuda}, {Yeh}, {Yonehara}, \&
  {Yuma}}]{HSC-SSP2018}
{Aihara}, H., {Arimoto}, N., {Armstrong}, R., {et~al.} 2018{\natexlab{b}},
  \pasj, 70, S4

\bibitem[{{Aihara} {et~al.}(2019){Aihara}, {AlSayyad}, {Ando}, {Armstrong},
  {Bosch}, {Egami}, {Furusawa}, {Furusawa}, {Goulding}, {Harikane}, {Hikage},
  {Ho}, {Hsieh}, {Huang}, {Ikeda}, {Imanishi}, {Ito}, {Iwata}, {Jaelani},
  {Kakuma}, {Kawana}, {Kikuta}, {Kobayashi}, {Koike}, {Komiyama}, {Li},
  {Liang}, {Lin}, {Luo}, {Lupton}, {Lust}, {MacArthur}, {Matsuoka}, {Mineo},
  {Miyatake}, {Miyazaki}, {More}, {Murata}, {Namiki}, {Nishizawa}, {Oguri},
  {Okabe}, {Okamoto}, {Okura}, {Ono}, {Onodera}, {Onoue}, {Osato}, {Ouchi},
  {Shibuya}, {Strauss}, {Sugiyama}, {Suto}, {Takada}, {Takagi}, {Takata},
  {Takita}, {Tanaka}, {Terai}, {Toba}, {Uchiyama}, {Utsumi}, {Wang}, {Wang}, \&
  {Yamada}}]{HSC2-data}
{Aihara}, H., {AlSayyad}, Y., {Ando}, M., {et~al.} 2019, \pasj, 71, 114

\bibitem[{{Aihara} {et~al.}(2021){Aihara}, {AlSayyad}, {Ando}, {Armstrong},
  {Bosch}, {Egami}, {Furusawa}, {Furusawa}, {Harasawa}, {Harikane}, {Hsieh},
  {Ikeda}, {Ito}, {Iwata}, {Kodama}, {Koike}, {Kokubo}, {Komiyama}, {Li},
  {Liang}, {Lin}, {Lupton}, {Lust}, {MacArthur}, {Mawatari}, {Mineo},
  {Miyatake}, {Miyazaki}, {More}, {Morishima}, {Murayama}, {Nakajima},
  {Nakata}, {Nishizawa}, {Oguri}, {Okabe}, {Okura}, {Ono}, {Osato}, {Ouchi},
  {Pan}, {Plazas Malag{\'o}n}, {Price}, {Reed}, {Rykoff}, {Shibuya},
  {Simunovic}, {Strauss}, {Sugimori}, {Suto}, {Suzuki}, {Takada}, {Takagi},
  {Takata}, {Takita}, {Tanaka}, {Tang}, {Taranu}, {Terai}, {Toba}, {Turner},
  {Uchiyama}, {Vijarnwannaluk}, {Waters}, {Yamada}, {Yamamoto}, \&
  {Yamashita}}]{HSC3-data}
---. 2021, arXiv:2108.13045

\bibitem[{{Alam} {et~al.}(2017){Alam}, {Miyatake}, {More}, {Ho}, \&
  {Mandelbaum}}]{2017MNRAS.465.4853A}
{Alam}, S., {Miyatake}, H., {More}, S., {Ho}, S., \& {Mandelbaum}, R. 2017,
  \mnras, 465, 4853

\bibitem[{{Amon} {et~al.}(2021){Amon}, {Gruen}, {Troxel}, {MacCrann},
  {Dodelson}, {Choi}, {Doux}, {Secco}, {Samuroff}, {Krause}, {Cordero},
  {Myles}, {DeRose}, {Wechsler}, {Gatti}, {Navarro-Alsina}, {Bernstein},
  {Jain}, {Blazek}, {Alarcon}, {Fert{\'e}}, {Raveri}, {Lemos}, {Campos},
  {Prat}, {S{\'a}nchez}, {Jarvis}, {Alves}, {Andrade-Oliveira}, {Baxter},
  {Bechtol}, {Becker}, {Bridle}, {Camacho}, {Campos}, {Carnero Rosell},
  {Carrasco Kind}, {Cawthon}, {Chang}, {Chen}, {Chintalapati}, {Crocce},
  {Davis}, {Diehl}, {Drlica-Wagner}, {Eckert}, {Eifler}, {Elvin-Poole},
  {Everett}, {Fang}, {Fosalba}, {Friedrich}, {Giannini}, {Gruendl}, {Harrison},
  {Hartley}, {Herner}, {Huang}, {Huff}, {Huterer}, {Kuropatkin}, {Leget},
  {Liddle}, {McCullough}, {Muir}, {Pandey}, {Park}, {Porredon}, {Refregier},
  {Rollins}, {Roodman}, {Rosenfeld}, {Ross}, {Rykoff}, {Sanchez},
  {Sevilla-Noarbe}, {Sheldon}, {Shin}, {Troja}, {Tutusaus}, {Varga},
  {Weaverdyck}, {Yanny}, {Yin}, {Zhang}, {Zuntz}, {Aguena}, {Allam}, {Annis},
  {Bacon}, {Bertin}, {Bhargava}, {Brooks}, {Buckley-Geer}, {Burke},
  {Carretero}, {Costanzi}, {da Costa}, {Pereira}, {De Vicente}, {Desai},
  {Dietrich}, {Doel}, {Ferrero}, {Flaugher}, {Frieman}, {Garc{\'\i}a-Bellido},
  {Gaztanaga}, {Gerdes}, {Giannantonio}, {Gschwend}, {Gutierrez}, {Hinton},
  {Hollowood}, {Honscheid}, {Hoyle}, {James}, {Kron}, {Kuehn}, {Lahav}, {Lima},
  {Lin}, {Maia}, {Marshall}, {Martini}, {Melchior}, {Menanteau}, {Miquel},
  {Mohr}, {Morgan}, {Ogando}, {Palmese}, {Paz-Chinch{\'o}n}, {Petravick},
  {Pieres}, {Plazas Malag{\'o}n}, {Romer}, {Sanchez}, {Scarpine}, {Schubnell},
  {Serrano}, {Smith}, {Soares-Santos}, {Suchyta}, {Tarle}, {Thomas}, {To}, \&
  {Weller}}]{DESY3-cosmicShearA}
{Amon}, A., {Gruen}, D., {Troxel}, M.~A., {et~al.} 2021, arXiv e-prints,
  arXiv:2105.13543

\bibitem[{{Antilogus} {et~al.}(2014){Antilogus}, {Astier}, {Doherty},
  {Guyonnet}, \& {Regnault}}]{Antilogus2014}
{Antilogus}, P., {Astier}, P., {Doherty}, P., {Guyonnet}, A., \& {Regnault}, N.
  2014, Journal of Instrumentation, 9, C03048

\bibitem[{{Asgari} {et~al.}(2021){Asgari}, {Lin}, {Joachimi}, {Giblin},
  {Heymans}, {Hildebrandt}, {Kannawadi}, {St{\"o}lzner}, {Tr{\"o}ster}, {van
  den Busch}, {Wright}, {Bilicki}, {Blake}, {de Jong}, {Dvornik}, {Erben},
  {Getman}, {Hoekstra}, {K{\"o}hlinger}, {Kuijken}, {Miller}, {Radovich},
  {Schneider}, {Shan}, \& {Valentijn}}]{KiDS1000-CS}
{Asgari}, M., {Lin}, C.-A., {Joachimi}, B., {et~al.} 2021, \aap, 645, A104

\bibitem[{{Bacon} {et~al.}(2000){Bacon}, {Refregier}, \&
  {Ellis}}]{2000MNRAS.318..625B}
{Bacon}, D.~J., {Refregier}, A.~R., \& {Ellis}, R.~S. 2000, \mnras, 318, 625

\bibitem[{{Bernstein}(2010)}]{modelBias-Bernstein10}
{Bernstein}, G.~M. 2010, \mnras, 406, 2793

\bibitem[{{Bernstein} {et~al.}(2016){Bernstein}, {Armstrong}, {Krawiec}, \&
  {March}}]{BFD-Bernstein2016}
{Bernstein}, G.~M., {Armstrong}, R., {Krawiec}, C., \& {March}, M.~C. 2016,
  \mnras, 459, 4467

\bibitem[{{Bernstein} \& {Jarvis}(2002)}]{shearEst-Bernstein2002}
{Bernstein}, G.~M., \& {Jarvis}, M. 2002, \aj, 123, 583

\bibitem[{{Bertin}(2011)}]{PSFEx11}
{Bertin}, E. 2011, in Astronomical Society of the Pacific Conference Series,
  Vol. 442, Astronomical Data Analysis Software and Systems XX, ed. I.~N.
  {Evans}, A.~{Accomazzi}, D.~J. {Mink}, \& A.~H. {Rots}, 435

\bibitem[{{Blake} {et~al.}(2016){Blake}, {Joudaki}, {Heymans}, {Choi}, {Erben},
  {Harnois-Deraps}, {Hildebrandt}, {Joachimi}, {Nakajima}, {van Waerbeke}, \&
  {Viola}}]{2016MNRAS.456.2806B}
{Blake}, C., {Joudaki}, S., {Heymans}, C., {et~al.} 2016, \mnras, 456, 2806

\bibitem[{{Bosch} {et~al.}(2018){Bosch}, {Armstrong}, {Bickerton}, {Furusawa},
  {Ikeda}, {Koike}, {Lupton}, {Mineo}, {Price}, {Takata}, {Tanaka}, {Yasuda},
  {AlSayyad}, {Becker}, {Coulton}, {Coupon}, {Garmilla}, {Huang}, {Krughoff},
  {Lang}, {Leauthaud}, {Lim}, {Lust}, {MacArthur}, {Mandelbaum}, {Miyatake},
  {Miyazaki}, {Murata}, {More}, {Okura}, {Owen}, {Swinbank}, {Strauss},
  {Yamada}, \& {Yamanoi}}]{HSC1-pipeline}
{Bosch}, J., {Armstrong}, R., {Bickerton}, S., {et~al.} 2018, \pasj, 70, S5

\bibitem[{{Bosch} {et~al.}(2019){Bosch}, {AlSayyad}, {Armstrong}, {Bellm},
  {Chiang}, {Eggl}, {Findeisen}, {Fisher-Levine}, {Guy}, {Guyonnet},
  {Ivezi{\'c}}, {Jenness}, {Kov{\'a}cs}, {Krughoff}, {Lupton}, {Lust},
  {MacArthur}, {Meyers}, {Moolekamp}, {Morrison}, {Morton}, {O'Mullane},
  {Parejko}, {Plazas}, {Price}, {Rawls}, {Reed}, {Schellart}, {Slater},
  {Sullivan}, {Swinbank}, {Taranu}, {Waters}, \&
  {Wood-Vasey}}]{LSSTpipe-Bosch2019}
{Bosch}, J., {AlSayyad}, Y., {Armstrong}, R., {et~al.} 2019, in Astronomical
  Society of the Pacific Conference Series, Vol. 523, Astronomical Data
  Analysis Software and Systems XXVII, ed. P.~J. {Teuben}, M.~W. {Pound}, B.~A.
  {Thomas}, \& E.~M. {Warner}, 521

\bibitem[{{Burke} {et~al.}(2018){Burke}, {Rykoff}, {Allam}, {Annis}, {Bechtol},
  {Bernstein}, {Drlica-Wagner}, {Finley}, {Gruendl}, {James}, {Kent},
  {Kessler}, {Kuhlmann}, {Lasker}, {Li}, {Scolnic}, {Smith}, {Tucker},
  {Wester}, {Yanny}, {Abbott}, {Abdalla}, {Benoit-L{\'e}vy}, {Bertin}, {Carnero
  Rosell}, {Carrasco Kind}, {Carretero}, {Cunha}, {D'Andrea}, {da Costa},
  {Desai}, {Diehl}, {Doel}, {Estrada}, {Garc{\'\i}a-Bellido}, {Gruen},
  {Gutierrez}, {Honscheid}, {Kuehn}, {Kuropatkin}, {Maia}, {March}, {Marshall},
  {Melchior}, {Menanteau}, {Miquel}, {Plazas}, {Sako}, {Sanchez}, {Scarpine},
  {Schindler}, {Sevilla-Noarbe}, {Smith}, {Smith}, {Soares-Santos}, {Sobreira},
  {Suchyta}, {Tarle}, {Walker}, \& {DES Collaboration}}]{FGCM-Burke2018}
{Burke}, D.~L., {Rykoff}, E.~S., {Allam}, S., {et~al.} 2018, \aj, 155, 41

\bibitem[{{Chang} {et~al.}(2013){Chang}, {Jarvis}, {Jain}, {Kahn}, {Kirkby},
  {Connolly}, {Krughoff}, {Peng}, \&
  {Peterson}}]{WLsurvey-neffective-Chang2013}
{Chang}, C., {Jarvis}, M., {Jain}, B., {et~al.} 2013, \mnras, 434, 2121

\bibitem[{{Coulton} {et~al.}(2018){Coulton}, {Armstrong}, {Smith}, {Lupton}, \&
  {Spergel}}]{HSC-BFE-Coulton2018}
{Coulton}, W.~R., {Armstrong}, R., {Smith}, K.~M., {Lupton}, R.~H., \&
  {Spergel}, D.~N. 2018, \aj, 155, 258

\bibitem[{{Dark Energy Survey Collaboration} {et~al.}(2016){Dark Energy Survey
  Collaboration}, {Abbott}, {Abdalla}, {Aleksi{\'c}}, {Allam}, {Amara},
  {Bacon}, {Balbinot}, {Banerji}, {Bechtol}, {Benoit-L{\'e}vy}, {Bernstein},
  {Bertin}, {Blazek}, {Bonnett}, {Bridle}, {Brooks}, {Brunner}, {Buckley-Geer},
  {Burke}, {Caminha}, {Capozzi}, {Carlsen}, {Carnero-Rosell}, {Carollo},
  {Carrasco-Kind}, {Carretero}, {Castander}, {Clerkin}, {Collett}, {Conselice},
  {Crocce}, {Cunha}, {D'Andrea}, {da Costa}, {Davis}, {Desai}, {Diehl},
  {Dietrich}, {Dodelson}, {Doel}, {Drlica-Wagner}, {Estrada}, {Etherington},
  {Evrard}, {Fabbri}, {Finley}, {Flaugher}, {Foley}, {Fosalba}, {Frieman},
  {Garc{\'{\i}}a-Bellido}, {Gaztanaga}, {Gerdes}, {Giannantonio}, {Goldstein},
  {Gruen}, {Gruendl}, {Guarnieri}, {Gutierrez}, {Hartley}, {Honscheid}, {Jain},
  {James}, {Jeltema}, {Jouvel}, {Kessler}, {King}, {Kirk}, {Kron}, {Kuehn},
  {Kuropatkin}, {Lahav}, {Li}, {Lima}, {Lin}, {Maia}, {Makler}, {Manera},
  {Maraston}, {Marshall}, {Martini}, {McMahon}, {Melchior}, {Merson}, {Miller},
  {Miquel}, {Mohr}, {Morice-Atkinson}, {Naidoo}, {Neilsen}, {Nichol}, {Nord},
  {Ogando}, {Ostrovski}, {Palmese}, {Papadopoulos}, {Peiris}, {Peoples},
  {Percival}, {Plazas}, {Reed}, {Refregier}, {Romer}, {Roodman}, {Ross},
  {Rozo}, {Rykoff}, {Sadeh}, {Sako}, {S{\'a}nchez}, {Sanchez}, {Santiago},
  {Scarpine}, {Schubnell}, {Sevilla-Noarbe}, {Sheldon}, {Smith}, {Smith},
  {Soares-Santos}, {Sobreira}, {Soumagnac}, {Suchyta}, {Sullivan}, {Swanson},
  {Tarle}, {Thaler}, {Thomas}, {Thomas}, {Tucker}, {Vieira}, {Vikram},
  {Walker}, {Wechsler}, {Weller}, {Wester}, {Whiteway}, {Wilcox}, {Yanny},
  {Zhang}, \& {Zuntz}}]{DES16}
{Dark Energy Survey Collaboration}, {Abbott}, T., {Abdalla}, F.~B., {et~al.}
  2016, \mnras, 460, 1270

\bibitem[{{de Jong} {et~al.}(2013){de Jong}, {Verdoes Kleijn}, {Kuijken}, \&
  {Valentijn}}]{KIDS13}
{de Jong}, J. T.~A., {Verdoes Kleijn}, G.~A., {Kuijken}, K.~H., \& {Valentijn},
  E.~A. 2013, Experimental Astronomy, 35, 25

\bibitem[{{de Vaucouleurs}(1948)}]{deVauProfile1948}
{de Vaucouleurs}, G. 1948, Annales d'Astrophysique, 11, 247

\bibitem[{{Fenech Conti} {et~al.}(2017){Fenech Conti}, {Herbonnet}, {Hoekstra},
  {Merten}, {Miller}, \& {Viola}}]{KIDS-shapeCalib-Conti2017}
{Fenech Conti}, I., {Herbonnet}, R., {Hoekstra}, H., {et~al.} 2017, \mnras,
  467, 1627

\bibitem[{{Furusawa} {et~al.}(2018{\natexlab{a}}){Furusawa}, {Koike}, {Takata},
  {Okura}, {Miyatake}, {Lupton}, {Bickerton}, {Price}, {Bosch}, {Yasuda},
  {Mineo}, {Yamada}, {Miyazaki}, {Nakata}, {Koshida}, {Komiyama}, {Utsumi},
  {Kawanomoto}, {Jeschke}, {Noumaru}, {Schubert}, {Iwata}, {Finet},
  {Fujiyoshi}, {Tajitsu}, {Terai}, \& {Lee}}]{HSC-hardware-Furusawa2018}
{Furusawa}, H., {Koike}, M., {Takata}, T., {et~al.} 2018{\natexlab{a}}, \pasj,
  70, S3

\bibitem[{{Furusawa} {et~al.}(2018{\natexlab{b}}){Furusawa}, {Koike}, {Takata},
  {Okura}, {Miyatake}, {Lupton}, {Bickerton}, {Price}, {Bosch}, {Yasuda},
  {Mineo}, {Yamada}, {Miyazaki}, {Nakata}, {Koshida}, {Komiyama}, {Utsumi},
  {Kawanomoto}, {Jeschke}, {Noumaru}, {Schubert}, {Iwata}, {Finet},
  {Fujiyoshi}, {Tajitsu}, {Terai}, \& {Lee}}]{onSiteSys-Furusawa2018}
---. 2018{\natexlab{b}}, \pasj, 70, S3

\bibitem[{{Gaia Collaboration} {et~al.}(2018){Gaia Collaboration}, {Brown},
  {Vallenari}, {Prusti}, {de Bruijne}, {Babusiaux}, {Bailer-Jones}, {Biermann},
  {Evans}, {Eyer}, {Jansen}, {Jordi}, {Klioner}, {Lammers}, {Lindegren},
  {Luri}, {Mignard}, {Panem}, {Pourbaix}, {Randich}, {Sartoretti}, {Siddiqui},
  {Soubiran}, {van Leeuwen}, {Walton}, {Arenou}, {Bastian}, {Cropper},
  {Drimmel}, {Katz}, {Lattanzi}, {Bakker}, {Cacciari}, {Casta{\~n}eda},
  {Chaoul}, {Cheek}, {De Angeli}, {Fabricius}, {Guerra}, {Holl}, {Masana},
  {Messineo}, {Mowlavi}, {Nienartowicz}, {Panuzzo}, {Portell}, {Riello},
  {Seabroke}, {Tanga}, {Th{\'e}venin}, {Gracia-Abril}, {Comoretto},
  {Garcia-Reinaldos}, {Teyssier}, {Altmann}, {Andrae}, {Audard},
  {Bellas-Velidis}, {Benson}, {Berthier}, {Blomme}, {Burgess}, {Busso},
  {Carry}, {Cellino}, {Clementini}, {Clotet}, {Creevey}, {Davidson}, {De
  Ridder}, {Delchambre}, {Dell'Oro}, {Ducourant},
  {Fern{\'a}ndez-Hern{\'a}ndez}, {Fouesneau}, {Fr{\'e}mat}, {Galluccio},
  {Garc{\'\i}a-Torres}, {Gonz{\'a}lez-N{\'u}{\~n}ez}, {Gonz{\'a}lez-Vidal},
  {Gosset}, {Guy}, {Halbwachs}, {Hambly}, {Harrison}, {Hern{\'a}ndez},
  {Hestroffer}, {Hodgkin}, {Hutton}, {Jasniewicz}, {Jean-Antoine-Piccolo},
  {Jordan}, {Korn}, {Krone-Martins}, {Lanzafame}, {Lebzelter}, {L{\"o}ffler},
  {Manteiga}, {Marrese}, {Mart{\'\i}n-Fleitas}, {Moitinho}, {Mora}, {Muinonen},
  {Osinde}, {Pancino}, {Pauwels}, {Petit}, {Recio-Blanco}, {Richards},
  {Rimoldini}, {Robin}, {Sarro}, {Siopis}, {Smith}, {Sozzetti}, {S{\"u}veges},
  {Torra}, {van Reeven}, {Abbas}, {Abreu Aramburu}, {Accart}, {Aerts},
  {Altavilla}, {{\'A}lvarez}, {Alvarez}, {Alves}, {Anderson}, {Andrei},
  {Anglada Varela}, {Antiche}, {Antoja}, {Arcay}, {Astraatmadja}, {Bach},
  {Baker}, {Balaguer-N{\'u}{\~n}ez}, {Balm}, {Barache}, {Barata}, {Barbato},
  {Barblan}, {Barklem}, {Barrado}, {Barros}, {Barstow}, {Bartholom{\'e}
  Mu{\~n}oz}, {Bassilana}, {Becciani}, {Bellazzini}, {Berihuete}, {Bertone},
  {Bianchi}, {Bienaym{\'e}}, {Blanco-Cuaresma}, {Boch}, {Boeche}, {Bombrun},
  {Borrachero}, {Bossini}, {Bouquillon}, {Bourda}, {Bragaglia}, {Bramante},
  {Breddels}, {Bressan}, {Brouillet}, {Br{\"u}semeister}, {Brugaletta},
  {Bucciarelli}, {Burlacu}, {Busonero}, {Butkevich}, {Buzzi}, {Caffau},
  {Cancelliere}, {Cannizzaro}, {Cantat-Gaudin}, {Carballo}, {Carlucci},
  {Carrasco}, {Casamiquela}, {Castellani}, {Castro-Ginard}, {Charlot},
  {Chemin}, {Chiavassa}, {Cocozza}, {Costigan}, {Cowell}, {Crifo}, {Crosta},
  {Crowley}, {Cuypers}, {Dafonte}, {Damerdji}, {Dapergolas}, {David}, {David},
  {de Laverny}, {De Luise}, {De March}, {de Martino}, {de Souza}, {de Torres},
  {Debosscher}, {del Pozo}, {Delbo}, {Delgado}, {Delgado}, {Di Matteo},
  {Diakite}, {Diener}, {Distefano}, {Dolding}, {Drazinos}, {Dur{\'a}n},
  {Edvardsson}, {Enke}, {Eriksson}, {Esquej}, {Eynard Bontemps}, {Fabre},
  {Fabrizio}, {Faigler}, {Falc{\~a}o}, {Farr{\`a}s Casas}, {Federici},
  {Fedorets}, {Fernique}, {Figueras}, {Filippi}, {Findeisen}, {Fonti},
  {Fraile}, {Fraser}, {Fr{\'e}zouls}, {Gai}, {Galleti}, {Garabato},
  {Garc{\'\i}a-Sedano}, {Garofalo}, {Garralda}, {Gavel}, {Gavras}, {Gerssen},
  {Geyer}, {Giacobbe}, {Gilmore}, {Girona}, {Giuffrida}, {Glass}, {Gomes},
  {Granvik}, {Gueguen}, {Guerrier}, {Guiraud}, {Guti{\'e}rrez-S{\'a}nchez},
  {Haigron}, {Hatzidimitriou}, {Hauser}, {Haywood}, {Heiter}, {Helmi}, {Heu},
  {Hilger}, {Hobbs}, {Hofmann}, {Holland}, {Huckle}, {Hypki}, {Icardi},
  {Jan{\ss}en}, {Jevardat de Fombelle}, {Jonker}, {Juh{\'a}sz}, {Julbe},
  {Karampelas}, {Kewley}, {Klar}, {Kochoska}, {Kohley}, {Kolenberg},
  {Kontizas}, {Kontizas}, {Koposov}, {Kordopatis}, {Kostrzewa-Rutkowska},
  {Koubsky}, {Lambert}, {Lanza}, {Lasne}, {Lavigne}, {Le Fustec}, {Le
  Poncin-Lafitte}, {Lebreton}, {Leccia}, {Leclerc}, {Lecoeur-Taibi},
  {Lenhardt}, {Leroux}, {Liao}, {Licata}, {Lindstr{\o}m}, {Lister}, {Livanou},
  {Lobel}, {L{\'o}pez}, {Managau}, {Mann}, {Mantelet}, {Marchal}, {Marchant},
  {Marconi}, {Marinoni}, {Marschalk{\'o}}, {Marshall}, {Martino}, {Marton},
  {Mary}, {Massari}, {Matijevi{\v{c}}}, {Mazeh}, {McMillan}, {Messina},
  {Michalik}, {Millar}, {Molina}, {Molinaro}, {Moln{\'a}r}, {Montegriffo},
  {Mor}, {Morbidelli}, {Morel}, {Morris}, {Mulone}, {Muraveva}, {Musella},
  {Nelemans}, {Nicastro}, {Noval}, {O'Mullane}, {Ord{\'e}novic},
  {Ord{\'o}{\~n}ez-Blanco}, {Osborne}, {Pagani}, {Pagano}, {Pailler},
  {Palacin}, {Palaversa}, {Panahi}, {Pawlak}, {Piersimoni}, {Pineau}, {Plachy},
  {Plum}, {Poggio}, {Poujoulet}, {Pr{\v{s}}a}, {Pulone}, {Racero}, {Ragaini},
  {Rambaux}, {Ramos-Lerate}, {Regibo}, {Reyl{\'e}}, {Riclet}, {Ripepi}, {Riva},
  {Rivard}, {Rixon}, {Roegiers}, {Roelens}, {Romero-G{\'o}mez}, {Rowell},
  {Royer}, {Ruiz-Dern}, {Sadowski}, {Sagrist{\`a} Sell{\'e}s}, {Sahlmann},
  {Salgado}, {Salguero}, {Sanna}, {Santana-Ros}, {Sarasso}, {Savietto},
  {Schultheis}, {Sciacca}, {Segol}, {Segovia}, {S{\'e}gransan}, {Shih},
  {Siltala}, {Silva}, {Smart}, {Smith}, {Solano}, {Solitro}, {Sordo}, {Soria
  Nieto}, {Souchay}, {Spagna}, {Spoto}, {Stampa}, {Steele},
  {Steidelm{\"u}ller}, {Stephenson}, {Stoev}, {Suess}, {Surdej}, {Szabados},
  {Szegedi-Elek}, {Tapiador}, {Taris}, {Tauran}, {Taylor}, {Teixeira},
  {Terrett}, {Teyssandier}, {Thuillot}, {Titarenko}, {Torra Clotet}, {Turon},
  {Ulla}, {Utrilla}, {Uzzi}, {Vaillant}, {Valentini}, {Valette}, {van Elteren},
  {Van Hemelryck}, {van Leeuwen}, {Vaschetto}, {Vecchiato}, {Veljanoski},
  {Viala}, {Vicente}, {Vogt}, {von Essen}, {Voss}, {Votruba}, {Voutsinas},
  {Walmsley}, {Weiler}, {Wertz}, {Wevers}, {Wyrzykowski}, {Yoldas},
  {{\v{Z}}erjal}, {Ziaeepour}, {Zorec}, {Zschocke}, {Zucker}, {Zurbach}, \&
  {Zwitter}}]{GAIA-DR2}
{Gaia Collaboration}, {Brown}, A.~G.~A., {Vallenari}, A., {et~al.} 2018, \aap,
  616, A1

\bibitem[{{Gatti} {et~al.}(2021){Gatti}, {Sheldon}, {Amon}, {Becker}, {Troxel},
  {Choi}, {Doux}, {MacCrann}, {Navarro-Alsina}, {Harrison}, {Gruen},
  {Bernstein}, {Jarvis}, {Secco}, {Fert{\'e}}, {Shin}, {McCullough}, {Rollins},
  {Chen}, {Chang}, {Pandey}, {Tutusaus}, {Prat}, {Elvin-Poole}, {Sanchez},
  {Plazas}, {Roodman}, {Zuntz}, {Abbott}, {Aguena}, {Allam}, {Annis}, {Avila},
  {Bacon}, {Bertin}, {Bhargava}, {Brooks}, {Burke}, {Carnero Rosell}, {Carrasco
  Kind}, {Carretero}, {Castander}, {Conselice}, {Costanzi}, {Crocce}, {da
  Costa}, {Davis}, {De Vicente}, {Desai}, {Diehl}, {Dietrich}, {Doel},
  {Drlica-Wagner}, {Eckert}, {Everett}, {Ferrero}, {Frieman},
  {Garc{\'\i}a-Bellido}, {Gerdes}, {Giannantonio}, {Gruendl}, {Gschwend},
  {Gutierrez}, {Hartley}, {Hinton}, {Hollowood}, {Honscheid}, {Hoyle}, {Huff},
  {Huterer}, {Jain}, {James}, {Jeltema}, {Krause}, {Kron}, {Kuropatkin},
  {Lima}, {Maia}, {Marshall}, {Miquel}, {Morgan}, {Myles}, {Palmese},
  {Paz-Chinch{\'o}n}, {Rykoff}, {Samuroff}, {Sanchez}, {Scarpine}, {Schubnell},
  {Serrano}, {Sevilla-Noarbe}, {Smith}, {Soares-Santos}, {Suchyta}, {Swanson},
  {Tarle}, {Thomas}, {To}, {Tucker}, {Varga}, {Wechsler}, {Weller}, {Wester},
  \& {Wilkinson}}]{DESY3-catalog}
{Gatti}, M., {Sheldon}, E., {Amon}, A., {et~al.} 2021, \mnras, 504, 4312

\bibitem[{{Giblin} {et~al.}(2021){Giblin}, {Heymans}, {Asgari}, {Hildebrandt},
  {Hoekstra}, {Joachimi}, {Kannawadi}, {Kuijken}, {Lin}, {Miller},
  {Tr{\"o}ster}, {van den Busch}, {Wright}, {Bilicki}, {Blake}, {de Jong},
  {Dvornik}, {Erben}, {Getman}, {Napolitano}, {Schneider}, {Shan}, \&
  {Valentijn}}]{KiDS1000-catalog}
{Giblin}, B., {Heymans}, C., {Asgari}, M., {et~al.} 2021, \aap, 645, A105

\bibitem[{{Hamana} {et~al.}(2020){Hamana}, {Shirasaki}, {Miyazaki}, {Hikage},
  {Oguri}, {More}, {Armstrong}, {Leauthaud}, {Mandelbaum}, {Miyatake},
  {Nishizawa}, {Simet}, {Takada}, {Aihara}, {Bosch}, {Komiyama}, {Lupton},
  {Murayama}, {Strauss}, \& {Tanaka}}]{HSC1-cs-real}
{Hamana}, T., {Shirasaki}, M., {Miyazaki}, S., {et~al.} 2020, \pasj, 72, 16

\bibitem[{{Hartlap} {et~al.}(2007){Hartlap}, {Simon}, \&
  {Schneider}}]{Cov-Hartlap}
{Hartlap}, J., {Simon}, P., \& {Schneider}, P. 2007, \aap, 464, 399

\bibitem[{{Heymans} {et~al.}(2021){Heymans}, {Tr{\"o}ster}, {Asgari}, {Blake},
  {Hildebrandt}, {Joachimi}, {Kuijken}, {Lin}, {S{\'a}nchez}, {van den Busch},
  {Wright}, {Amon}, {Bilicki}, {de Jong}, {Crocce}, {Dvornik}, {Erben},
  {Fortuna}, {Getman}, {Giblin}, {Glazebrook}, {Hoekstra}, {Joudaki},
  {Kannawadi}, {K{\"o}hlinger}, {Lidman}, {Miller}, {Napolitano}, {Parkinson},
  {Schneider}, {Shan}, {Valentijn}, {Verdoes Kleijn}, \&
  {Wolf}}]{2021A&A...646A.140H}
{Heymans}, C., {Tr{\"o}ster}, T., {Asgari}, M., {et~al.} 2021, \aap, 646, A140

\bibitem[{{Hikage} {et~al.}(2019{\natexlab{a}}){Hikage}, {Oguri}, {Hamana},
  {More}, {Mandelbaum}, {Takada}, {K{\"o}hlinger}, {Miyatake}, {Nishizawa},
  {Aihara}, {Armstrong}, {Bosch}, {Coupon}, {Ducout}, {Ho}, {Hsieh},
  {Komiyama}, {Lanusse}, {Leauthaud}, {Lupton}, {Medezinski}, {Mineo},
  {Miyama}, {Miyazaki}, {Murata}, {Murayama}, {Shirasaki}, {Sif{\'o}n},
  {Simet}, {Speagle}, {Spergel}, {Strauss}, {Sugiyama}, {Tanaka}, {Utsumi},
  {Wang}, \& {Yamada}}]{cosmicShear_HSC1_Chiaki2019}
{Hikage}, C., {Oguri}, M., {Hamana}, T., {et~al.} 2019{\natexlab{a}}, \pasj,
  71, 43

\bibitem[{{Hikage} {et~al.}(2019{\natexlab{b}}){Hikage}, {Oguri}, {Hamana},
  {More}, {Mandelbaum}, {Takada}, {K{\"o}hlinger}, {Miyatake}, {Nishizawa},
  {Aihara}, {Armstrong}, {Bosch}, {Coupon}, {Ducout}, {Ho}, {Hsieh},
  {Komiyama}, {Lanusse}, {Leauthaud}, {Lupton}, {Medezinski}, {Mineo},
  {Miyama}, {Miyazaki}, {Murata}, {Murayama}, {Shirasaki}, {Sif{\'o}n},
  {Simet}, {Speagle}, {Spergel}, {Strauss}, {Sugiyama}, {Tanaka}, {Utsumi},
  {Wang}, \& {Yamada}}]{HSC1-cs-fourier}
---. 2019{\natexlab{b}}, \pasj, 71, 43

\bibitem[{{Hildebrandt} {et~al.}(2017{\natexlab{a}}){Hildebrandt}, {Viola},
  {Heymans}, {Joudaki}, {Kuijken}, {Blake}, {Erben}, {Joachimi}, {Klaes},
  {Miller}, {Morrison}, {Nakajima}, {Verdoes Kleijn}, {Amon}, {Choi}, {Covone},
  {de Jong}, {Dvornik}, {Fenech Conti}, {Grado}, {Harnois-D{\'e}raps},
  {Herbonnet}, {Hoekstra}, {K{\"o}hlinger}, {McFarland}, {Mead}, {Merten},
  {Napolitano}, {Peacock}, {Radovich}, {Schneider}, {Simon}, {Valentijn}, {van
  den Busch}, {van Uitert}, \& {Van Waerbeke}}]{2017MNRAS.465.1454H}
{Hildebrandt}, H., {Viola}, M., {Heymans}, C., {et~al.} 2017{\natexlab{a}},
  \mnras, 465, 1454

\bibitem[{{Hildebrandt} {et~al.}(2017{\natexlab{b}}){Hildebrandt}, {Viola},
  {Heymans}, {Joudaki}, {Kuijken}, {Blake}, {Erben}, {Joachimi}, {Klaes},
  {Miller}, {Morrison}, {Nakajima}, {Verdoes Kleijn}, {Amon}, {Choi}, {Covone},
  {de Jong}, {Dvornik}, {Fenech Conti}, {Grado}, {Harnois-D{\'e}raps},
  {Herbonnet}, {Hoekstra}, {K{\"o}hlinger}, {McFarland}, {Mead}, {Merten},
  {Napolitano}, {Peacock}, {Radovich}, {Schneider}, {Simon}, {Valentijn}, {van
  den Busch}, {van Uitert}, \& {Van Waerbeke}}]{KiDS450-CS}
---. 2017{\natexlab{b}}, \mnras, 465, 1454

\bibitem[{{Hildebrandt} {et~al.}(2017{\natexlab{c}}){Hildebrandt}, {Viola},
  {Heymans}, {Joudaki}, {Kuijken}, {Blake}, {Erben}, {Joachimi}, {Klaes},
  {Miller}, {Morrison}, {Nakajima}, {Verdoes Kleijn}, {Amon}, {Choi}, {Covone},
  {de Jong}, {Dvornik}, {Fenech Conti}, {Grado}, {Harnois-D{\'e}raps},
  {Herbonnet}, {Hoekstra}, {K{\"o}hlinger}, {McFarland}, {Mead}, {Merten},
  {Napolitano}, {Peacock}, {Radovich}, {Schneider}, {Simon}, {Valentijn}, {van
  den Busch}, {van Uitert}, \& {Van Waerbeke}}]{cosmicShear-KiDS2017}
---. 2017{\natexlab{c}}, \mnras, 465, 1454

\bibitem[{{Hirata} \& {Seljak}(2003)}]{Regaussianization}
{Hirata}, C., \& {Seljak}, U. 2003, \mnras, 343, 459

\bibitem[{{Hsieh} \& {Yee}(2014)}]{DEmPZ-Hsieh2014}
{Hsieh}, B.~C., \& {Yee}, H.~K.~C. 2014, \apj, 792, 102

\bibitem[{{Huff} \& {Mandelbaum}(2017)}]{metacal-Huff2017}
{Huff}, E., \& {Mandelbaum}, R. 2017, ArXiv e-prints, arXiv:1702.02600

\bibitem[{{Ilbert} {et~al.}(2009){Ilbert}, {Capak}, {Salvato}, {Aussel},
  {McCracken}, {Sanders}, {Scoville}, {Kartaltepe}, {Arnouts}, {Le Floc'h},
  {Mobasher}, {Taniguchi}, {Lamareille}, {Leauthaud}, {Sasaki}, {Thompson},
  {Zamojski}, {Zamorani}, {Bardelli}, {Bolzonella}, {Bongiorno}, {Brusa},
  {Caputi}, {Carollo}, {Contini}, {Cook}, {Coppa}, {Cucciati}, {de la Torre},
  {de Ravel}, {Franzetti}, {Garilli}, {Hasinger}, {Iovino}, {Kampczyk},
  {Kneib}, {Knobel}, {Kovac}, {Le Borgne}, {Le Brun}, {Le F{\`e}vre}, {Lilly},
  {Looper}, {Maier}, {Mainieri}, {Mellier}, {Mignoli}, {Murayama}, {Pell{\`o}},
  {Peng}, {P{\'e}rez-Montero}, {Renzini}, {Ricciardelli}, {Schiminovich},
  {Scodeggio}, {Shioya}, {Silverman}, {Surace}, {Tanaka}, {Tasca}, {Tresse},
  {Vergani}, \& {Zucca}}]{COSMOSZ-Ilbert2009}
{Ilbert}, O., {Capak}, P., {Salvato}, M., {et~al.} 2009, \apj, 690, 1236

\bibitem[{{Ivezi{\'c}} {et~al.}(2019){Ivezi{\'c}}, {Kahn}, {Tyson}, {Abel},
  {Acosta}, {Allsman}, {Alonso}, {AlSayyad}, {Anderson}, {Andrew}, {Angel},
  {Angeli}, {Ansari}, {Antilogus}, {Araujo}, {Armstrong}, {Arndt}, {Astier},
  {Aubourg}, {Auza}, {Axelrod}, {Bard}, {Barr}, {Barrau}, {Bartlett}, {Bauer},
  {Bauman}, {Baumont}, {Bechtol}, {Bechtol}, {Becker}, {Becla}, {Beldica},
  {Bellavia}, {Bianco}, {Biswas}, {Blanc}, {Blazek}, {Blandford}, {Bloom},
  {Bogart}, {Bond}, {Booth}, {Borgland}, {Borne}, {Bosch}, {Boutigny},
  {Brackett}, {Bradshaw}, {Brandt}, {Brown}, {Bullock}, {Burchat}, {Burke},
  {Cagnoli}, {Calabrese}, {Callahan}, {Callen}, {Carlin}, {Carlson},
  {Chandrasekharan}, {Charles-Emerson}, {Chesley}, {Cheu}, {Chiang}, {Chiang},
  {Chirino}, {Chow}, {Ciardi}, {Claver}, {Cohen-Tanugi}, {Cockrum}, {Coles},
  {Connolly}, {Cook}, {Cooray}, {Covey}, {Cribbs}, {Cui}, {Cutri}, {Daly},
  {Daniel}, {Daruich}, {Daubard}, {Daues}, {Dawson}, {Delgado}, {Dellapenna},
  {de Peyster}, {de Val-Borro}, {Digel}, {Doherty}, {Dubois},
  {Dubois-Felsmann}, {Durech}, {Economou}, {Eifler}, {Eracleous}, {Emmons},
  {Fausti Neto}, {Ferguson}, {Figueroa}, {Fisher-Levine}, {Focke}, {Foss},
  {Frank}, {Freemon}, {Gangler}, {Gawiser}, {Geary}, {Gee}, {Geha}, {Gessner},
  {Gibson}, {Gilmore}, {Glanzman}, {Glick}, {Goldina}, {Goldstein}, {Goodenow},
  {Graham}, {Gressler}, {Gris}, {Guy}, {Guyonnet}, {Haller}, {Harris},
  {Hascall}, {Haupt}, {Hernandez}, {Herrmann}, {Hileman}, {Hoblitt}, {Hodgson},
  {Hogan}, {Howard}, {Huang}, {Huffer}, {Ingraham}, {Innes}, {Jacoby}, {Jain},
  {Jammes}, {Jee}, {Jenness}, {Jernigan}, {Jevremovi{\'c}}, {Johns}, {Johnson},
  {Johnson}, {Jones}, {Juramy-Gilles}, {Juri{\'c}}, {Kalirai}, {Kallivayalil},
  {Kalmbach}, {Kantor}, {Karst}, {Kasliwal}, {Kelly}, {Kessler}, {Kinnison},
  {Kirkby}, {Knox}, {Kotov}, {Krabbendam}, {Krughoff}, {Kub{\'a}nek},
  {Kuczewski}, {Kulkarni}, {Ku}, {Kurita}, {Lage}, {Lambert}, {Lange},
  {Langton}, {Le Guillou}, {Levine}, {Liang}, {Lim}, {Lintott}, {Long},
  {Lopez}, {Lotz}, {Lupton}, {Lust}, {MacArthur}, {Mahabal}, {Mandelbaum},
  {Markiewicz}, {Marsh}, {Marshall}, {Marshall}, {May}, {McKercher}, {McQueen},
  {Meyers}, {Migliore}, {Miller}, {Mills}, {Miraval}, {Moeyens}, {Moolekamp},
  {Monet}, {Moniez}, {Monkewitz}, {Montgomery}, {Morrison}, {Mueller},
  {Muller}, {Mu{\~n}oz Arancibia}, {Neill}, {Newbry}, {Nief}, {Nomerotski},
  {Nordby}, {O'Connor}, {Oliver}, {Olivier}, {Olsen}, {O'Mullane}, {Ortiz},
  {Osier}, {Owen}, {Pain}, {Palecek}, {Parejko}, {Parsons}, {Pease},
  {Peterson}, {Peterson}, {Petravick}, {Libby Petrick}, {Petry},
  {Pierfederici}, {Pietrowicz}, {Pike}, {Pinto}, {Plante}, {Plate}, {Plutchak},
  {Price}, {Prouza}, {Radeka}, {Rajagopal}, {Rasmussen}, {Regnault}, {Reil},
  {Reiss}, {Reuter}, {Ridgway}, {Riot}, {Ritz}, {Robinson}, {Roby}, {Roodman},
  {Rosing}, {Roucelle}, {Rumore}, {Russo}, {Saha}, {Sassolas}, {Schalk},
  {Schellart}, {Schindler}, {Schmidt}, {Schneider}, {Schneider}, {Schoening},
  {Schumacher}, {Schwamb}, {Sebag}, {Selvy}, {Sembroski}, {Seppala}, {Serio},
  {Serrano}, {Shaw}, {Shipsey}, {Sick}, {Silvestri}, {Slater}, {Smith},
  {Smith}, {Sobhani}, {Soldahl}, {Storrie-Lombardi}, {Stover}, {Strauss},
  {Street}, {Stubbs}, {Sullivan}, {Sweeney}, {Swinbank}, {Szalay}, {Takacs},
  {Tether}, {Thaler}, {Thayer}, {Thomas}, {Thornton}, {Thukral}, {Tice},
  {Trilling}, {Turri}, {Van Berg}, {Vanden Berk}, {Vetter}, {Virieux},
  {Vucina}, {Wahl}, {Walkowicz}, {Walsh}, {Walter}, {Wang}, {Wang}, {Warner},
  {Wiecha}, {Willman}, {Winters}, {Wittman}, {Wolff}, {Wood-Vasey}, {Wu},
  {Xin}, {Yoachim}, \& {Zhan}}]{LSSTOverviwe2019}
{Ivezi{\'c}}, {\v{Z}}., {Kahn}, S.~M., {Tyson}, J.~A., {et~al.} 2019, \apj,
  873, 111

\bibitem[{{Jarvis} {et~al.}(2004){Jarvis}, {Bernstein}, \&
  {Jain}}]{TreeCorr-Jarvis2004}
{Jarvis}, M., {Bernstein}, G., \& {Jain}, B. 2004, \mnras, 352, 338

\bibitem[{{Jarvis} {et~al.}(2016){Jarvis}, {Sheldon}, {Zuntz}, {Kacprzak},
  {Bridle}, {Amara}, {Armstrong}, {Becker}, {Bernstein}, {Bonnett}, {Chang},
  {Das}, {Dietrich}, {Drlica-Wagner}, {Eifler}, {Gangkofner}, {Gruen},
  {Hirsch}, {Huff}, {Jain}, {Kent}, {Kirk}, {MacCrann}, {Melchior}, {Plazas},
  {Refregier}, {Rowe}, {Rykoff}, {Samuroff}, {S{\'a}nchez}, {Suchyta},
  {Troxel}, {Vikram}, {Abbott}, {Abdalla}, {Allam}, {Annis}, {Benoit-L{\'e}vy},
  {Bertin}, {Brooks}, {Buckley-Geer}, {Burke}, {Capozzi}, {Carnero Rosell},
  {Carrasco Kind}, {Carretero}, {Castander}, {Clampitt}, {Crocce}, {Cunha},
  {D'Andrea}, {da Costa}, {DePoy}, {Desai}, {Diehl}, {Doel}, {Fausti Neto},
  {Flaugher}, {Fosalba}, {Frieman}, {Gaztanaga}, {Gerdes}, {Gruendl},
  {Gutierrez}, {Honscheid}, {James}, {Kuehn}, {Kuropatkin}, {Lahav}, {Li},
  {Lima}, {March}, {Martini}, {Miquel}, {Mohr}, {Neilsen}, {Nord}, {Ogando},
  {Reil}, {Romer}, {Roodman}, {Sako}, {Sanchez}, {Scarpine}, {Schubnell},
  {Sevilla-Noarbe}, {Smith}, {Soares-Santos}, {Sobreira}, {Swanson}, {Tarle},
  {Thaler}, {Thomas}, {Walker}, \& {Wechsler}}]{Jarvis2016MNRAS}
{Jarvis}, M., {Sheldon}, E., {Zuntz}, J., {et~al.} 2016, \mnras, 460, 2245

\bibitem[{{Jarvis} {et~al.}(2021){Jarvis}, {Bernstein}, {Amon}, {Davis},
  {L{\'e}get}, {Bechtol}, {Harrison}, {Gatti}, {Roodman}, {Chang}, {Chen},
  {Choi}, {Desai}, {Drlica-Wagner}, {Gruen}, {Gruendl}, {Hernandez},
  {MacCrann}, {Meyers}, {Navarro-Alsina}, {Pandey}, {Plazas}, {Secco},
  {Sheldon}, {Troxel}, {Vorperian}, {Wei}, {Zuntz}, {Abbott}, {Aguena},
  {Allam}, {Avila}, {Bhargava}, {Bridle}, {Brooks}, {Carnero Rosell}, {Carrasco
  Kind}, {Carretero}, {Costanzi}, {da Costa}, {De Vicente}, {Diehl}, {Doel},
  {Everett}, {Flaugher}, {Fosalba}, {Frieman}, {Garc{\'\i}a-Bellido},
  {Gaztanaga}, {Gerdes}, {Gutierrez}, {Hinton}, {Hollowood}, {Honscheid},
  {James}, {Kent}, {Kuehn}, {Kuropatkin}, {Lahav}, {Maia}, {March}, {Marshall},
  {Melchior}, {Menanteau}, {Miquel}, {Ogando}, {Paz-Chinch{\'o}n}, {Rykoff},
  {Sanchez}, {Scarpine}, {Schubnell}, {Serrano}, {Sevilla-Noarbe}, {Smith},
  {Suchyta}, {Swanson}, {Tarle}, {Varga}, {Walker}, {Wester}, {Wilkinson}, \&
  {DES Collaboration}}]{Jarvis2021}
{Jarvis}, M., {Bernstein}, G.~M., {Amon}, A., {et~al.} 2021, \mnras, 501, 1282

\bibitem[{{Kaiser} \& {Squires}(1993)}]{KS93}
{Kaiser}, N., \& {Squires}, G. 1993, \apj, 404, 441

\bibitem[{{Kaiser} {et~al.}(1995){Kaiser}, {Squires}, \&
  {Broadhurst}}]{KSB-Kaiser1995}
{Kaiser}, N., {Squires}, G., \& {Broadhurst}, T. 1995, \apj, 449, 460

\bibitem[{{Kannawadi} {et~al.}(2019){Kannawadi}, {Hoekstra}, {Miller}, {Viola},
  {Fenech Conti}, {Herbonnet}, {Erben}, {Heymans}, {Hildebrandt}, {Kuijken},
  {Vakili}, \& {Wright}}]{KIDS-ImgSim-Kannawadi2019}
{Kannawadi}, A., {Hoekstra}, H., {Miller}, L., {et~al.} 2019, \aap, 624, A92

\bibitem[{{Koekemoer} {et~al.}(2007){Koekemoer}, {Aussel}, {Calzetti}, {Capak},
  {Giavalisco}, {Kneib}, {Leauthaud}, {Le F{\`e}vre}, {McCracken}, {Massey},
  {Mobasher}, {Rhodes}, {Scoville}, \& {Shopbell}}]{HST-ACSpipe}
{Koekemoer}, A.~M., {Aussel}, H., {Calzetti}, D., {et~al.} 2007, pjs, 172, 196

\bibitem[{{Laureijs} {et~al.}(2011){Laureijs}, {Amiaux}, {Arduini},
  {Augu{\`e}res}, {Brinchmann}, {Cole}, {Cropper}, {Dabin}, {Duvet}, {Ealet},
  \& et~al.}]{Euclid2011}
{Laureijs}, R., {Amiaux}, J., {Arduini}, S., {et~al.} 2011, ArXiv e-prints,
  arXiv:1110.3193

\bibitem[{{Leauthaud} {et~al.}(2007){Leauthaud}, {Massey}, {Kneib}, {Rhodes},
  {Johnston}, {Capak}, {Heymans}, {Ellis}, {Koekemoer}, {Le F{\`e}vre},
  {Mellier}, {R{\'e}fr{\'e}gier}, {Robin}, {Scoville}, {Tasca}, {Taylor}, \&
  {Van Waerbeke}}]{HST-shapeCatalog-Alexie2007}
{Leauthaud}, A., {Massey}, R., {Kneib}, J.-P., {et~al.} 2007, \apjs, 172, 219

\bibitem[{{Li} \& {Zhang}(2020)}]{FQ-Li2021}
{Li}, H., \& {Zhang}, J. 2020, arXiv e-prints, arXiv:2012.10899

\bibitem[{{Li} {et~al.}(2018){Li}, {Katayama}, {Oguri}, \&
  {More}}]{FPFS-Li2018}
{Li}, X., {Katayama}, N., {Oguri}, M., \& {More}, S. 2018, \mnras, 481, 4445

\bibitem[{{Li} {et~al.}(2021){Li}, {Li}, \& {Massey}}]{FPFS-Li2021}
{Li}, X., {Li}, Y., \& {Massey}, R. 2021, arXiv e-prints, arXiv:2110.01214

\bibitem[{Li {et~al.}(2020)Li, Oguri, Katayama, Luo, Wang, Han, Miyatake,
  Nakamura, \& More}]{FPFSHSC1-Li2020}
Li, X., Oguri, M., Katayama, N., {et~al.} 2020, The Astrophysical Journal
  Supplement Series, 251, 19

\bibitem[{{Lu} {et~al.}(2017){Lu}, {Zhang}, {Dong}, {Li}, {Liu}, {Fu}, {Li}, \&
  {Fan}}]{LuPSF17}
{Lu}, T., {Zhang}, J., {Dong}, F., {et~al.} 2017, \aj, 153, 197

\bibitem[{{Lupton} {et~al.}(2001){Lupton}, {Gunn}, {Ivezi{\'c}}, {Knapp}, \&
  {Kent}}]{SDSS-pipeline-Lupton2001}
{Lupton}, R., {Gunn}, J.~E., {Ivezi{\'c}}, Z., {Knapp}, G.~R., \& {Kent}, S.
  2001, in Astronomical Society of the Pacific Conference Series, Vol. 238,
  Astronomical Data Analysis Software and Systems X, ed. F.~R. {Harnden}, Jr.,
  F.~A. {Primini}, \& H.~E. {Payne}, 269

\bibitem[{{MacCrann} {et~al.}(2020){MacCrann}, {Becker}, {McCullough}, {Amon},
  {Gruen}, {Jarvis}, {Choi}, {Troxel}, {Sheldon}, {Yanny}, {Herner},
  {Dodelson}, {Zuntz}, {Eckert}, {Rollins}, {Varga}, {Bernstein}, {Gruendl},
  {Harrison}, {Hartley}, {Sevilla-Noarbe}, {Pieres}, {Bridle}, {Myles},
  {Alarcon}, {Everett}, {S{\'a}nchez}, {Huff}, {Tarsitano}, {Gatti}, {Secco},
  {Abbott}, {Aguena}, {Allam}, {Annis}, {Bacon}, {Bertin}, {Brooks}, {Burke},
  {Carnero Rosell}, {Carrasco Kind}, {Carretero}, {Costanzi}, {Crocce},
  {Pereira}, {De Vicente}, {Desai}, {Diehl}, {Dietrich}, {Doel}, {Eifler},
  {Ferrero}, {Fert{\'e}}, {Flaugher}, {Fosalba}, {Frieman},
  {Garc{\'\i}a-Bellido}, {Gaztanaga}, {Gerdes}, {Giannantonio}, {Gschwend},
  {Gutierrez}, {Hinton}, {Hollowood}, {Honscheid}, {James}, {Lahav}, {Lima},
  {Maia}, {March}, {Marshall}, {Martini}, {Melchior}, {Menanteau}, {Miquel},
  {Mohr}, {Morgan}, {Muir}, {Ogando}, {Palmese}, {Paz-Chinch{\'o}n}, {Plazas},
  {Rodriguez-Monroy}, {Roodman}, {Samuroff}, {Sanchez}, {Scarpine}, {Serrano},
  {Smith}, {Soares-Santos}, {Suchyta}, {Swanson}, {Tarle}, {Thomas}, {To}, \&
  {Wilkinson}}]{DESY3-BlendshearCalib-MacCrann2021}
{MacCrann}, N., {Becker}, M.~R., {McCullough}, J., {et~al.} 2020, arXiv
  e-prints, arXiv:2012.08567

\bibitem[{{Mandelbaum}(2018)}]{revRachel17}
{Mandelbaum}, R. 2018, \araa, 56, 393

\bibitem[{{Mandelbaum} {et~al.}(2013){Mandelbaum}, {Slosar}, {Baldauf},
  {Seljak}, {Hirata}, {Nakajima}, {Reyes}, \&
  {Smith}}]{SDSS-gglensDR7-Mandelbaum2013}
{Mandelbaum}, R., {Slosar}, A., {Baldauf}, T., {et~al.} 2013, \mnras, 432, 1544

\bibitem[{{Mandelbaum} {et~al.}(2005){Mandelbaum}, {Hirata}, {Seljak}, {Guzik},
  {Padmanabhan}, {Blake}, {Blanton}, {Lupton}, \&
  {Brinkmann}}]{SDSS-shape-Mandelbaum2005}
{Mandelbaum}, R., {Hirata}, C.~M., {Seljak}, U., {et~al.} 2005, \mnras, 361,
  1287

\bibitem[{{Mandelbaum} {et~al.}(2018{\natexlab{a}}){Mandelbaum}, {Miyatake},
  {Hamana}, {Oguri}, {Simet}, {Armstrong}, {Bosch}, {Murata}, {Lanusse},
  {Leauthaud}, {Coupon}, {More}, {Takada}, {Miyazaki}, {Speagle}, {Shirasaki},
  {Sif{\'o}n}, {Huang}, {Nishizawa}, {Medezinski}, {Okura}, {Okabe}, {Czakon},
  {Takahashi}, {Coulton}, {Hikage}, {Komiyama}, {Lupton}, {Strauss}, {Tanaka},
  \& {Utsumi}}]{HSC1-shape}
{Mandelbaum}, R., {Miyatake}, H., {Hamana}, T., {et~al.} 2018{\natexlab{a}},
  \pasj, 70, S25

\bibitem[{{Mandelbaum} {et~al.}(2018{\natexlab{b}}){Mandelbaum}, {Lanusse},
  {Leauthaud}, {Armstrong}, {Simet}, {Miyatake}, {Meyers}, {Bosch}, {Murata},
  \& {Miyazaki}}]{HSC1-GREAT3Sim}
{Mandelbaum}, R., {Lanusse}, F., {Leauthaud}, A., {et~al.} 2018{\natexlab{b}},
  \mnras, 481, 3170

\bibitem[{{Massey} {et~al.}(2007{\natexlab{a}}){Massey}, {Rhodes}, {Ellis},
  {Scoville}, {Leauthaud}, {Finoguenov}, {Capak}, {Bacon}, {Aussel}, {Kneib},
  {Koekemoer}, {McCracken}, {Mobasher}, {Pires}, {Refregier}, {Sasaki},
  {Starck}, {Taniguchi}, {Taylor}, \& {Taylor}}]{Massey2007}
{Massey}, R., {Rhodes}, J., {Ellis}, R., {et~al.} 2007{\natexlab{a}}, \nat,
  445, 286

\bibitem[{{Massey} {et~al.}(2007{\natexlab{b}}){Massey}, {Heymans},
  {Berg{\'e}}, {Bernstein}, {Bridle}, {Clowe}, {Dahle}, {Ellis}, {Erben},
  {Hetterscheidt}, {High}, {Hirata}, {Hoekstra}, {Hudelot}, {Jarvis},
  {Johnston}, {Kuijken}, {Margoniner}, {Mandelbaum}, {Mellier}, {Nakajima},
  {Paulin-Henriksson}, {Peeples}, {Roat}, {Refregier}, {Rhodes}, {Schrabback},
  {Schirmer}, {Seljak}, {Semboloni}, \& {van Waerbeke}}]{galsim-STEP2}
{Massey}, R., {Heymans}, C., {Berg{\'e}}, J., {et~al.} 2007{\natexlab{b}},
  \mnras, 376, 13

\bibitem[{{Miyatake} {et~al.}(2019){Miyatake}, {Battaglia}, {Hilton},
  {Medezinski}, {Nishizawa}, {More}, {Aiola}, {Bahcall}, {Bond}, \&
  {Calabrese}}]{massCalib-actpolCluster-Miyatake2019}
{Miyatake}, H., {Battaglia}, N., {Hilton}, M., {et~al.} 2019, \apj, 875, 63

\bibitem[{{Miyatake} {et~al.}(2021){Miyatake}, {Sugiyama}, {Takada},
  {Nishimichi}, {Shirasaki}, {Kobayashi}, {Mandelbaum}, {More}, {Oguri},
  {Osato}, {Park}, {Takahashi}, {Coupon}, {Hikage}, {Hsieh}, {Leauthaud}, {Li},
  {Luo}, {Lupton}, {Miyazaki}, {Murayama}, {Nishizawa}, {Price}, {Simet},
  {Speagle}, {Strauss}, {Tanaka}, \& {Yoshida}}]{HSC1-ggLensClustII}
{Miyatake}, H., {Sugiyama}, S., {Takada}, M., {et~al.} 2021, arXiv e-prints,
  arXiv:2111.02419

\bibitem[{{Miyazaki} {et~al.}(2018{\natexlab{a}}){Miyazaki}, {Oguri}, {Hamana},
  {Shirasaki}, {Koike}, {Komiyama}, {Umetsu}, {Utsumi}, {Okabe}, {More},
  {Medezinski}, {Lin}, {Miyatake}, {Murayama}, {Ota}, \&
  {Mitsuishi}}]{2018PASJ...70S..27M}
{Miyazaki}, S., {Oguri}, M., {Hamana}, T., {et~al.} 2018{\natexlab{a}}, \pasj,
  70, S27

\bibitem[{{Miyazaki} {et~al.}(2018{\natexlab{b}}){Miyazaki}, {Komiyama},
  {Kawanomoto}, {Doi}, {Furusawa}, {Hamana}, {Hayashi}, {Ikeda}, {Kamata},
  {Karoji}, {Koike}, {Kurakami}, {Miyama}, {Morokuma}, {Nakata}, {Namikawa},
  {Nakaya}, {Nariai}, {Obuchi}, {Oishi}, {Okada}, {Okura}, {Tait}, {Takata},
  {Tanaka}, {Tanaka}, {Terai}, {Tomono}, {Uraguchi}, {Usuda}, {Utsumi},
  {Yamada}, {Yamanoi}, {Aihara}, {Fujimori}, {Mineo}, {Miyatake}, {Oguri},
  {Uchida}, {Tanaka}, {Yasuda}, {Takada}, {Murayama}, {Nishizawa}, {Sugiyama},
  {Chiba}, {Futamase}, {Wang}, {Chen}, {Ho}, {Liaw}, {Chiu}, {Ho}, {Lai},
  {Lee}, {Jeng}, {Iwamura}, {Armstrong}, {Bickerton}, {Bosch}, {Gunn},
  {Lupton}, {Loomis}, {Price}, {Smith}, {Strauss}, {Turner}, {Suzuki},
  {Miyazaki}, {Muramatsu}, {Yamamoto}, {Endo}, {Ezaki}, {Ito}, {Kawaguchi},
  {Sofuku}, {Taniike}, {Akutsu}, {Dojo}, {Kasumi}, {Matsuda}, {Imoto}, {Miwa},
  {Suzuki}, {Takeshi}, \& {Yokota}}]{2018PASJ...70S...1M}
{Miyazaki}, S., {Komiyama}, Y., {Kawanomoto}, S., {et~al.} 2018{\natexlab{b}},
  \pasj, 70, S1

\bibitem[{{Miyazaki} {et~al.}(2018{\natexlab{c}}){Miyazaki}, {Komiyama},
  {Kawanomoto}, {Doi}, {Furusawa}, {Hamana}, {Hayashi}, {Ikeda}, {Kamata},
  {Karoji}, {Koike}, {Kurakami}, {Miyama}, {Morokuma}, {Nakata}, {Namikawa},
  {Nakaya}, {Nariai}, {Obuchi}, {Oishi}, {Okada}, {Okura}, {Tait}, {Takata},
  {Tanaka}, {Tanaka}, {Terai}, {Tomono}, {Uraguchi}, {Usuda}, {Utsumi},
  {Yamada}, {Yamanoi}, {Aihara}, {Fujimori}, {Mineo}, {Miyatake}, {Oguri},
  {Uchida}, {Tanaka}, {Yasuda}, {Takada}, {Murayama}, {Nishizawa}, {Sugiyama},
  {Chiba}, {Futamase}, {Wang}, {Chen}, {Ho}, {Liaw}, {Chiu}, {Ho}, {Lai},
  {Lee}, {Jeng}, {Iwamura}, {Armstrong}, {Bickerton}, {Bosch}, {Gunn},
  {Lupton}, {Loomis}, {Price}, {Smith}, {Strauss}, {Turner}, {Suzuki},
  {Miyazaki}, {Muramatsu}, {Yamamoto}, {Endo}, {Ezaki}, {Ito}, {Kawaguchi},
  {Sofuku}, {Taniike}, {Akutsu}, {Dojo}, {Kasumi}, {Matsuda}, {Imoto}, {Miwa},
  {Suzuki}, {Takeshi}, \& {Yokota}}]{HSC-hardware-Miyazaki2018}
---. 2018{\natexlab{c}}, \pasj, 70, S1

\bibitem[{{More} {et~al.}(2015){More}, {Miyatake}, {Mandelbaum}, {Takada},
  {Spergel}, {Brownstein}, \& {Schneider}}]{2015ApJ...806....2M}
{More}, S., {Miyatake}, H., {Mandelbaum}, R., {et~al.} 2015, \apj, 806, 2

\bibitem[{Murata {et~al.}(2019)Murata, Oguri, Nishimichi, Takada, Mandelbaum,
  More, Shirasaki, Nishizawa, \& Osato}]{massCalib-CAMIRA-Murata2019}
Murata, R., Oguri, M., Nishimichi, T., {et~al.} 2019, Publications of the
  Astronomical Society of Japan, 71,
  https://academic.oup.com/pasj/article-pdf/71/5/107/30161534/psz092.pdf, 107

\bibitem[{{Nishizawa} {et~al.}(2020){Nishizawa}, {Hsieh}, {Tanaka}, \&
  {Takata}}]{HSC2-photoz}
{Nishizawa}, A.~J., {Hsieh}, B.-C., {Tanaka}, M., \& {Takata}, T. 2020, arXiv
  e-prints, arXiv:2003.01511

\bibitem[{{Nishizawa} {et~al.}({in prep.})}]{dnnz-Nishizawa2021}
{Nishizawa}, T., {et~al.} {in prep.}

\bibitem[{{Oguri} {et~al.}(2018){Oguri}, {Miyazaki}, {Hikage}, {Mandelbaum},
  {Utsumi}, {Miyatake}, {Takada}, {Armstrong}, {Bosch}, {Komiyama},
  {Leauthaud}, {More}, {Nishizawa}, {Okabe}, \& {Tanaka}}]{HSC1-wlmap}
{Oguri}, M., {Miyazaki}, S., {Hikage}, C., {et~al.} 2018, \pasj, 70, S26

\bibitem[{{Percival} {et~al.}(2021){Percival}, {Friedrich}, {Sellentin}, \&
  {Heavens}}]{cov-Percival2021}
{Percival}, W.~J., {Friedrich}, O., {Sellentin}, E., \& {Heavens}, A. 2021,
  arXiv e-prints, arXiv:2108.10402

\bibitem[{{Planck Collaboration} {et~al.}(2020){Planck Collaboration},
  {Aghanim}, {Akrami}, {Ashdown}, {Aumont}, {Baccigalupi}, {Ballardini},
  {Banday}, {Barreiro}, {Bartolo}, {Basak}, {Battye}, {Benabed}, {Bernard},
  {Bersanelli}, {Bielewicz}, {Bock}, {Bond}, {Borrill}, {Bouchet}, {Boulanger},
  {Bucher}, {Burigana}, {Butler}, {Calabrese}, {Cardoso}, {Carron},
  {Challinor}, {Chiang}, {Chluba}, {Colombo}, {Combet}, {Contreras}, {Crill},
  {Cuttaia}, {de Bernardis}, {de Zotti}, {Delabrouille}, {Delouis}, {Di
  Valentino}, {Diego}, {Dor{\'e}}, {Douspis}, {Ducout}, {Dupac}, {Dusini},
  {Efstathiou}, {Elsner}, {En{\ss}lin}, {Eriksen}, {Fantaye}, {Farhang},
  {Fergusson}, {Fernandez-Cobos}, {Finelli}, {Forastieri}, {Frailis},
  {Fraisse}, {Franceschi}, {Frolov}, {Galeotta}, {Galli}, {Ganga},
  {G{\'e}nova-Santos}, {Gerbino}, {Ghosh}, {Gonz{\'a}lez-Nuevo}, {G{\'o}rski},
  {Gratton}, {Gruppuso}, {Gudmundsson}, {Hamann}, {Handley}, {Hansen},
  {Herranz}, {Hildebrandt}, {Hivon}, {Huang}, {Jaffe}, {Jones}, {Karakci},
  {Keih{\"a}nen}, {Keskitalo}, {Kiiveri}, {Kim}, {Kisner}, {Knox},
  {Krachmalnicoff}, {Kunz}, {Kurki-Suonio}, {Lagache}, {Lamarre}, {Lasenby},
  {Lattanzi}, {Lawrence}, {Le Jeune}, {Lemos}, {Lesgourgues}, {Levrier},
  {Lewis}, {Liguori}, {Lilje}, {Lilley}, {Lindholm}, {L{\'o}pez-Caniego},
  {Lubin}, {Ma}, {Mac{\'\i}as-P{\'e}rez}, {Maggio}, {Maino}, {Mandolesi},
  {Mangilli}, {Marcos-Caballero}, {Maris}, {Martin}, {Martinelli},
  {Mart{\'\i}nez-Gonz{\'a}lez}, {Matarrese}, {Mauri}, {McEwen}, {Meinhold},
  {Melchiorri}, {Mennella}, {Migliaccio}, {Millea}, {Mitra},
  {Miville-Desch{\^e}nes}, {Molinari}, {Montier}, {Morgante}, {Moss}, {Natoli},
  {N{\o}rgaard-Nielsen}, {Pagano}, {Paoletti}, {Partridge}, {Patanchon},
  {Peiris}, {Perrotta}, {Pettorino}, {Piacentini}, {Polastri}, {Polenta},
  {Puget}, {Rachen}, {Reinecke}, {Remazeilles}, {Renzi}, {Rocha}, {Rosset},
  {Roudier}, {Rubi{\~n}o-Mart{\'\i}n}, {Ruiz-Granados}, {Salvati}, {Sandri},
  {Savelainen}, {Scott}, {Shellard}, {Sirignano}, {Sirri}, {Spencer},
  {Sunyaev}, {Suur-Uski}, {Tauber}, {Tavagnacco}, {Tenti}, {Toffolatti},
  {Tomasi}, {Trombetti}, {Valenziano}, {Valiviita}, {Van Tent}, {Vibert},
  {Vielva}, {Villa}, {Vittorio}, {Wandelt}, {Wehus}, {White}, {White},
  {Zacchei}, \& {Zonca}}]{2020A&A...641A...6P}
{Planck Collaboration}, {Aghanim}, N., {Akrami}, Y., {et~al.} 2020, \aap, 641,
  A6

\bibitem[{{Plazas} \& {Bernstein}(2012)}]{Plazas2012}
{Plazas}, A.~A., \& {Bernstein}, G. 2012, \pasp, 124, 1113

\bibitem[{{Plazas} {et~al.}(2014){Plazas}, {Bernstein}, \&
  {Sheldon}}]{Plazas2014}
{Plazas}, A.~A., {Bernstein}, G.~M., \& {Sheldon}, E.~S. 2014, Journal of
  Instrumentation, 9, C04001

\bibitem[{{Pujol} {et~al.}(2019){Pujol}, {Kilbinger}, {Sureau}, \&
  {Bobin}}]{preciseSim-Pujol2019}
{Pujol}, A., {Kilbinger}, M., {Sureau}, F., \& {Bobin}, J. 2019, \aap, 621, A2

\bibitem[{{Refregier} {et~al.}(2012){Refregier}, {Kacprzak}, {Amara}, {Bridle},
  \& {Rowe}}]{noiseBiasRefregier2012}
{Refregier}, A., {Kacprzak}, T., {Amara}, A., {Bridle}, S., \& {Rowe}, B. 2012,
  \mnras, 425, 1951

\bibitem[{{Reid} {et~al.}(2016{\natexlab{a}}){Reid}, {Ho}, {Padmanabhan},
  {Percival}, {Tinker}, {Tojeiro}, {White}, {Eisenstein}, {Maraston}, {Ross},
  {S{\'a}nchez}, {Schlegel}, {Sheldon}, {Strauss}, {Thomas}, {Wake}, {Beutler},
  {Bizyaev}, {Bolton}, {Brownstein}, {Chuang}, {Dawson}, {Harding}, {Kitaura},
  {Leauthaud}, {Masters}, {McBride}, {More}, {Olmstead}, {Oravetz}, {Nuza},
  {Pan}, {Parejko}, {Pforr}, {Prada}, {Rodr{\'\i}guez-Torres},
  {Salazar-Albornoz}, {Samushia}, {Schneider}, {Sc{\'o}ccola}, {Simmons}, \&
  {Vargas-Magana}}]{LOWZ_CMASS-Reid2016}
{Reid}, B., {Ho}, S., {Padmanabhan}, N., {et~al.} 2016{\natexlab{a}}, \mnras,
  455, 1553

\bibitem[{{Reid} {et~al.}(2016{\natexlab{b}}){Reid}, {Ho}, {Padmanabhan},
  {Percival}, {Tinker}, {Tojeiro}, {White}, {Eisenstein}, {Maraston}, {Ross},
  {S{\'a}nchez}, {Schlegel}, {Sheldon}, {Strauss}, {Thomas}, {Wake}, {Beutler},
  {Bizyaev}, {Bolton}, {Brownstein}, {Chuang}, {Dawson}, {Harding}, {Kitaura},
  {Leauthaud}, {Masters}, {McBride}, {More}, {Olmstead}, {Oravetz}, {Nuza},
  {Pan}, {Parejko}, {Pforr}, {Prada}, {Rodr{\'\i}guez-Torres},
  {Salazar-Albornoz}, {Samushia}, {Schneider}, {Sc{\'o}ccola}, {Simmons}, \&
  {Vargas-Magana}}]{BOSS-cmass}
---. 2016{\natexlab{b}}, \mnras, 455, 1553

\bibitem[{{Rhodes} {et~al.}(2001){Rhodes}, {Refregier}, \&
  {Groth}}]{2001ApJ...552L..85R}
{Rhodes}, J., {Refregier}, A., \& {Groth}, E.~J. 2001, \apjl, 552, L85

\bibitem[{{Rowe}(2010)}]{Rowe2010MNRAS}
{Rowe}, B. 2010, \mnras, 404, 350

\bibitem[{{Rowe} {et~al.}(2015){Rowe}, {Jarvis}, {Mandelbaum}, {Bernstein},
  {Bosch}, {Simet}, {Meyers}, {Kacprzak}, {Nakajima}, {Zuntz}, {Miyatake},
  {Dietrich}, {Armstrong}, {Melchior}, \& {Gill}}]{Galsim}
{Rowe}, B.~T.~P., {Jarvis}, M., {Mandelbaum}, R., {et~al.} 2015, Astronomy and
  Computing, 10, 121

\bibitem[{{Secco} {et~al.}(2021){Secco}, {Samuroff}, {Krause}, {Jain},
  {Blazek}, {Raveri}, {Campos}, {Amon}, {Chen}, {Doux}, {Choi}, {Gruen},
  {Bernstein}, {Chang}, {DeRose}, {Myles}, {Fert{\'e}}, {Lemos}, {Huterer},
  {Prat}, {Troxel}, {MacCrann}, {Liddle}, {Kacprzak}, {Fang}, {S{\'a}nchez},
  {Pandey}, {Dodelson}, {Chintalapati}, {Hoffmann}, {Alarcon}, {Alves},
  {Andrade-Oliveira}, {Baxter}, {Bechtol}, {Becker}, {Brandao-Souza},
  {Camacho}, {Carnero Rosell}, {Carrasco Kind}, {Cawthon}, {Cordero}, {Crocce},
  {Davis}, {Di Valentino}, {Drlica-Wagner}, {Eckert}, {Eifler}, {Elidaiana},
  {Elsner}, {Elvin-Poole}, {Everett}, {Fosalba}, {Friedrich}, {Gatti},
  {Giannini}, {Gruendl}, {Harrison}, {Hartley}, {Herner}, {Huang}, {Huff},
  {Jarvis}, {Jeffrey}, {Kuropatkin}, {Leget}, {Muir}, {Mccullough}, {Navarro
  Alsina}, {Omori}, {Park}, {Porredon}, {Rollins}, {Roodman}, {Rosenfeld},
  {Ross}, {Rykoff}, {Sanchez}, {Sevilla-Noarbe}, {Sheldon}, {Shin}, {Tutusaus},
  {Varga}, {Weaverdyck}, {Wechsler}, {Yanny}, {Yin}, {Zhang}, {Zuntz},
  {Abbott}, {Aguena}, {Allam}, {Annis}, {Bacon}, {Bertin}, {Bhargava},
  {Bridle}, {Brooks}, {Buckley-Geer}, {Burke}, {Carretero}, {Costanzi}, {da
  Costa}, {De Vicente}, {Diehl}, {Dietrich}, {Doel}, {Ferrero}, {Flaugher},
  {Frieman}, {Garc{\'\i}a-Bellido}, {Gaztanaga}, {Gerdes}, {Giannantonio},
  {Gschwend}, {Gutierrez}, {Hinton}, {Hollowood}, {Honscheid}, {Hoyle},
  {James}, {Jeltema}, {Kuehn}, {Lahav}, {Lima}, {Lin}, {Maia}, {Marshall},
  {Martini}, {Melchior}, {Menanteau}, {Miquel}, {Mohr}, {Morgan}, {Ogando},
  {Palmese}, {Paz-Chinch{\'o}n}, {Petravick}, {Pieres}, {Plazas Malag{\'o}n},
  {Rodriguez-Monroy}, {Romer}, {Sanchez}, {Scarpine}, {Schubnell}, {Scolnic},
  {Serrano}, {Smith}, {Soares-Santos}, {Suchyta}, {Swanson}, {Tarle}, {Thomas},
  \& {To}}]{DESY3-cosmicshearS}
{Secco}, L.~F., {Samuroff}, S., {Krause}, E., {et~al.} 2021, arXiv e-prints,
  arXiv:2105.13544

\bibitem[{{Sheldon} {et~al.}(2020){Sheldon}, {Becker}, {MacCrann}, \&
  {Jarvis}}]{metaDet-Sheldon2020}
{Sheldon}, E.~S., {Becker}, M.~R., {MacCrann}, N., \& {Jarvis}, M. 2020, \apj,
  902, 138

\bibitem[{{Sheldon} \& {Huff}(2017)}]{metacal-Sheldon2017}
{Sheldon}, E.~S., \& {Huff}, E.~M. 2017, \apj, 841, 24

\bibitem[{{Shirasaki} {et~al.}(2019){Shirasaki}, {Hamana}, {Takada},
  {Takahashi}, \& {Miyatake}}]{Shirasakietal:2019}
{Shirasaki}, M., {Hamana}, T., {Takada}, M., {Takahashi}, R., \& {Miyatake}, H.
  2019, \mnras, 486, 52

\bibitem[{{Shirasaki} {et~al.}(2017){Shirasaki}, {Takada}, {Miyatake},
  {Takahashi}, {Hamana}, {Nishimichi}, \&
  {Murata}}]{wl-covariance-Shirasaki2017}
{Shirasaki}, M., {Takada}, M., {Miyatake}, H., {et~al.} 2017, \mnras, 470, 3476

\bibitem[{{Shirasaki} {et~al.}({in prep.})}]{HSC3-mock}
{Shirasaki}, M., {et~al.} {in prep.}

\bibitem[{{Spergel} {et~al.}(2015){Spergel}, {Gehrels}, {Baltay}, {Bennett},
  {Breckinridge}, {Donahue}, {Dressler}, {Gaudi}, {Greene}, {Guyon}, {Hirata},
  {Kalirai}, {Kasdin}, {Macintosh}, {Moos}, {Perlmutter}, {Postman},
  {Rauscher}, {Rhodes}, {Wang}, {Weinberg}, {Benford}, {Hudson}, {Jeong},
  {Mellier}, {Traub}, {Yamada}, {Capak}, {Colbert}, {Masters}, {Penny},
  {Savransky}, {Stern}, {Zimmerman}, {Barry}, {Bartusek}, {Carpenter}, {Cheng},
  {Content}, {Dekens}, {Demers}, {Grady}, {Jackson}, {Kuan}, {Kruk}, {Melton},
  {Nemati}, {Parvin}, {Poberezhskiy}, {Peddie}, {Ruffa}, {Wallace}, {Whipple},
  {Wollack}, \& {Zhao}}]{WFIRST15}
{Spergel}, D., {Gehrels}, N., {Baltay}, C., {et~al.} 2015, ArXiv e-prints,
  arXiv:1503.03757

\bibitem[{{Suzuki} {et~al.}(2012){Suzuki}, {Rubin}, {Lidman}, {Aldering},
  {Amanullah}, {Barbary}, {Barrientos}, {Botyanszki}, {Brodwin}, {Connolly},
  {Dawson}, {Dey}, {Doi}, {Donahue}, {Deustua}, {Eisenhardt}, {Ellingson},
  {Faccioli}, {Fadeyev}, {Fakhouri}, {Fruchter}, {Gilbank}, {Gladders},
  {Goldhaber}, {Gonzalez}, {Goobar}, {Gude}, {Hattori}, {Hoekstra}, {Hsiao},
  {Huang}, {Ihara}, {Jee}, {Johnston}, {Kashikawa}, {Koester}, {Konishi},
  {Kowalski}, {Linder}, {Lubin}, {Melbourne}, {Meyers}, {Morokuma}, {Munshi},
  {Mullis}, {Oda}, {Panagia}, {Perlmutter}, {Postman}, {Pritchard}, {Rhodes},
  {Ripoche}, {Rosati}, {Schlegel}, {Spadafora}, {Stanford}, {Stanishev},
  {Stern}, {Strovink}, {Takanashi}, {Tokita}, {Wagner}, {Wang}, {Yasuda},
  {Yee}, \& {Supernova Cosmology Project}}]{2012ApJ...746...85S}
{Suzuki}, N., {Rubin}, D., {Lidman}, C., {et~al.} 2012, \apj, 746, 85

\bibitem[{{Takahashi} {et~al.}(2017){Takahashi}, {Hamana}, {Shirasaki},
  {Namikawa}, {Nishimichi}, {Osato}, \& {Shiroyama}}]{FullSkySim-Takahashi2017}
{Takahashi}, R., {Hamana}, T., {Shirasaki}, M., {et~al.} 2017, \apj, 850, 24

\bibitem[{{Tanaka}(2015)}]{MizukiZ-Tanaka2015}
{Tanaka}, M. 2015, \apj, 801, 20

\bibitem[{{Troxel} {et~al.}(2018){Troxel}, {MacCrann}, {Zuntz}, {Eifler},
  {Krause}, {Dodelson}, {Gruen}, {Blazek}, {Friedrich}, {Samuroff}, {Prat},
  {Secco}, {Davis}, {Fert{\'e}}, {DeRose}, {Alarcon}, {Amara}, {Baxter},
  {Becker}, {Bernstein}, {Bridle}, {Cawthon}, {Chang}, {Choi}, {De Vicente},
  {Drlica-Wagner}, {Elvin-Poole}, {Frieman}, {Gatti}, {Hartley}, {Honscheid},
  {Hoyle}, {Huff}, {Huterer}, {Jain}, {Jarvis}, {Kacprzak}, {Kirk}, {Kokron},
  {Krawiec}, {Lahav}, {Liddle}, {Peacock}, {Rau}, {Refregier}, {Rollins},
  {Rozo}, {Rykoff}, {S{\'a}nchez}, {Sevilla-Noarbe}, {Sheldon}, {Stebbins},
  {Varga}, {Vielzeuf}, {Wang}, {Wechsler}, {Yanny}, {Abbott}, {Abdalla},
  {Allam}, {Annis}, {Bechtol}, {Benoit-L{\'e}vy}, {Bertin}, {Brooks},
  {Buckley-Geer}, {Burke}, {Carnero Rosell}, {Carrasco Kind}, {Carretero},
  {Castander}, {Crocce}, {Cunha}, {D'Andrea}, {da Costa}, {DePoy}, {Desai},
  {Diehl}, {Dietrich}, {Doel}, {Fernandez}, {Flaugher}, {Fosalba},
  {Garc{\'\i}a-Bellido}, {Gaztanaga}, {Gerdes}, {Giannantonio}, {Goldstein},
  {Gruendl}, {Gschwend}, {Gutierrez}, {James}, {Jeltema}, {Johnson}, {Johnson},
  {Kent}, {Kuehn}, {Kuhlmann}, {Kuropatkin}, {Li}, {Lima}, {Lin}, {Maia},
  {March}, {Marshall}, {Martini}, {Melchior}, {Menanteau}, {Miquel}, {Mohr},
  {Neilsen}, {Nichol}, {Nord}, {Petravick}, {Plazas}, {Romer}, {Roodman},
  {Sako}, {Sanchez}, {Scarpine}, {Schindler}, {Schubnell}, {Smith}, {Smith},
  {Soares-Santos}, {Sobreira}, {Suchyta}, {Swanson}, {Tarle}, {Thomas},
  {Tucker}, {Vikram}, {Walker}, {Weller}, {Zhang}, \& {DES
  Collaboration}}]{2018PhRvD..98d3528T}
{Troxel}, M.~A., {MacCrann}, N., {Zuntz}, J., {et~al.} 2018, \prd, 98, 043528

\bibitem[{{Van Waerbeke} {et~al.}(2000){Van Waerbeke}, {Mellier}, {Erben},
  {Cuillandre}, {Bernardeau}, {Maoli}, {Bertin}, {McCracken}, {Le F{\`e}vre},
  {Fort}, {Dantel-Fort}, {Jain}, \& {Schneider}}]{2000A&A...358...30V}
{Van Waerbeke}, L., {Mellier}, Y., {Erben}, T., {et~al.} 2000, \aap, 358, 30

\bibitem[{{Weinberg} {et~al.}(2013){Weinberg}, {Mortonson}, {Eisenstein},
  {Hirata}, {Riess}, \& {Rozo}}]{2013PhR...530...87W}
{Weinberg}, D.~H., {Mortonson}, M.~J., {Eisenstein}, D.~J., {et~al.} 2013,
  \physrep, 530, 87

\bibitem[{{Zhang} \& {Komatsu}(2011)}]{Z11}
{Zhang}, J., \& {Komatsu}, E. 2011, \mnras, 414, 1047

\bibitem[{{Zhang} {et~al.}(2017){Zhang}, {Zhang}, \& {Luo}}]{Z17}
{Zhang}, J., {Zhang}, P., \& {Luo}, W. 2017, \apj, 834, 8

\bibitem[{{Zuntz} {et~al.}(2018){Zuntz}, {Sheldon}, {Samuroff}, {Troxel},
  {Jarvis}, {MacCrann}, {Gruen}, {Prat}, {S{\'a}nchez}, {Choi}, {Bridle},
  {Bernstein}, {Dodelson}, {Drlica-Wagner}, {Fang}, {Gruendl}, {Hoyle}, {Huff},
  {Jain}, {Kirk}, {Kacprzak}, {Krawiec}, {Plazas}, {Rollins}, {Rykoff},
  {Sevilla-Noarbe}, {Soergel}, {Varga}, {Abbott}, {Abdalla}, {Allam}, {Annis},
  {Bechtol}, {Benoit-L{\'e}vy}, {Bertin}, {Buckley-Geer}, {Burke}, {Carnero
  Rosell}, {Carrasco Kind}, {Carretero}, {Castander}, {Crocce}, {Cunha},
  {D'Andrea}, {da Costa}, {Davis}, {Desai}, {Diehl}, {Dietrich}, {Doel},
  {Eifler}, {Estrada}, {Evrard}, {Fausti Neto}, {Fernandez}, {Flaugher},
  {Fosalba}, {Frieman}, {Garc{\'\i}a-Bellido}, {Gaztanaga}, {Gerdes},
  {Giannantonio}, {Gschwend}, {Gutierrez}, {Hartley}, {Honscheid}, {James},
  {Jeltema}, {Johnson}, {Johnson}, {Kuehn}, {Kuhlmann}, {Kuropatkin}, {Lahav},
  {Li}, {Lima}, {Maia}, {March}, {Martini}, {Melchior}, {Menanteau}, {Miller},
  {Miquel}, {Mohr}, {Neilsen}, {Nichol}, {Ogando}, {Roe}, {Romer}, {Roodman},
  {Sanchez}, {Scarpine}, {Schindler}, {Schubnell}, {Smith}, {Smith},
  {Soares-Santos}, {Sobreira}, {Suchyta}, {Swanson}, {Tarle}, {Thomas},
  {Tucker}, {Vikram}, {Walker}, {Wechsler}, {Zhang}, \& {DES
  Collaboration}}]{Zuntz2018}
{Zuntz}, J., {Sheldon}, E., {Samuroff}, S., {et~al.} 2018, \mnras, 481, 1149

\end{thebibliography}


\appendix
\section{$\chi^2$ and $p$ values for null tests}
\label{app:1}

\begin{table*}
\caption{
    The $\chi^2$ and $p$ values for null hypothesis of stacked tangential shear
    profiles.
    }
\begin{center}
\begin{tabular}{cllll}
  \hline
  Field & Random & GAIA bright & GAIA intermediate & GAIA faint \\
  & $\chi^2$ ($p$ value) & $\chi^2$ ($p$ value) & $\chi^2$ ($p$ value) & $\chi^2$ ($p$ value) \\ \hline
     XMM & 24.76 (0.74) & 17.26 (0.75) & 27.01 (0.46) & 48.97 (0.02) \\
 GAMA09H & 40.85 (0.09) & 21.67 (0.48) & 24.91 (0.58) & 25.01 (0.72) \\
 WIDE12H & 27.79 (0.58) & 27.78 (0.18) & 26.60 (0.48) & 27.50 (0.60) \\
 GAMA15H & 25.60 (0.70) & 22.42 (0.43) & 28.70 (0.38) & 25.41 (0.70) \\
    VVDS & 29.38 (0.50) & 11.92 (0.96) & 25.57 (0.49) & 24.56 (0.75) \\
HECTOMAP & 22.72 (0.83) & 28.51 (0.16) & 30.54 (0.29) & 26.49 (0.65) \\
     ALL & 28.86 (0.53) & 18.28 (0.69) & 25.81 (0.53) & 21.96 (0.86) \\
     \hline
\end{tabular}
\end{center}
\label{table:stack_e}
\end{table*}

\begin{table*}
\caption{
    The $\chi^2$ and $p$ values for null hypothesis of stacked
    cross shear profiles.
    }
\begin{center}
\begin{tabular}{clllll}
  \hline
  Field & CMASS & Random & GAIA bright & GAIA intermediate & GAIA faint \\
  & $\chi^2$ ($p$ value) & $\chi^2$ ($p$ value) & $\chi^2$ ($p$ value)
  & $\chi^2$ ($p$ value) & $\chi^2$ ($p$ value) \\ \hline
     XMM & 26.85 (0.63) & 26.29 (0.66) & 23.85 (0.35) & 33.63 (0.18) & 29.85 (0.47) \\
 GAMA09H & 32.07 (0.36) & 27.32 (0.61) & 17.65 (0.73) & 44.73 (0.02) & 17.96 (0.96) \\
 WIDE12H & 28.25 (0.56) & 24.68 (0.74) & 19.98 (0.58) & 17.25 (0.93) & 32.28 (0.35) \\
 GAMA15H & 25.64 (0.69) & 29.51 (0.49) & 18.46 (0.68) & 23.69 (0.65) & 39.82 (0.11) \\
    VVDS & 27.22 (0.61) & 29.93 (0.47) & 22.66 (0.42) & 47.86 (0.01) & 19.91 (0.92) \\
HECTOMAP & 17.60 (0.96) & 29.60 (0.49) & 19.96 (0.58) & 14.85 (0.97) & 15.34 (0.99) \\
     ALL & 31.88 (0.37) & 48.70 (0.02) & 25.35 (0.28) & 32.02 (0.23) & 22.67 (0.83) \\
     \hline
\end{tabular}
\end{center}
\label{table:stack_b}
\end{table*}

\end{document}